\documentstyle[12pt,bezier,german]{report}

\let\a=\alpha \let\b=\beta \let\g=\gamma \let\d=\delta \let\e=\epsilon
  \let\h=\eta \let\th=\theta
 \let\l=\lambda \let\m=\mu \let\n=\nu \let\x=\xi

\def\k{\vec{k}} \def\p{\vec{p}} \def\q{\vec{q}} 

\let\r=\rho \let\s=\sigma 
\def\t{\tilde{t}} \let\o=\omega 
    \let\O=\Omega
\let\S=\Sigma \let\Th=\Theta \let\L=\Lambda \let\G=\Gamma
\let\D=\Delta  
\def\2{{1\over2}} \def\4{{1\over4}} \def\52{{5\over2}} \def\6{\partial }

\def\CL{{\cal L}}
\def\CP{{\cal P}}

\def\CV{{\cal V}}

\def\({\left(} \def\){\right)} \def\<{\langle } \def\>{\rangle }

\def\beg{\begin{equation}}
\def\begar{\begin{eqnarray}}
\def\ee{\end{equation}}
\def\ea{\end{eqnarray}} \def\nn{\nonumber}

\renewcommand{\Im}{\mbox{Im}}       
\newcommand{\pss}{p\!\!\!/\,'}      
\newcommand{\ps}{p\!\!\!/}         

\input epsf
\newcommand{\pref}[1]{\ref{#1}}                
\newcommand{\plabel}[1]{\label{#1}}              
\newcommand{\pcite}[1]{\cite{#1}}                
\newcommand{\pbib}[1]{\bibitem{#1}}              

\begin{document}            
\begin{titlepage}
\begin{center}
{\Large
DISSERTATION} \\
\vfill
\vfill
zum Thema \\
\vfill
\vfill
\begin{LARGE}
\begin{bf}
Quantum Field Theory Near \\
Thresholds of Ultra-Heavy Particles: Top-Antitop and Beyond.
\end{bf}
\end{LARGE}
\vfill
\vfill
ausgef\"uhrt zum Zwecke der Erlangung des akademischen Grades\\
Doktor der technischen Wissenschaften
\vfill
eingereicht an der Technischen Universit\"at Wien \\
Technisch-Naturwissenschaftliche Fakult\"at \\
\vfill
von \\
\vfill
Dipl.-Ing.~Wolfgang M\"odritsch \\
\vfill
Tullnerbachstr. 11/2/39, 3002 Purkersdorf \\
Matrikelnummer 8626176 \\
geboren am 6. Oktober 1966 in Wien
\end{center}
\vfill
\vfill
\vfill
Wien, im August 1995 \hfill \hfill \dotfill \hfill \\
\end{titlepage}

\newpage

\vspace*{9cm}
\thispagestyle{empty}
\centerline{\it F\"ur meine Familie}

\newpage
\Large
Kurzfassung\\
\\
\normalsize 
\begin{sloppypar}

Das Standardmodell der elektroschwachen und starken Wechselwirkung (SM)
be\-schreibt bis heute fast alle experimentellen Daten innerhalb der 
experimentellen und theo\-retischen Unsicherheiten. Zu seiner inneren 
Konsistenz, dh. im speziellen zur Anomalie\-freiheit ben\"otigt das SM
ein sechstes, zu Beginn dieser Arbeit noch nicht entdecktes Quark, das
Topquark ($t$). In den letzten beiden Jahren (1994-95) wurde dann die Evidenz
f\"ur seine tats\"achliche Existenz durch Experimente am Hadronbeschleuniger 
Tevatron immer st\"arker, soda\3 heute der Nachweis bei einer Masse 
von ca. 180 GeV  als gesichert gilt.

Der ideale Platz f\"ur eine genauere Untersuchung der Eigenschaften des 
Topquarks w\"are aber ein $e^+e^-$-Beschleuniger. Dort k\"onnte man 
Top-Antitop Paare ($t\bar{t}$) nahe der Schwelle erzeugen und so seine Masse, 
Zerfallsbreite und die Kopplungen zu anderen Elementar\-teilchen genau 
bestimmen. Durch die hohe Masse und den raschen Zerfall in $W^+$ und 
$b$ besitzt das Topquark die einzigartige Eigenschaft unter den Quarks, 
da\3 sein Verhalten rein st\"orungstheoretisch bestimmt werden kann.
Die st\"orungstheoretische Behandlung der Bindungszustandseffekte nahe der
Schwelle f\"ur die Erzeugung ist Gegenstand der vorliegenden Arbeit.

Eine der einfachsten Erweiterungen des SM ist das sog. Minimale 
Supersymmetrische SM (MSSM). Die darin vorhergesagten skalaren Partner zu 
den existierenden Fermionen k\"onnten in \"ahnlicher Weise erzeugt werden 
und werden daher in dieser Arbeit auch behandelt.

Im Kapitel 2 werden zuerst sowohl f\"ur instabile Fermionen
als auch f\"ur Skalare Bindungszustandsgleichungen angegeben, die als Grundlage 
f\"ur eine systematische St\"orungstheorie dienen. Als Beispiel wird 
das Massenspektrum zweier gebundener Skalare zu $O(\a^4)$ berechnet. 
Dann wird das $t\bar{t}$-Potential in konsistenter Weise zu numerischer 
Ordnung $O(\a_s^4)$ be\-stimmt. Dabei wurden ein neuer Box-graph 
berechnet sowie Beitr\"age der elektroschwachen Wechselwirkung 
ber\"ucksichtigt.
Im Rest des Kapitels wird dann die Frage einer m\"oglicher\-weisen gro\3en
Korrektur zum Wirkungsquerschnitt auf Grund einer "laufenden" 
Zerfallsbreite behandelt. Es kann gezeigt werden, da\3 ein bisher nicht
ber\"ucksichtigter Beitrag zu gro\3en Kompensationen f\"uhrt. Durch die
Einf\"uhrung der neuen Gleichung f\"ur instabile Fermionen und der 
Verwendung einer Wardidentit\"at l\"a\3t sich schlie\3lich die Rechnung
soweit vereinfachen, da\3 der allgemeine Mechanismus dieser Kompensationen
sichtbar wird. Dies f\"uhrt zu dem Theorem, da\3 f\"ur alle nicht durch 
Annihilation zerfallenden schwach gebundenen Systeme die 
Bindungszustandskorrekturen zur Zerfallsbreite allein durch die 
Zeitdilatation beschrieben werden k\"onne.

Kapitel 3 besch\"aftigt sich mit der Berechnug der Wirkungsquerschnitte 
nahe der Schwelle mit numerischen Methoden. Zuerst wird der allgemeine 
Formalismus beschrieben und der totale Wirkungsquerschnitt f\"ur die 
$t\bar{t}$ Produktion diskutiert. Dabei zeigt sich unter anderem die 
Wichtigkeit der gro\3en Zerfallsbreite f\"ur eine st\"orungstheoretische 
Behandlung einerseits sowie die Bedeutung der Resummierung im 
Gluonpropagator anderer\-seits. Dann wird die 
Vorw\"artsr\"uckwertsasymmetrie
nahe der Schwelle erstmals ohne die Einf\"uhrung eines k\"unstlichen 
Cut-offs berechnet. Die dabei verwendete Methode l\"a\3t sich auch auf
die Berechnung des axialen Beitrags zum totalen Wirkungsquerschnitts beim
$t\bar{t}$-System und die Bestimmung des f\"uhrenden Beitrags zur
Stop-Antistopproduktion anwenden.

Konnten in dieser Arbeit wichtige Aspekte der Produktion schwerer, 
stark wechsel\-wirkender Teilchen nahe der Schwelle gekl\"art werden, 
so bleiben doch noch einige theo\-retische Herausforderungen f\"ur
die Zukunft. Dabei sei speziell die Anwendung der ri\-gorosen Methoden 
aus Kap.2 auf die Berechnung physikalischer Observablen wie in Kap.3
erw\"ahnt. Dies wird notwendig werden, um Theorie und Experiment mit 
Genauigkeiten von kleiner als 1\% zu vergleichen.
\end{sloppypar}

\newpage

\vspace*{7cm}
An dieser Stelle m\"ochte ich dem Institut f\"ur Theoretische Physik
f\"ur die freundliche Aufnahme und meinen Freunden und Kollegen
f\"ur viele wertvolle Diskussionen danken.
Ganz besonderer Dank gilt auch meinem Lehrer Herrn O.Univ.Prof. Dr.
W.Kummer f\"ur seine Unterst\"utzung und viele wertvolle
Diskussionen.

Meinen Eltern und Schwiegereltern sowie meiner Frau Elaine m"ochte ich 
ganz herzlich f"ur ihre Unterst"utzung und Geduld danken.           
\originalTeX             

\tableofcontents

\chapter{Introduction}

Perturbative expansions in the coupling constant in quantum field theory
possess two types of applications, the calculation of scattering
amplitudes
and the computation of processes involving weakly bound systems.  Many of
the successes of quantum electrodynamics (QED) are, in fact, related to
positronium, i.e. to the second one of the aforementioned applications.
The proper starting point for any bound-state calculation in quantum field
theory is an integral equation, comprising an infinite sum of Feynman
graphs. The Bethe-Salpeter (BS) equation
\pcite{Bethe} fulfills this task and it is well known that in the limit of
binding energies of $O(\a^2m)$ the Schr\"odinger equation with static
Coulomb attraction is obtained. The computation of higher order
corrections to the Bohr-levels, however, turned out to be far from
trivial. It was recognized, though, relatively late that, at least
conceptually, substantial progress with respect to a systematic treatment
results from a consistent use of a perturbation theory geared to the
original BS equation \pcite{Lep77}. As an additional bonus one
avoids nonrelativistic expansions as implied by Hamiltonian approaches with
successive Fouldy-Wouthuysen transformations \pcite{Fein}.
Within the BS-technique, however, it is desirable to have an exactly
solvable zero order equation different from the Schr\"odinger equation,
because otherwise e.g. the approximation procedure for the wave function lacks
sufficient transparency, especially in higher orders and may even break 
down in certain cases, one of which is treated here.
One of the
advantages of the BS approach to perturbation theory is the freedom to
select a different zero order equation. Of course, in that case, certain
corrections included already at the zero level are to be properly
subtracted out in higher orders. An especially useful zero order equation
has been proposed some time ago by Barbieri and Remiddi (BR equation
\pcite{BR}). Still, one of the most
annoying features of all bound-state calculations remains the pivotal rule
played by the Coulomb gauge. In other gauges, e.g. already the (in QED
vanishing) corrections to $O(\a^3m)$ of the Bohr levels imply to take into
account an infinite set of Feynman graphs \pcite{Love}. Only in very
special cases, when certain subsets of graphs can be shown to represent
a gauge-independent correction, another more suitable gauge may
be chosen.\\
By contrast to QED the vast literature on bound state
problems in quantum chromodynamics (QCD) adheres to a description of the
quark-antiquark system by the Schr\"odinger equation with corrections
'motivated' by QCD \pcite{ph}. As long as a relatively small number of
parameters suffices for an adequate phenomenological description of
observed quantum levels, this approach undoubtedly has an ample practical
justification. Furthermore even for the relatively heavy flavors charm
and bottom nonperturbative effects can be shown to be of the same order
of magnitude as the leading perturbative ones. Apart from
lattice calculations there exists no method to do bound state calculations
in a
non-perturbative way. Only if the non-perturbative contributions are
smaller than the leading perturbative ones it is known how to include them
in a systematic BS pertubation theory \pcite{Leut,Shif}.
In the case of QCD this requires particles with masses in the range of
100GeV. Especially the recent evidence for the top quark with a mass
around 170 GeV seems to indicate that this particle is the ideal testing
ground for perturbative QCD bound state calculations. The cleanest
environment to observe the top quark clearly would be a next generation
$e^+ e^-$ linear collider. On such a machine other heavy particles with
strong interaction could be discovered as well. If supersymmetry
is realized in nature it seems very likely that one of the scalar partners
of either top or bottom could be observed. Since it may well be that
even in, say, a 500GeV collider those new scalar particles (and also the top
quark) will be produced near their threshold, a reliable prediction of the production
cross section slightly above threshold seems very important.
The signal of the top quark is  expected to be best suited for an exact
determination of its mass, its width and an independent measurement of the
strong coupling constant.

Thus also for this reason a return to more rigorous QCD arguments remains as desirable
as ever. The standard literature on quarkonia (see e.g. \pcite{Buchm})
is almost exclusively
based on nonrelativistic expansions \pcite{Fein} or on the calculation of
purely static forces \pcite{Fisch}. Moreover, very often potentials
with higher order corrections as determined from on-shell quarkonia
scattering are used \pcite{scatt}. In these cases relevant off-shell
effects which are typical for higher order corrections may even be lost
altogether .
On the other hand, from the point of view of relativistic quantum field
theory as elaborated in the abelian case of positronium, a similar, more
systematic approach seems desirable, the more so because the basic
techniques are well developed. In addition, at least in one case, namely
the decay of S-wave quarkonium, the result of a full BS-perturbation
calculation \pcite{KumW}, including the QCD corrections to the bound state
wave function, yields a result very different from the one which
took into account only the corrections to the quark antiquark annihilation
alone \pcite{BC}.

In this context the relatively large size of the running coupling
constant even at high energies represents a well known problem, together
with large coefficients from a perturbative expansion. Therefore
e.g. problems arise in the comparison of the coupling constant as
determined from scattering experiments within the minimal subtraction
scheme ($\overline{MS}$), with the coupling constant to be used in a
consistent weak bound-state approach. The philosophy within our present
work will be that the orders of magnitude, as determined from
$\a_{\overline{MS}}$ will be used for estimates, but that we shall imply a
determination of $\a_s$ by some physical observable of the quarkonium system itself.
In that way delicate correlations of 'genuine' orders of $\a_s$ from {\it
basically} different types of experiments are avoided.

From high
precision electro-weak experiments of the LEP collaborations, the mass
range of the top quark now seems to be established to lie in the range
160-200$GeV$ \pcite{ALEPH}, in agreement with the direct searches at the
tevatron \pcite{CDF}. Thus for the first time a nonabelian bound
state quarkonium system seems to fulfill the high mass criterion
required for a genuine field theoretical approach. Unfortunately the
drawback of this
situation is that the weak decay $t \to b+W$ broadens the energy levels
\pcite{Kuehn} for increasing mass
$m_t$ so that above $m_t \approx 120 GeV$ individual levels effectively
disappear.

But this disadvantage turns into a virtue since it completely eliminates
nonperturbative and renormalon contributions. The reason for this can be stated
as follows. The time scale $\L_{QCD}^{-1}$ necessary to allow strong
interactions to become effective and to produce e.g. hadrons with open top
cannot compete with the decay time $\G_0^{-1}$. Therefore the top quark decays
before nonperturbative strong interactions can act on it.
In this situation "pure"
quantum chromodynamic (QCD) perturbation theory is sufficient for a
complete description of a strongly interacting system. This applies as
well to Coulombic QCD bound-state effects in toponium although also there
the hydrogen-like levels are strongly smeared out by $\G_0$. Nevertheless,
the computation of those levels represents a necessary first step since
the order of magnitude of these corrections can be used as a order
parameter for the determination of the quark-antiquark potential.
Furthermore we will see in chapter 2 of
this work that a lot of insight into the physics of toponium can be gained
from the study of the properties of the four point function at the bound state
poles.

After all, this is the first instance where the application of quantum
field theory for weakly bound nonabelian systems is justified because even
nonperturbative effects can be taken into account in an equally
well-defined perturbative manner.

A correct formulation of QCD in Coulomb gauge entails not only
Faddeev-Popov-ghost terms but also the inclusion of nonlocal interaction
terms \pcite{SCL}. Therefore, the full Lagrangian reads ($a$=1,...,8 for
SU(3)):
\begin{equation} \plabel{Lag}
{\cal L} = -\frac{1}{4} F_{\m\n}^{a} F^{a \m\n} + \sum_{j=1}^{f}
\bar{\Psi}_j (i \g D -m_j) \Psi_j +
B^a (\6_j A_j^a)-\bar{\h}^a \6^i (\d_{ab} \6_i + g f_{abc} A_i^c) \h^b +
v_1 + v_2
\end{equation}
where the Lagrange multiplier $B^a$ imposes the Coulomb gauge condition,
and where
\begar
D_{\m} &=& \6_{\m} - i g T^a A_{\m}^a, \\ F_{\m \n}^a &=& \6_{\m} A_{\n}^a
-\6_{\n} A_{\m}^a + g f_{abc} A_{\m}^b A_{\n}^c.
\ea
$v_1$ and $v_2$ are given in \pcite{SCL} and will be discussed more
explicitly below.  The above Lagrangian will include all effects of the
strong interaction, but, as we will show, QED and weak corrections may
also give contributions within the numerical order of our main interest.
This effects will be discussed within the framework of the Standard Model
(SM). With respect to possible extensions of the SM we will restrict
ourselves
to the Minimal Supersymmetric SM (MSSM) where we especially focus on the
occurrence of new heavy scalars like the stop and the lightest (uncharged)
higgs boson.

The thesis is organized as follows. In chapter 2 we first construct a solvable
zero order equation for decaying particles (sect. (2.1)-(2.2)).
Then in sect. (2.4) we calculate within the framework of Bethe-Salpeter perturbation
theory (sect.2.3) the potential for the top-antitop ($t\bar{t}$) system to
numerical order $O(\a_s^4)$ .
As a further application of these methods we derive the bound state
corrections to the toponium decay width in sect. (2.5).

Chapter 3 is devoted to the calculation of cross sections to be measured
in a next generation linear collider. Within the Green-function approach
we study some consequences of the results of chapter 2 for the total cross
section (sect. 3.1.1). Furthermore in sect. 3.1.2 we propose
a method to circumvent an unphysical singularity in the calculation of the
forward-backward asymmetry. The same method may also be applied to the
calculation of the axial contribution to the $t\bar{t}$ total cross
section, as well as to the production of heavy scalar particles near
threshold (sect. 3.2).

{\bf Acknowledgement}: 
This work has been supported partly by the Austrian Science
Foundation (FWF), project P10063-PHY within the framework of the EEC- Program
"Human Capital and Mobility", Network "Physics at High Energy Colliders",
contract CHRX-CT93-0357 (DG 12 COMA). I would like to greatfully
acknowledge helpful discussions with M.~Je\.zabek, V.A.~Khoze, J.H.~K\"uhn, 
T.~Teubner, Y.~Sumino and a collaboration with A.~Vairo.

\chapter{Application of Rigorous Bound State Methods to
         Toponium}

\section{A Relativistic Equation for Decaying Fermions}
\plabel{Form}

In this section we will present a solvable relativistic equation for
decaying fermions, similar in form to that for stable fermions of
Barbieri and Remiddi \pcite{BR}.
While this equation is applicable to variety of
bound state problems we will focus especially on the $t\bar{t}$ system.

The Bethe-Salpeter (BS) approach for weakly bound systems starts from
the BS-equation
\newline
\unitlength1.0cm
\begin{picture}(18,5)
  \put(.5,3.5){\line(1,0){1.5}}                   \put(3,3.5){\line(1,0){1.5}}
  \put(.5,2.5){\line(1,0){1.5}} \put(2.5,3){\circle{1.5}} \put(3,2.5){\line(1,0){1.5}}
  \put(1.5,3.5){\vector(-1,0){0.3}}               \put(4,3.5){\vector(-1,0){0.3}}
  \put(1,2.5){\vector(1,0){0.3}}               \put(3.5,2.5){\vector(1,0){0.3}}
  \put(1.1,3.7){$p_1$}                            \put(3.7,3.7){$\bar{p_1}$}
  \put(1.1,2.2){$p_2$}                            \put(3.7,2.2){$\bar{p_2}$}
  \put(0.2,3.4){$i$}          \put(4.6,3.4){$\bar{i}$}
  \put(0.2,2.4){$j$}          \put(4.6,2.4){$\bar{j}$} \put(5,2.9){=}

  \put(5.5,3.5){\line(1,0){1}} \put(6.75,3.5){\circle{.5}} \put(7,3.5){\line(1,0){1}}
  \put(5.5,2.5){\line(1,0){1}} \put(6.75,2.5){\circle{.5}} \put(7,2.5){\line(1,0){1}}
  \put(6.3,3.5){\vector(-1,0){0.3}}               \put(7.7,3.5){\vector(-1,0){0.3}}
  \put(5.7,2.5){\vector(1,0){0.3}}               \put(7.4,2.5){\vector(1,0){0.3}}
  \put(5.9,3.7){$p_1$}                            \put(7.5,3.7){$\bar{p_1}$}
  \put(5.9,2.2){$p_2$}                            \put(7.5,2.2){$\bar{p_2}$}
  \put(5.2,3.4){$i$}          \put(8.1,3.4){$\bar{i}$}
  \put(5.2,2.4){$j$}          \put(8.1,2.4){$\bar{j}$} \put(8.75,2.9){+}

  \put(9.5,3.5){\line(1,0){.5}} \put(10.25,3.5){\circle{.5}} \put(10.5,3.5){\line(1,0){1}}
  \put(9.5,2.5){\line(1,0){.5}} \put(10.25,2.5){\circle{.5}} \put(10.5,2.5){\line(1,0){1}}
  \put(10,3.5){\vector(-1,0){0.3}}               \put(11.2,3.5){\vector(-1,0){0.3}}
  \put(9.5,2.5){\vector(1,0){0.3}}               \put(10.7,2.5){\vector(1,0){0.3}}
  \put(12.5,3.5){\line(1,0){1}} \put(12,3){\circle{1.5}} \put(12.5,2.5){\line(1,0){1}}
  \put(14.5,3.5){\line(1,0){1}} \put(14,3){\circle{1.5}} \put(14.5,2.5){\line(1,0){1}}
  \put(13.2,3.5){\vector(-1,0){0.3}}               \put(15.2,3.5){\vector(-1,0){0.3}}
  \put(12.7,2.5){\vector(1,0){0.3}}               \put(14.7,2.5){\vector(1,0){0.3}}
  \put(9.6,3.7){$p_1$}                            \put(14.7,3.7){$\bar{p_1}$}
  \put(9.6,2.2){$p_2$}                            \put(14.7,2.2){$\bar{p_2}$}
  \put(9.3,3.4){$i$}          \put(15.6,3.4){$\bar{i}$}
  \put(9.3,2.4){$j$}          \put(15.6,2.4){$\bar{j}$}
  \put(2.5,1){G} \put(5,1){=}\put(6.75,1){D} \put(8.75,1){+}\put(10.25,1){D}
  \put(12,1){K}\put(14,1){G}    \put(11.9,2.85){K}
\end{picture}
\beg \plabel{BS} \ee
for the two fermion Green function $G(P,p,p')$,
where $D$ is the product of the two (full) propagators and $K$ represents
the 2pi BS-kernel. All four point functions in \pref{BS} depend on the
total momentum $P = (P_0,\vec{0})$; the incoming (outgoing) lines carry
momentum $P/2 \pm p$ ($P/2 \pm p'$).

Near a bound state pole the Green function assumes the form
\beg \plabel{Pol}
  G_{ij,\bar{i}\bar{j}}(p_1,p_2,\bar{p_1},\bar{p_2}) = (2\pi)^4 \d^4(\bar{P}-P)
     \sum_{l} \chi_{ij}(p,n,l)\frac{i}{P_0 - M_n} 
\bar{\chi}_{\bar{i}\bar{j}}(\bar{p},n,l) + G_{reg}
\ee
where $\chi_{ij}(p,n,l)$ is called BS- or bound state wave function.

Perturbation theory starts from an equation  similar to eq. \pref{BS}
with $G_0,D_0$ and $K_0$ chosen in such a way that the exact solution is known.
If $D_0$ consists of the nonrelativistic propagators only and if $K_0$ is
the Coulomb kernel the zero order equation corresponding to \pref{BS}
(after integrating out
$p_0$, $p'_0$ etc. ) simply reduces to the Schr\"odinger
equation. A very convenient zero order equation which already includes
the relativistic free fermion propagators and still remains solvable
has been found by Barbieri and Remiddi (BR) some time ago \pcite{BR}.
However if the decay width of the constituents becomes comparable to the
binding energy of $O(\a ^2 m)$, perturbation theory runs into troubles because of the
occurrence of terms $\G/(\a^2 m)$ in graphs like fig. \pref{dmultide},
containing a chain of subgraphs which are responsible for the decay.
To circumvent these difficulties one has to include at least a part
of the exact self energy function in the zero order equation.

\begin{center}
\leavevmode
\epsfxsize=8cm
 \epsfbox{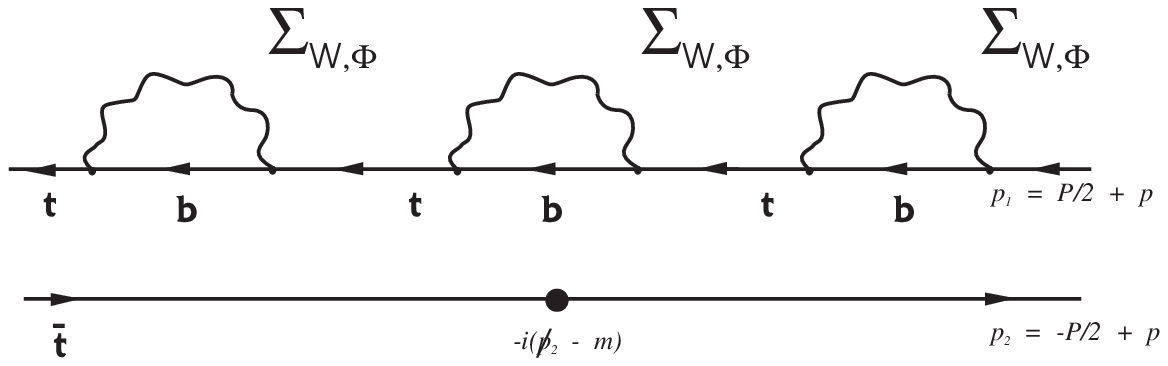}
\\
\centerline{ Fig. \refstepcounter{figure} \plabel{dmultide} \thefigure}
\end{center}

The BS equation in terms of Feynman amplitudes for $K$ and $S$ reads for
a bound state wave function $\chi$
\beg \plabel{allg}
\chi_{ij}^{BS}(p;P) = -i S_{ii'}(\frac{P}{2}+p)S_{j'j}(-\frac{P}{2}+p)
\int \frac{d^4p'}{(2\pi)^4}         K_{i'j',i''j''}(P,p,p')
\chi_{i''j''}^{BS}(p';P),
\ee
where $S$ is the exact fermion propagator,
and $K$ is the sum of all two fermion irreducible graphs. Furthermore, we
have introduced relative momenta $p$ and $p'$, a total momentum
$P=p_1-p_2$, and we choose the center of mass (CM) frame where
$P=(P_0,\vec{0})=(2m+E,\vec{0})$. The Green function
reads in terms of field operators
\begar
G_{ij,\bar{i}\bar{j}}(p_1,p_2,\bar{p_1},\bar{p_2}) &:=& \int d^4x_1 d^4\bar{x_1} d^4x_2 d^4\bar{x_2}
         e^{i(p_1x_1+\bar{p_2}\bar{x_2}-p_2x_2-\bar{p_1}\bar{x_1})}
G_{i,j;\bar{i},\bar{j}}(x_1,x_2,\bar{x_1},\bar{x_2}) \nonumber \\
 G_{ij,\bar{i}\bar{j}}(x_1,x_2,\bar{x_1},\bar{x_2}) &:=& \<0|T \bar{\Psi}_i(x_1) \Psi_j(x_2) \Psi_{\bar{i}}(\bar{x_1})
                                  \bar{\Psi}_{\bar{j}}(\bar{x_2})|0\>
\ea
In terms of Feynman amplitudes and the relative coordinates this can be written as
\begar
G_{ij,\bar{i}\bar{j}}(p_1,p_2,\bar{p_1},\bar{p_2}) &=&
 i (2\pi)^4 \d(P-\bar{P}) G_{ij,\bar{i}\bar{j}}(P,p,p') \plabel{Gfeyn} \\
G_{ij,\bar{i}\bar{j}}(P,p,p') &=& \int d^4X d^4x d^4\bar{x}
 e^{i(XP+xp-\bar{x}\bar{p})} \<0|T \bar{\Psi}_i(X+\frac{x}{2})
\Psi_j(X-\frac{x}{2}) \Psi_{\bar{i}}(-\frac{\bar{x}}{2})
                           \bar{\Psi}_{\bar{j}}(\frac{\bar{x}}{2})|0\>
\nonumber
\ea
The notation can be read off the
pictorial representation in fig \pref{fig1}. $i,j$ are collective indices for spin
($\s,\r$) and color (noted $\a,\b$).

\unitlength1.0cm
\begin{center}
\begin{picture}(12.5,4)
  \put(.5,2.5){\line(1,0){1.5}}                        \put(3.24,2.05){\line(1,0){1.05}}
  \put(.5,1.5){\line(1,0){1.5}} \put(2.5,2){\circle{1.5}}\put(3.24,1.95){\line(1,0){1.05}}
  \put(1.5,2.5){\vector(-1,0){0.3}}    \put(2.4,1.9){$\chi$}
  \put(1,1.5){\vector(1,0){0.3}}
  \put(1.1,2.7){$p_1$}                            \put(3.7,2.4){$P$}
  \put(1.1,1.2){$p_2$}                        \put(4,2.2){\vector(-1,0){0.3}}
  \put(0.2,2.4){$i$}
  \put(0.2,1.4){$j$}           \put(5,1.9){=}

  \put(6.5,2.5){\line(1,0){.5}} \put(7.25,2.5){\circle{.5}} \put(7.5,2.5){\line(1,0){1}}
  \put(6.5,1.5){\line(1,0){.5}} \put(7.25,1.5){\circle{.5}} \put(7.5,1.5){\line(1,0){1}}
  \put(7,2.5){\vector(-1,0){0.3}}               \put(8.2,2.5){\vector(-1,0){0.3}}
  \put(6.5,1.5){\vector(1,0){0.3}}               \put(7.7,1.5){\vector(1,0){0.3}}
  \put(9.5,2.5){\line(1,0){1}} \put(9,2){\circle{1.5}} \put(9.5,1.5){\line(1,0){1}}
                               \put(8.9,1.9){$K$}
  \put(11.74,2.05){\line(1,0){1.05}} \put(11,2){\circle{1.5}} \put(11.74,1.95){\line(1,0){1.05}}
                                     \put(10.9,1.9){$\chi$}
  \put(10.2,2.5){\vector(-1,0){0.3}}               \put(12.3,2.2){\vector(-1,0){0.3}}
  \put(9.7,1.5){\vector(1,0){0.3}}                \put(12,2.3){$P$}
  \put(6.6,2.7){$p_1$}                            \put(9.7,2.7){$p_1'$}
  \put(6.6,1.2){$p_2$}                            \put(9.7,1.2){$p_2'$}
  \put(6.3,2.4){$i$}          \put(10.15,2.7){$i'$}
  \put(6.3,1.4){$j$}          \put(10.15,1.2){$j'$}
   \end{picture}

\vspace{-.5cm}
Fig. \refstepcounter{figure} \plabel{fig1} \thefigure
\end{center}

It is well known that the dominant part in $K$ for weak binding ($\a \to
0$) is the one-Coulomb gluon exchange which results in an ordinary
Schr\"odinger equation with static Coulomb potential. With the help of
the nonrelativistic scaling argument \pcite{Kum}
\begar p_0 &\approx& O(m\a^2), |\p|
\approx O(m\a),  \plabel{scal}\\
P_0 &\approx& 2m-O(m\a^2) \nn
\ea
this result is
even independent of the chosen gauge: \beg \plabel{Kc}   K
\to K_c \g^0 \otimes \g^0 := -\frac{4 \pi \a}{(\p-\p\,')^2} \g^0_{\s \s'}
\g^0_{\r' \r} \ee Focusing on QCD we notice the fact that only color
singlet states can form bound states because the Coulomb potential is
repulsive for color octets. The color trace will always be understood
to be already done, leading to the definition \beg  \a \equiv \frac{4}{3}
\frac{g^2}{4\pi} = \frac{4}{3} \a_s\plabel{alpha} \ee to be used in the
following, in terms of the usual strong coupling constant $\a_s$.
Because the above mentioned nonrelativistic limit of the BR equation
contains the projection operators $\l^\pm$, defined below, it is awkward
to calculate the so-called relativistic corrections in a straightforward
way within the framework of BS perturbation theory, starting from
\pref{allg} with \pref{Kc}. Therefore we will use the generalized  BR
equation as described below instead of the Schr\"odinger
equation.

Moreover, for our present case, we need a generalization of the equation
given in \pcite{BR} for unstable fermions, described by complex
$m \to \tilde{m}=m-i \hat{\G}_0/2$ where $\hat{\G}_0/2$ represents
the imaginary part of the self energy graph responsible for the weak
decay of the free top quark. Since the real part of $\tilde{m}$ should
still determine the sum of polarizations in the numerator of
the propagator, we take
by analogy to the stable case in the zero order approximation to
\pref{allg} the relativistic free propagator ($ E_p = \sqrt{\p\,^2+m^2}, \h = m
\hat{\G}_0/2 E_p$)

\begar \plabel{Sg}
  S(\pm \frac{P_0}{2}+p_0) &\to& [(\pm \frac{P_0}{2}+p)\g - \tilde{m}]^{-1}=\\
             & & =  \frac{\L^+ \g_0}{\pm \frac{P_0}{2}+p_0-E_p+i \h} +
       \frac{\L^- \g_0}{\pm \frac{P_0}{2}+p_0-E_p-i \h} + O(\frac{\hat{\G}_0}{m})
\ea
with the relativistic projectors
\beg
 \L^{\pm}(\p) \equiv \frac{E_p \pm (\vec{\a} \p + \b m)}{2E_p}. \plabel{L}
\ee
If furthermore $\6 K_0/\6 p_0 = 0$ both sides of \pref{allg} may be
integrated with respect to $p_0$. On the r.h.s. the product of the two
propagators $S$ with \pref{Sg} yield four terms, with Cauchy poles determined
by $i\h$, two of which give no contribution.
Generically we obtain
\beg \plabel{D}
 \int \frac{d p_0}{2 \pi i} S \otimes S = \frac{\L^+ \g_0 \otimes \L^-
\g_0}{P_0-2E_p+2 i \h} -
          \frac{\L^- \g_0 \otimes \L^+ \g_0 }{P_0+2E_p-2i\h}
\ee
so that $ K_{0} (\p, \p\,') \Phi_{0}$ with
\beg
  \Phi_{0} = \frac{1}{2 \pi} \int \chi_{0} dp_0
\ee
remains to be inserted at the places of "$\otimes$". Written with indices the
direct product notation can be explained as: $A \otimes B := A_{\a \a'} B_{\beta' \beta}$.
As in \pcite{BR} the kernel is now chosen as
\beg  \plabel{K0}
K_{0} \Phi_{0} = [\g_0 \L^+ \l^+ \L^+] \Phi_{0}  [\L'^- \l^- \L^- \g_0 ] \tilde{K}
\ee
so as to annihilate the second term in \pref{D}. In
\beg
(P_0 -2 E_p + 2 i \h) \Phi_0 = \int \frac{d^3p'}{(2\pi)^3}
         \tilde{K}(\p,\p\,',P_0,i\h) \L^+ \l^+ \L^+ \Phi_{0} \tilde{\L'}^- \l^- \tilde{\L}^-
\ee
instead of $\L^-$ the projector $\tilde{\L}^- := \L^-(-\p)$ appears.
Thus expanding $\Phi_{0} = \sum\limits_{A,B=\pm} \L^{A} \tilde{\Phi}_{0}^{(AB)} \tilde{\L}^{B}$
only $\Phi^{(+-)}_{0}$ is found to differ from zero and obeys
\beg  \plabel{LL}
(P_0 -2 E_p + 2 i \h) \L^+ \Phi_0^{(+-)} \tilde{\L}^- = \int \frac{d^3p'}{(2\pi)^3}
         \tilde{K}(\p,\p\,',P_0,i\h) \L^+ \l^+ \Phi_{0}^{(+-)}(\p\,')  \l^- \tilde{\L}^-.
\ee
The nonrelativistic projection of \pref{LL} onto $\l^+ \otimes \l^-$ with
\begar
  \l^+ \L^{+} \l^+ &=& \frac{\l^+}{\m} \nonumber \\
  \l^- \tilde{\L}^{-} \l^- &=& \frac{\l^-}{\m} \\
   \m = \frac{2 E_p}{E_p+m}, \nonumber
\ea
and the introduction of appropriate factors in $\tilde{K}$ relatively to the
Coulomb kernel \pref{Kc}
\begar
  \tilde{K} &=& \m \m' \n \n' K_c \plabel{Ktil} \\
     \n ^2 &=& 4/(P_0 + 2 E_p +2 i \h) \nonumber
\ea
and in the wave function
\beg
 \Phi_{0}(\p) \propto \n^{-1} \m^{-1} \phi (\p)
\ee
also for the case of an unstable fermion lead to a Schr\"odinger equation
for the wave-functions $\Phi (\p)$ in momentum space ($\tilde{E}=
P_0-2m +2 i \h = E+2 i\h$):
\begar
[\frac{\p\,^2}{m} - \hat{E}_n]  \phi (\p) &=& \frac{\a}{(2 \pi)^2} \int \frac{d^3 p'}{(\p -\p\,')^2}
                    \phi(\p\,')  \plabel{Schr}
\ea
The eigenvalues for $\hat{E}_n=\tilde{E} + \frac{\tilde{E}^2}{4m}$ in \pref{Schr}
clearly occur at the real Bohr levels $E_n$, i.e.
\beg
 \tilde{E}_n = M_n^{0} - 2m = -\frac{m \a^2}{4 n^2} -\frac{m \a^4}{64 n^4} +
              O(\a^6); \qquad M_n^0 = 2m\sqrt{1-\s_n^2} \qquad 
              \s_n^2=\frac{\a^2}{4 n^2}
\ee
In addition to the selection of the "large" components
by the choice \pref{K0}, obviously also the sign of $i\h$ in \pref{Ktil}
was crucial for the dependence of \pref{Schr} on the combination
$E+2 i \h$ alone. Still, for the bound-state argument at complex values
of the energy, the independence of $\h$ with respect to the momentum
$\p$ is essential. Going back to \pref{Sg} we observe that with the
choice
\beg  \plabel{gh}
   \hat{\G}_0 = \frac{E_p}{m} \G,
\ee
where $\G = const$ is the (c.m.) decay rate of a single top quark, the disturbing $\p$-dependence
in $\h$ is cancelled. This leads to
$$
 \h= \frac{\G}{2}.
$$

The full BS-wave function (color singlet, BS-normalized \pcite{Luri}) for the BR-kernel
can be obtained by going backwards to $\chi^{(+-)}_{0}:= \chi(p)$.
It appears as the real residue of the complex pole.
The wave function $(\o_n = E_p - M_n^0/2-i\e)$
\beg
\chi_n (p,\e)= \g_0 \bar{\chi}_n^*(p,-\e) \g_0 = i \frac{\L^+ S \tilde{\L}^- }{(p_0^2 - \o_n^2) }
 \frac{\m(p)}{\n(p)} \frac{2 \o_n}{\sqrt{P_0}} \phi(\p). \plabel{xBR}
\ee
is identical to the BR wave-function of stable quarks and belongs to the
spectrum of bound states $ P_n =M_n^0 - i \G_t $. The $i \e$ has been
introduced to determine the integration around that pole.

In eqs. \pref{xBR} $S$ is a constant $4\times4$ matrix which
represents the spin state of the particle-antiparticle system:
\beg \plabel{spin}
      S = \left\{ \begin{array}{l@{\quad:\quad}l}
                   \g_5 \l^- &  \mbox{singlet} \\
                \vec{a}_m \vec{\g} \l^- &  \mbox{triplet}.
                   \end{array}  \right.
\ee
$\phi$ is simply the normalized solution of the Schr\"odinger equation in
momentum space, depending on the usual quantum numbers $(n,l,m)$ \pcite{Wf},
$a_{\pm 1}, a_0$ in \pref{spin} describe the triplet states.
In the following it will
often be sufficient to use the nonrelativistic approximations of
eqs. \pref{xBR} ($\o_n \approx \tilde{\o}_n = (\p\,^2/m+E_n)/2 -i\e $)
\begar
  \chi(p)^{nr} &=& \frac{\sqrt{2} i \tilde{\o}_n}{p_0^2-\tilde{\o}_n^2} \phi(\p)  S
  = -\g_0 \bar{\chi}(p)^{nr} \g_0 \plabel{xapp}
\ea
A similar calculation is needed in order to obtain the solution for
the zero order approximation to the BS-equation itself.
Written in terms of Feynman amplitudes this equation reads:
\beg \plabel{bsf}
 iG_0 = -D_0 + D_0 K_0 G_0,
\ee

It can be shown that the solution to this equation  is
\begar
 G_0 &=& i (2\pi)^4 \d (p-p') D_0 +
     \frac{2 \o}{p_0^2-\o^2} [\L^+ \l^+ (\L^+)' \g_0 ] \otimes
         [ \g_0 \L^- \l^- (\L^-)' ] \times  \plabel{G0}\\
 &  &\times \m \m' \left[ \frac{(2 \pi)^3 \d(\p-\p\,')}{2 \o} -
       \frac{G_C(\widehat{E},\p,\p\,')}{m \n \n'}  \right]
     \frac{2 \o'}{(p_0')^2-(\o')^2}  \nonumber
\ea
where
\begar
 \widehat{E} &=& \frac{(P_0+i\G)^2-4m^2}{4m}, \\
 \o &=& E_k - \frac{P_0 + i \G}{2},
\ea
and  $G_C$ denotes the Coulomb Green function, given e.g. in
\pcite{Schw}.

In the following we will calculate corrections to the position of
the bound state poles of this Green function. One may therefore
ask whether it is possible to write it as a sum over all resonances
and the continuum. The Coulomb Green function $G_C$ clearly can be
decomposed in that way, but inserting this spectral representation
into \pref{G0} leads to an expansion with $P_0$ dependent wave
functions
\beg \plabel{Gbrsum}
G_0 = \sum_{n} \frac{\chi_n(P_0 ,p) \bar{\chi}_n(P_0,p')}{P_0^2 -P_n^{2}}
       + \tilde{D}
\ee
with
\begar
\tilde{D} &=& i (2\pi)^4 \d (p-p') D_0 + \plabel{Dt} \\
   & &(2 \pi)^3 \d(\p-\p\,')  \frac{2 \o}{(p_0^2-\o^2)^2} [\L^+ \l^+ (\L^+)' \g_0 ] \otimes
         [ \g_0 \L^- \l^- (\L^-)' ] \m^2  \nn \\
\chi_n(P_0 ,p) &=&  i \frac{\L^+ S \tilde{\L}^- }{(p_0^2 - \o^2) }
 \frac{\m(p)}{\n(p)} 2\sqrt{2} \o  \phi_n(\p). \plabel{chi2}
\ea
The summation over $n$ represents the sum over all quantum numbers:
discrete $(n,l,m,S,M_S)$ and continuous $(k,l,m,S,M_S)$. We emphasize that
$\o$ now depends on the total momentum $P_0$ and not on the fixed
energy $P_n$. Furthermore also the normalization of the wave functions
had to be changed due to the different denominator in \pref{Gbrsum}
compared to \pref{Pol}.

It is
interesting to note that the wave functions \pref{chi2}, in contrast
to the wave functions \pref{xBR}, also fulfill an orthogonality
relation for different principal quantum numbers.
\beg
tr [ \int \frac{d^4p}{(2 \pi)^4} \bar{\chi}_n(P_0 ,p)
      \frac{2 D_0^{-1}}{P_0^2 - E_p^2} \chi_m(P_0 ,p) ] = \d_{nm}
\ee

The net result of this subsection is that the width of a particle can
also consistently be included for a {\it relativistic} zero order equation
similar to the nonrelativistic case where the replacement $E \to E+i\G$ in the
Green function has been proposed first \pcite{Fadin}.
Here we have the replacement $P_0 \to P_0 +i \G$ except in the
free propagator for the small components.

\section{An Exactly Solvable Zero Order Equation for
              Scalar-Scalar Bound States}

In the last section we developed the bound state formalism for
decaying particles. Until today only fermionic
matter fields have been discovered, but in supersymmetric theories
for each fermion two scalar partners are required.
Since some of them, probably stop or sbottom, could have masses within 
the reach of the next generation of $e^+ e^-$ accelerators, even
the observation of bound states  of those particles seems
possible. These objects and systems built of scalar composite particles in atomic
physics underline the need of an equally clear and transparent
approach as presented above for the fermionic case. A recent attempt in
this direction \pcite{Owen} splits the boson propagator in a particle and
anti-particle propagator to be able to treat them like fermions.
The spectrum is then obtained by constructing the Hamiltonian via a Foldy -
Wouthuysen transformation and a perturbation theory \`a la Salpeter.
This approach suffers from several drawbacks. First it will break down in
higher orders due to the appearance of higher powers in $\p$ and second
the introduction of a "new" propagator raises questions on whether one is
really calculating consequences of the original field theory. A difference
in our result to that of ref. \pcite{Owen} seems to confirm this
suspicion.

To the best of our knowledge there exists no attempt in the literature to construct a
solvable zero order equation for the BS equation containing two
charged scalars interacting via a vector field.
Thus we will generalize the approach of the foregoing chapter to this case.

\subsection{Stable Particles}

As starting
point we present here an exactly solvable equation
for stable scalar particles which interact via a vector field and
show how to calculate the fine structure of such a system
within our approach.

Starting from eq. \pref{allg} we would like to use beside the relativistic
scalar propagators the kernel due to the Coulomb interaction
\beg \plabel{kcs}
K_C(p,p') =  4\pi \a \frac{(P_0+p_0+p_0')(P_0-p_0-p_0')}{(\p-\p\,')^2}.
\ee
This kernel has the drawback that it is $p_0$ dependent and the
exact solution of eq. \pref{allg} with \pref{kcs} is not known. However,
in the nonrelativistic regime by the scaling argument (eq. \pref{scal})
we can start with an instantaneous
approximation to the kernel since $p_0$ is of $O(\a^2 m)$ in this region
and may be included in the corrections afterwards.
Then we can perform the zero component integration on the propagator
\begar
&-i& \int \frac{dp_0}{2  \pi} \frac{1}{[(\frac{P_0}{2}+p_0)^2-E_p^2+i\e]
 [(-\frac{P_0}{2}+p_0)^2-E_p^2+i\e]} =  \nn \\
&=& \frac{1}{2 E_p P_0}\left[\frac{1}{2E_p-P_0}-\frac{1}{2E_p+P_0} \right] =
 \frac{1}{E_p (4  E_p^2-P_0^2)}  \plabel{sprop}
\ea
and it is quite easy to show that
\beg \plabel{k0s}
K_0(p,p') = 4\pi \a \frac{4 m \sqrt{E_p E_{p'}}}{\q\,^2}
\ee
gives a solvable equation with the normalized solutions
\begar
\chi(p) &=& i \frac{\sqrt{E_p} (P_0^2-4E_p^2)}{\sqrt{2P_0}
          [(\frac{P_0}{2}+p_0)^2-E_p^2+i\e][(-\frac{P_0}{2}+p_0)^2-E_p^2+i\e]}
          \phi(\p) \\
\bar{\chi}(p,\e) &=& -\chi^*(p,-\e)  \plabel{chisbar}
\ea
to the eigenvalues
\beg
 P_0 = M_n^{(0)} = 2m \sqrt{1-\s_n^2}. \plabel{Pn}
\ee
Eq. \pref{chisbar} is dictated by the requirement that $\bar{\chi}$
should acquire the same analytic properties as the underlying field
correlators
\begar
 \chi(p) &=& \int e^{i p x} \< 0| T \Phi^{\dagger}(\frac{x}{2})
                        \Phi(-\frac{x}{2})|P_n\>, \\
 \bar{\chi}(p)&=& \int e^{-i p x} \<P_n| T \Phi(\frac{x}{2})
                        \Phi^{\dagger}(-\frac{x}{2})|P_n\>.
\ea
Using the integral representation for the step function which is
included in the time ordered product, one derives eq. \pref{chisbar}.

Taking the equation for the Green function \pref{bsf} instead of that for the BS
wave function and using again \pref{k0s} we find

\begar \plabel{Gscal}
G_0 = - F(p) \frac{G_C(\widehat{E},\p,\p\,')}{4 m}  F(p')
\ea
with
\beg
 \widehat{E} = \frac{P_0^2-4m^2}{4m}
\ee
and
\begar
 F(p) &=& \frac{\sqrt{E_p} (P_0^2-4E_p^2)}
  {[(\frac{P_0}{2}+p_0)^2-E_p^2+i\e][(-\frac{P_0}{2}+p_0)^2-E_p^2+i\e]}
\ea
These solutions can be used for a systematic BS perturbation theory for
scalar constituents, as will be demonstrated in the next section.

\subsection{Unstable Particles}

As in the fermionic case it is desirable to construct a bound state
equation for decaying particles. This can be done by the replacement
\beg \plabel{root}
 E_p \to \sqrt{E_p^2-i \G m}.
\ee
While \pref{root} leads to expressions for the BS wave functions which contain
very unpleasant expressions for the particle poles it has the advantage
the propagator has the form as expected from the phase space of a
unstable particle. Furthermore the above calculation remains essentially
unchanged if we define the square root in \pref{root} to be that with
the positive imaginary part. Only the energy in the resulting
Schr\"odinger equation and thus in \pref{Gscal} changes to
\beg
  \widehat{E} = \frac{P_0^2-4m^2}{4m}+i\G.
\ee
The eigenvalues for $P_0$ are
\beg
 P_{0,n} = 2 m \sqrt{1-\s_n^2-i \frac{\G}{m}} \approx M_n^{(0)}-i\G.
\ee

In the case of the fermions we managed to construct wave functions
independent of $\G$. This was possible because the small components
of the propagator containing $P_0-i\G$ instead of $P_0+i\G$ were projected
to zero with the help of an appropriate kernel.
This cannot be achieved in the scalar case and thus, surprisingly 
enough, the
scalar wave functions look more complicated than the fermionic
ones. To illustrate this we will
present here an approach in close analogy to the fermionic case.
Let us consider eq. \pref{sprop}.  To obtain an expression like
eq. \pref{D} we define the zero order propagator
\begar \plabel{Del0}
 D_0(P_0,p) &=& \D_{+}(P_0,p)\D_{-}(P_0,p) \\
 \D_{\pm}(P_0,p) &=& \frac{1}{2 E_p} \left[
              \frac{1}{\pm \frac{P_0}{2}+p_0-E_p^2+i\frac{\G}{2}} -
              \frac{1}{\pm \frac{P_0}{2}+p_0+E_p^2-i\frac{\G}{2}}
    \right].  \nn
\ea
In order to be able to use the wave functions of the Schr\"odinger equation
with  a Coulomb potential we choose a kernel
\beg
 K_0 = \sqrt{ \frac{p_+ p'_+}{\n_s \n'_s} }
       \frac{4m \sqrt{\r \r'} }{\q\,^2E_p E_{p'}}
\ee
with
\begar
 \r = E_p-i\frac{\G}{2} \\
 p_{\pm} = 2E_p \pm P_0 -i \G,  \\
 \n_{s} = 2E_p + P_0 +i \G.
\ea
This leads to the eigenvalue equation
\beg
 \chi(\p) = \frac{\r}{E_p p_+ p_-}\int \frac{d^3p'}{(2\pi)^3}
         \sqrt{ \frac{p_+ p'_+}{\n_s \n'_s}} \frac{4m \sqrt{\r \r'} }
        {\q\,^2E_p E_{p'}}  \chi(\p\,')
\ee
which is solved by the (already BS-normalized \pcite{Luri,dipl}) wave function
\begar
  \chi(\p) &=& i D_0(M_n^{(0)}-i\G,p) \sqrt{\frac{2 E_p+M_n^{(0)}}{2M_n^{(0)}(2 E_p+M_n^{(0)})
     + 2i\G (M_n^{(0)}-2E_p) (E_p-i\frac{\G}{2})}} E_p  \times \nn \\
       & & \times  (2 E_p-M_n^{(0)}) (2E_p+M_n^{(0)}-2i\G ) \phi(\p),  
         \plabel{chiscal} \\
  \bar{\chi}(\p) &=& \chi(\p).
\ea
The latter equation is the same as eq. \pref{chisbar} if we use real
solutions of the Schr\"odinger equation. It is preferable within
the present context because of the nasty complex expressions in 
\pref{chiscal}.

\section{Perturbation Theory}
\plabel{BSpert}

In view of the calculation of the $t\bar{t}$ potential we observe that
the BR wave function for decaying fermions is the same as for stable ones,
which means that, as far as the systematic determination of the
QCD potential is concerned, we may just consider perturbation
theory for "nonabelian positronium" for its determination.
Perturbation theory for the BS equation starts
from the BR equation for the Green function
$G_{BR}=G_0$ of the scattering of two fermions \pcite{BR2} which is
exactly solvable.
$D_0$ is the product of two zero order propagators, $K_0$ the corresponding kernel.
The exact Green function may be represented as
\beg \plabel{Reihe}
 G =\sum_{l} \chi_{nl}^{BS} \frac{1}{P_0 - P_n} \bar{\chi}_{nl}^{BS} + G_{reg}=G_0 \sum_{\n =0}^{\infty} (H G_0)^{\n} ,
\ee
where the corrections are contained in the insertions $H$.
Bound state poles $P_n=M_n-i\G_n$ occur, of course, only for $M_n<2m$.
It is easy to show that
$H$ can be expressed by the full kernel $K$ and the full propagators $D$:
\beg \plabel{H}
 H = -K + K_0 +iD^{-1}-iD_0^{-1}.
\ee
Since the corrections to the external propagators contribute
only to $O(\a^5)$ to the energy displacement \pcite{Male}, the perturbation kernel is essentially
the negative difference of the exact BS-kernel and of the
zero order approximation.

Expanding both sides of equation \pref{Reihe} in powers of $P_0-P_n$,
the mass shift is obtained \pcite{Lep77}:
\beg \plabel{dM}
 \D M - i\frac{\G}{2}= \< h_0 \> (1+\< h_1 \> ) + \< h_0 g_1 h_0 \> + O(h^3) .
\ee
Here the BS-expectation values are defined as e.g.
\begar
 \<h\> &\equiv& \int \frac{d^4p}{(2\pi)^4} \int \frac{d^4p'}{(2\pi)^4}
           \bar{\chi}_{ij}(p) h_{ii'jj'}(p,p') \chi_{i'j'}(p'), \plabel{erww}
\ea
We emphasize the four-dimensional p-integrations which correspond to the generic
case, rather than the usual three dimensional ones in a completely
nonrelativistic expansion. Of course, \pref{erww} reduces to an
ordinary "expectation value" involving $d^3p$ and $\Phi(\p)$, whenever
$h$ does not depend on $p_0$ and $p_0'$.

In \pref{dM} $h_i$ and $g_i$ represent the expansion coefficients of $H$ and $G_0$,
respectively, i.e.
\begar
 H&=& \sum_{n=0}^{\infty} h_n (P_0-P_n)^n \\
 G_0&=& \sum_{n=0}^{\infty} g_n (P_0-P_n)^{n-1} \plabel{gent}
\ea
Similar corrections arise for the wave functions \pcite{Lep77}:
\beg  \plabel{chi1}
\chi^{(1)} = ( g_1 h_0 + \2  \< h_1 \> ) \chi^{(0)}
\ee

To illustrate the use of this perturbation theory as well as the new zero
order equation for scalar particles presented in the last section we will
present here the calculation of the fine structure of two stable scalar
particles interacting via a vector particle. Existing calculations
\pcite{Owen} rely on a mix of Fouldy-Wouthuysen transformation and the
iterated Salpeter perturbation theory. Our present approach is much more
transparent and allows in principle the inclusion of any higher order
effect in a straightforward manner.
To be definite let us consider the stop--anti-stop system. We will
calculate the spectrum up to order $\a_s^4$.

Since in the zero order equation we have replaced the exact one Coulomb
exchange \pref{kcs} by $K_0$ as given in \pref{k0s} we have now to
calculate the contribution of $-K_C+K_0$ to the energy levels. This is
shown in fig. \pref{dfig2}a .
With
\begar
\<\<-K_c\>\> &=& - 4\pi \a \int \frac{d^4p}{(2\pi)^4} \frac{d^4p'}{(2\pi)^4}
                \bar{\chi}(p)  \frac{(P_0+p_0+p_0')
            (P_0-p_0-p_0')}{(\p-\p\,')^2} \chi(p') = \nn \\
         &=& \< \frac{P_0^2+2E_p^2+2E_{p'}^2}{4 P_0 \sqrt{E_p E_{p'}}}
             \frac{4 \pi \a}{\q^2} \> =  \plabel{p0int} \\
         &=& \< \left( \frac{4m}{2P_0} - \frac{\s_n^2}{2} \right)
             \frac{4 \pi \a}{\q^2} \> \nn \\
\<\<K_0\>\> &=& -\frac{4m}{2P_0} \< \frac{4 \pi \a}{\q^2} \>
\ea
we obtain
\begar \plabel{relko}
 \D M_{C} &=& \<\<-K_c+K_{0}\>\> = -\frac{\s_n^2}{2} \< \frac{4 \pi
               \a}{\q^2} \> = -\frac{m \a^4}{16 n^4}.
\ea
The fact that the p-integrations are well behaved and the result
is of $O(\a^4)$ proves the usefulness of our zero order kernel.

\begin{center}
\leavevmode
\epsfxsize=13cm
\epsfbox{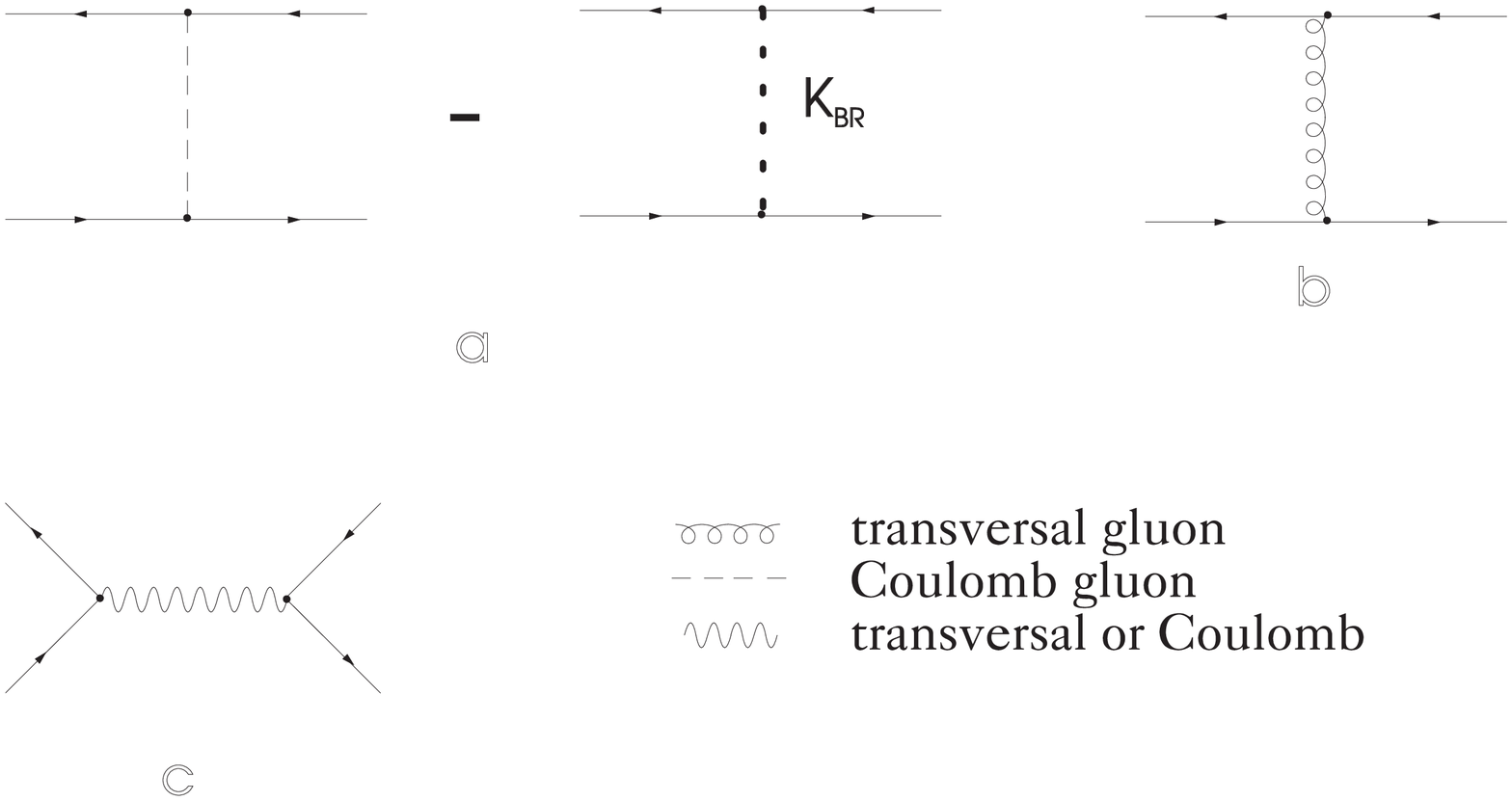}
\\
\centerline{ Fig. \refstepcounter{figure} \label{dfig2} \thefigure}
\end{center}

The transverse gluon of fig. \pref{dfig2}b gives rise to a kernel
\beg
H_T = \frac{4 \pi \a}{q^2} \left( (\p+\p\,')^2 - \frac{(\p^2-\p\,'^2)^2}{\q^2}
      \right) 
\ee
Performing the zero component integrations exactly and expanding in terms
of the three momenta one obtains to leading order (c.f. \pcite{Land})
\begar
 \D M_T = \<\< H_T \>\> &=&  \frac{4 \pi \a}{m^2} \<
                   \frac{\p\,^2}{\q\,^2}-\frac{(\p \q)^2}{\q\,^4} \>\\
&=& m \a^4 \left(\frac{1}{8 n^4} + \frac{\d_{l0}}{8 n^3} - \frac{3}{16 n^3
(l+\2)} \right)
\ea
The annihilation graph into one gluon vanishes due to the color trace 
since the bound states are color singlets.

The net result for the spectrum of two scalars bound by an abelian gauge
field differs from that of ref \pcite{Owen}. The difference can be traced
back to the relativistic correction \pref{relko}. We have also checked
the derivative $\6K_0/\6 P_0$ contributing to $h_1$ and the X-graphs
of fig. \pref{dfig3}.g for possible contributions. Our estimates only 
yield contributions to higher order. 

So far we have calculated the spectrum of two scalars interacting with a
abelian gauge field. It has been pointed out long ago \pcite{Duncan} that
in the case of a nonabelian gauge field further corrections arise due
to the gluon splitting vertices. The vertex correction shown below in
\pref{dfig3}.d has been calculated in \pcite{Duncan} for the fermionic
case.

Here we will give a calculation of the same contribution for scalar
constituents. 
After performing the color trace the perturbation kernel for the second
graph in \pref{dfig3}.e reads  
\beg
 H_{\pref{dfig3}.d,2} = -8ig^4 \int \frac{d^4 k}{(2 \pi)^4}
 \frac{(P_0+p_0+p_0'-k_0)(-P_0+p_0+p_0')}{\q\,^2 (\vec{k}-\q)^2
  [(\frac{P}{2}+p+k)^2-m^2] k^2} \left( -(\p \q) + \frac{(\p \vec{k})(\q
   \vec{k})}{\vec{k}\,^2} \right).
\ee
Performing the $k_0$ integration and using the scaling
\begar
 P_0 &\to& 2m + O(\a^2) \\
 p_0 &\to&  \a^2 p_0 \\
 \vec{k} &\to& \a \vec{k}
\ea
to extract the leading contribution in $\a$ we can show that
\beg
  H_{\pref{dfig3}.d,2} = - \frac{g^2 m}{2} \frac{\p\q}{|\q|^3}.
\ee
Adding the similar contribution from the first graph in \pref{dfig3}.d
gives
\beg
  H_{\pref{dfig3}.d} = - \frac{9\pi^2\a^2 m}{|\q|}.
\ee
This result differs by a factor $4m^2$ from the fermionic result which
is compensated by a corresponding difference in the wave functions to give
eventually precisely the same result as in the fermionic case
\beg
 \D M = \< \frac{9\pi^2\a^2}{4m|\q|} \> = \frac{9 m \a^4}{32 n ^3 (l+\2)} .
\ee
The corrections to the gluon propagator are trivially identical to the
fermionic case as can be seen by inspecting the only "critical" step, the
$p_0$ integration \pref{p0int}. In view of the fact that the result 
depends only on the angular momentum and not on the spin this seems 
reasonable.
Thus the difference in the spectrum of the scalar
bound state to $O(\a^4)$ is entirely due to the tree graphs discussed above.


\section{The $t\bar{t}$-Potential}

In the last section we showed how the BS-wave function for decaying particles
may be reduced to one for stable particles. Therefore, contributions from the 
width can only
arise from graphs with internal fermion lines. We will discuss such a case
in detail in sect. \pref{decay}. For the determination of the potential the
effect of the width turns out to be negligible. Especially we can
extract the contributions to the potential from the level shifts calculated
in \pcite{dipl} for the tree and one loop QCD graphs.
Our main philosophy will be the following. We estimate the magnitude of an
interaction by its effect on the level shift. That is to say our
potential to $O(\a^4)$ will contain all terms which would give rise to level
shifts to that order, keeping in mind that the only level that could be
measured is the $1S$ level. With the at present favored top mass value 
of $\approx 175$GeV this would be possible if $m_t$ is known to a 
sufficient high precision from other experiments by measuring
the rising edge of the total cross section $e^+e^- \to t\bar{t}$.
This will be discussed in sect. \pref{ttcross}.
The contribution of the tree and one loop QCD graphs to the potential
to $O(\a^4)$ reads \pcite{KM1}:
\begar
 \CV_1 &=& -\frac{\p^4}{4 m^3} - \frac{\a \pi}{m^2} \d (\vec{r})
      - \frac{\a }{2 m^2 r} \left( \p\,^2 + \frac{\vec{r} (\vec{r} \p) \p}
       {r^2} \right) + \nonumber\\
     & & + \frac{3 \a }{2 m^2 r^3} \vec{L} \vec{S} +  \frac{\a }{2 m^2 r^3} \left( \frac{3  (\vec{r} \vec{S})^2}
       {r^2} - \vec{S}\,^2 \right) + \frac{4 \pi \a }{3 m^2} \vec{S}\,^2 \d (\vec{r}) \nonumber \\
     & & - \frac{33 \a^2}{8 \pi r} ( \g + \ln \m r) +\frac{\a^2}{4\pi r} \sum_{j=1}^{5} [\mbox{Ei}(-r m_j e^{\frac{5}{6}})
             -\frac{5}{6} + \2 \ln (\frac{\m^2}{m_j^2} + e^{\frac{5}{3}})]  \plabel{result}\\
     & & + \frac{9 \a^2}{8 m r^2} . \nn
\ea
In eq. \pref{result} the first two lines arise from the relativistic Coulomb
correction and the exchange of a transverse gluon. They are, therefore,
almost identical to the positronium case if one includes the color factor
$4/3$ in the definition of $\a$ (c.f eq. \pref{alpha}). The only
difference is the absence of the annihilation contribution due to the
tracelessnes of the color matrix at the annihilation vertex. The first
term in the third line comes from the gluonic one loop correction to
the Coulomb gluon (fig. \pref{dfig3}.a, \pref{dfig3}.b). Fermion loops are 
included with their explicit mass
dependence in the next term (fig. \pref{dfig3}.c). The fourth line 
contains the contribution
of the nonabelian vertex correction fig. \pref{dfig3}.d. All other graphs
shown in fig. \pref{dfig3}.e-h do not contribute to the required order in Coulomb 
gauge.

\begin{center}
\leavevmode
\epsfysize=12cm
\epsfbox{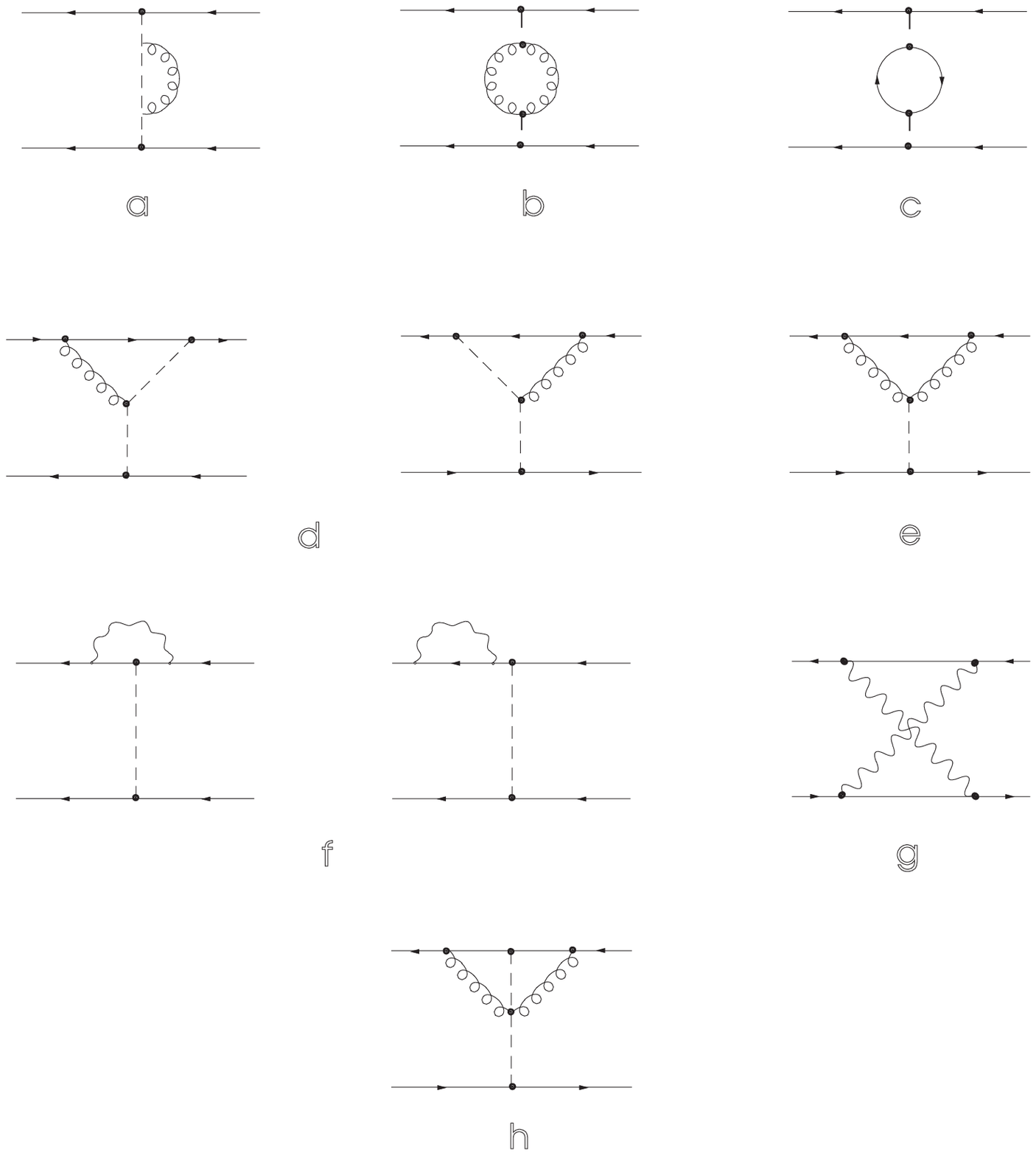}
\\
\centerline{ Fig. \refstepcounter{figure} \plabel{dfig3} \thefigure}
\end{center}

In the following section we will calculate contributions to the potential
from two loop QCD graphs and from the electroweak interaction. These
contributions will turn out to give effects to the same numerical order
of magnitude as the ones above. Therefore, the technique often used to
calculate a potential to a given number of loops is shown to be incorrect.

\subsection{Two Loop Vacuum Polarization}

As can be seen from the dependence of the potential \pref{result} from the
masses of the light flavors, the usual renormalization
group arguments relying on massless quarks in the running coupling
constant do not consistently include the effect of 'realistic'
quark masses in the toponium system, when a systematic BS
perturbation is attempted. On the other hand, in a full calculation of effects of $O(\a^4)$ two loops with gluons
cannot be neglected.
However, in the one loop case finite
quark masses yield terms of numerical order $O(\a^4)$, therefore
it can be expected that in a two loop calculation quark masses will
lead to corrections of $O(\a^5)$.

Furthermore, since two loop calculations with massive flavors are very
cumbersome we circumvent these problems, for the time being, by the
following argument, which also includes the three 'massless'
quarks u,d,s.
Because of the Ward identity for the Coulomb-vertex \pcite{Fein}, it is
clear from the theory of the renormalization group that the same corrections
can be obtained by expanding
the running coupling constant with a two loop (gluons+u,d,s) input
for the latter which provides also the first 'nonleading' logarithmic
contributions.
The beta function to two loops is renormalization scheme
independent for massless quarks \pcite{Muta} and its two loop part
has been calculated some time ago \pcite{Jon}:
\begar
 \b(g) &=& -\b_0 g^3 - \b_1 g^5 - ...\\
  \b_0 &=& \frac{1}{(4 \pi)^2} (11-\frac{2}{3} n_f)\\
  \b_1 &=& \frac{1}{(4 \pi)^4} (102-\frac{38}{3} n_f).
\ea
Here $n_f$ is the number of effectively massless flavors and $\b(g)$ is the
solution of
\begar
 \ln \frac{\sqrt{-q^2}}{\m} &=& \int_g^{\bar{g}} \frac{dg'}{\b(g)},
\ea
which reads up to two loops
\beg
  \ln \frac{\sqrt{-q^2}}{\m} = \frac{1}{2 \b_0} \left[
   \frac{1}{\bar{g}^2}-\frac{1}{g^2} + \frac{\b_1}{\b_0}
   \ln \frac{\bar{g}^2 (\b_0+\b_1 g^2)}{g^2 (\b_0+\b_1 \bar{g}^
   2)} \right].
\ee
Considering this as an equation for $\bar{g}=g(\q\,^2)$  we 'undo'
the renormalization group improvement by expanding with 'small'
$g^2 \propto \a$ (cf. eq. \pref{alpha} ):
\begar
 \a(\q\,^2)&=& \a \left\{ 1- \a \frac{33-2 n_f}{16 \pi}\ln \frac{\q\,^2}{\m^2}+ \right.\\
  & &  \qquad + \left. \frac{\a^2}{(16 \pi)^2}[(33-2 n_f)^2 \ln^2 \frac{\q\,^2}{\m^2}-
      9(102-\frac{38}{3}n_f)\ln \frac{\q\,^2}{\m^2}] \right\} \nonumber
\ea
Clearly the one loop term agrees with eq. \pref{result}
in the limit $m_j \to 0$. That limit, however, is not appropriate
here because in this way we would loose terms of numerical  $O(m\a^4)$.
For the computation of the rest we need the expectation
value of $(\ln^2 \frac{\q\,^2}{\m^2})/ \q\,^2$. This integral
can be done analytically (Appendix A) and the result is:
\begar
\<\frac{ \ln^2 \frac{\q\,^2}{\m^2}}{\q\,^2} \> &=& \frac{m \a}{2 \pi n^2}
       \{ \frac{\pi^2}{12} + \Psi_2(n+l+1)+s_{nl}+[\Psi_1(n+l+1)+\g+
       \ln \frac{\m n}{\a m}]^2 \}  \plabel{eln2}
\ea
with
\begar
s_{nl} &=& \frac{2 (n-l-1)!}{(n+l)!} \sum_{k=0}^{n-l-2} \frac{(2l+1+k)!}
       {k! (n-l-1-k)^2}. \nonumber
\ea
With eq. \pref{eln2} we obtain for the mass shift, induced by the
leading logs of the two loop vacuum polarization of the Coulomb gluon
a contribution:
\begar
 \D M_{2loop} &=& -\frac{m \a^4}{128 \pi^2 n^2} \left\{ 27^2[ \frac{\pi^2}{12} + \Psi_2(n+l+1)+s_{nl}+(\Psi_1(n+l+1)+\g+
       \ln \frac{\m n}{\a m})^2] \right. + \nonumber \\
       & & \qquad \qquad \qquad+ \left.288 (\Psi_1(n+l+1) + \g +\ln \frac{\m n}{\a m}) \right\}.
       \plabel{dM2l}
\ea
In this expression we have set $n_f=3$ as suggested by the number of sufficiently light quarks.
Whether eq. \pref{dM2l} really represents the full two loop quark-
gluon vacuum polarization, numerically consistent with other terms
$O(\a^4)$, must still be checked in a calculation of the Coulomb gluon's self-energy
to two loop order in the Coulomb gauge, i.e. going beyond the sample
calculation in \pcite{KM1}. We, nevertheless, indicate the corresponding potential
\begar
 \CV_{2loop} =- 2 \frac{\a^3}{(16 \pi)^2 r} \left\{ (33-2n_f)^2[\frac{\pi^2}{6}+
               2(\g+\ln \m r)^2] + 9 (102-\frac{38}{3} n_f)(\g+\ln \m r)
               \right\} \plabel{v2l}
\ea

\subsection{QCD 2-Loop Box Graphs}

It would be incorrect to extrapolate from the QED case the absence
of corrections to $O(\a^4)$, other than the abelian tree graphs
because gluon splitting allows new types of graphs.
Therefore we will discuss in this section possible sources of new
corrections.

Our first example of a QCD box graph is fig. \pref{dfig5}.a. Between the nonrelativistic
projectors $\l^{\pm}$ of the wave functions \pref{xapp} the perturbation kernel
from this graph can be written effectively as
\begar
 -i H_{\pref{dfig5}.a} &=&12 i g^6 \int \frac{d^4t}{(2\pi)^4} \frac{d^4k}{(2\pi)^4}
          \frac{(p_1^0-t_0 +m)(p_2^0-k_0-m)}{[(p_1-t)^2-m^2][(p_2-k)^2-m^2]}  \times  \nonumber \\
         & & \quad \times \frac{1}{(t-k)^2 \vec{k}\,^2
          (\q-\vec{k})^2 \vec{t}\,^2 (\q-\vec{t})^2} P(\q, \vec{k},
           \vec{t})    \plabel{Hgraph}
\ea
with
\beg
   P(\q, \vec{k}, \vec{t}):= (\q-\vec{k})\vec{k}-
         \frac{[(\vec{t}-\vec{k})(\q-\vec{k})][\vec{k} (\vec{t}-\vec{k})]}
             {(\vec{t}-\vec{k})^2} .
\ee
After performing the integrations over the zero components $t^0$ and $k^0$
we can use the scaling argument \pref{scal} to extract the leading
contribution. This is justified a posteriori by the (infrared) finiteness of the
remaining terms. The resulting expression will thus only depend on $\q$:
\begar
 H_{\pref{dfig5}.a} &=& 6 g^6 \int \frac{d^3t}{(2\pi)^3} \frac{d^3k}{(2\pi)^3}
         \frac{1}{\vec{k}\,^2 (\q-\vec{k})^2 \vec{t}\,^2 (\q-\vec{t})^2
         (\vec{t}-\vec{k})^2} P(\q, \vec{k}, \vec{t})  \plabel{Hgraph2}
\ea
The trick to simplify this expression is to write the inner products
in terms of quadratic expressions that cancel partly the denominator
\footnote{ I am grateful to Antonio Vairo for  bringing this trick to my
attention}.
\begar
P(\q, \vec{k}, \vec{t}) &=& \frac{1}{4}[ 2\q^2 - \vec{k}\,^2 -
(\q-\vec{k})^2 -\vec{t}\,^2 \plabel{trick} \\
& & - \frac{\vec{t}\,^2 (\q-\vec{k})^2}{(\vec{k}-\vec{t})^2} +
      \frac{\vec{t}\,^2 (\q-\vec{t})^2}{(\vec{k}-\vec{t})^2} +
      \frac{\vec{k}\,^2 (\q-\vec{k})^2}{(\vec{k}-\vec{t})^2} -
      \frac{\vec{k}\,^2 (\q-\vec{t})^2}{(\vec{k}-\vec{t})^2} +
       (\vec{k}-\vec{t})^2 - (\q-\vec{t})^2 ] \nonumber
\ea
While the original integral \pref{Hgraph2} is finite, the integrals one
obtains from a single summand in \pref{trick} are not. But they are
easily evaluated with the help of known results within dimensional
regularization \pcite{Muta}. A convenient check of the lengthy 
computations is that all divergent terms must cancel in
the final answer which can be written as
\beg
 H_{\pref{dfig5}.a} = -\frac{81 \pi}{128}(12- \pi^2) \frac{\a^3}{ |\q|^2}. \plabel{H5a}
\ee
The nonabelian box graph fig. \pref{dfig5}.a gives therefore rise to an $O(\a^4)$
correction which leads to a slight enhancement of the attractive Coulomb
force. This is the only contribution of this type in our
renormalization scheme which is defined as follows: 
$m_t$ is the pole mass,
$Z_{Gluon}$ and $Z_{1F}$ are defined by a subtraction at $\m$.
However, the usual predictions for scattering
measurements are in the $\overline{\mbox{MS}}$ scheme. In a calculation of the
potential in this scheme a host of contributions of the form of eq.
\pref{H5a} are to be expected. Since they are not known yet, and the
numerical coefficient in the result \pref{H5a} is small, this at present is
more of theoretical than of phenomenological interest. However, it provides
an example of a nontrivial contribution to $O(\a^4)$ of a two loop
graph which certainly is not included e.g. in the running coupling
constant. The result \pref{H5a} is new, but a qualitative 
estimate was mentioned already in
\pcite{Fein}, \pcite{Fisch} and \pcite{Duncan}.

\begin{center}
\leavevmode
\epsfysize=8cm
\epsfbox{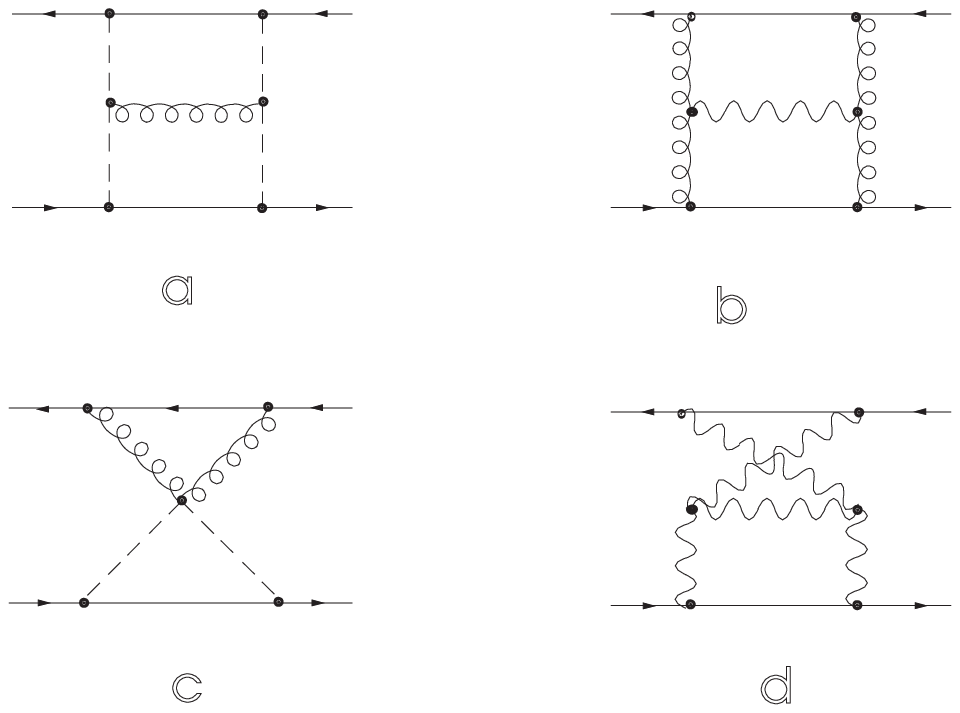}
\\
\centerline{ Fig. \refstepcounter{figure} \plabel{dfig5} \thefigure}
\end{center}

Considering the graph \pref{dfig5}.b we perform the same steps as for the graph 5.a.
One notes that either the spatial gamma matrices at the vertices or the
zero component integrations give rise to two additional powers in the
spatial momenta and consequently in powers of $\a$. Therefore, this graph
contributes only to higher order. It is, however, a spin dependent
interaction.

Graphs like fig. \pref{dfig5}.a and \pref{dfig5}.b with crossed Coulomb lines
(fig. \pref{dfig5}.d) are irrelevant because they vanish due to group theoretical factors.

By connecting the quark lines between the interactions in graphs
\pref{dfig3}.d and \pref{dfig5}.a as shown in fig. \pref{dfig5a} one obtains
graphs which could be named nonabelian Lamb shift graphs.
We have checked whether these graphs could give rise to further
$O(\a^4)$ effects - but we did not find any. Thus all these graphs will
contribute at the usual Lamb shift level $O(\a^5 \ln \a)$.

\begin{center}
\leavevmode
\epsfxsize=12cm
\epsfbox{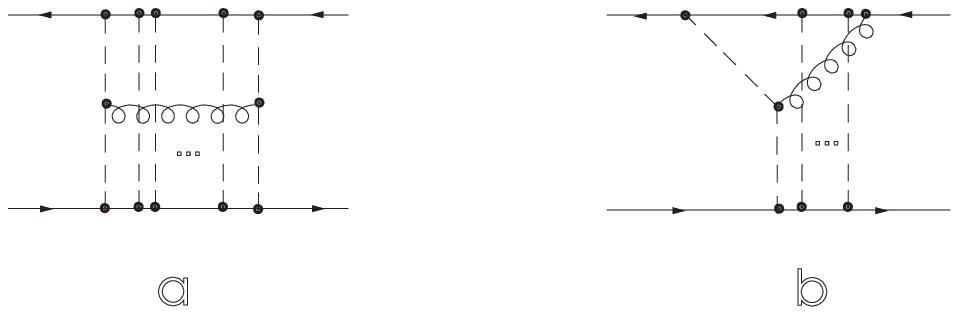}
\\
\centerline{ Fig. \refstepcounter{figure} \plabel{dfig5a} \thefigure}
\end{center}

The 'X` graph in fig. \pref{dfig5}.c can be checked more easily for possible
new contributions.
As in the calculation of fig. \pref{dfig5}.a it can be simplified  to
give
\begar
 -i H_{\pref{dfig5}.c} &=& 3 i g^6 \int \frac{d^4t}{(2\pi)^4} \frac{d^4k}{(2\pi)^4}
          \frac{(k_0+p_1^0+m)(t^0+p_2^0-m)}{[(k+p_1)^2-m^2][(t+p_2)^2-m^2] \vec{k}\,^2
          (\q-\vec{k})^2} \times  \nonumber \\
         & & \quad \times \frac{1}{t^2 (q-t)^2} \left( 1+ \frac{[\vec{t}(\q-\vec{t})]^2}
             {\vec{t}\,^2 (\q-\vec{t})^2} \right).  \plabel{Xgraph}
\ea
The integration over $k$ yields a divergence $1/|\q|$ if $\q$ and $\p$ tend to
zero. Contributions within the order of interest may only result from
possible poles after the $t$-integration. For simplicity we consider
the part of eq. \pref{Xgraph} from the factor one in the second
line:
\begar
 I_t &\equiv& \int \frac{d^4t}{(2\pi)^4} \frac{t^0+p_2^0+m}{[(t+p_2)^2-m^2]
               t^2 (q-t)^2} = \nonumber\\
     &=& \frac{-i}{(4\pi)^2} \int_0^1dx \int_0^{1-x} dy \frac{y q^0 -x p_2^0+p^0+m}
          {(y q-x p)^2-y q^2 -x(p^2-m^2)} \approx \nonumber\\
     &\approx&  \frac{im}{(4\pi)^2} \int_0^1dx \int_0^{1-x} dy \frac{x}
                {x^2 m^2+ y(1-x-y)\q\,^2} + O(\a)= \nonumber\\
     &=& \frac{-i}{(4\pi)^2m} \ln \frac{|\q|}{m} + O(\a).
\ea
Thus the part of graph \pref{dfig5}.c, specified above, has a leading term from
$$
  H_{\pref{dfig5}.c,1} = \frac{3 g^6}{32(4\pi)^2 |\q| m} \ln \frac{|\q|}{m}
$$
as $\q \to 0$ and therefore contributes to $O(\a^5 \ln \a)$. The second
part of graph \pref{dfig5}.c gives a similar result with a different numerical
factor.

We conclude that - opposite to the QED case \pcite{Fult}- box graphs
may be important to $O(\a^4)$. They could, in fact, be responsible
for changes of the zero order coupling of the Coulomb gluon.
As a matter of fact, indeed substantially different values of that coupling seem to be
required in phenomenological fits of lighter quarkonia.

As illustrated by the
explicit calculations above, to $O(\a^5)$ beside abelian QED type
corrections \cite{Male,Fult}, a host of further non-abelian contributions
can be foreseen.

\subsection{QED Correction}

As a rule, the consideration of electromagnetic effects in
QCD calculations is not necessary, but at high energies the strong
coupling decreases, and in the case of toponium we expect
$\a_s^2$ to be of the same order as $\a_{QED}$.
We may obtain this contribution by simply solving the BR equation for the
sum of a QED and a QCD Coulomb exchange. This results in the
energy levels
\begar
 P_0 =M_n^0 &=& 2m \sqrt{1-\frac{(\a+\a_{QED} Q^2)^2}{4n^2}} \approx  \\
    &\approx& 2m - m \frac{\a^2}{4n^2} - \frac{m \a \a_{QED} Q^2}{2 n^2}
    - m \frac{\a^4}{64n^4}  -\frac{m \a_{QED}^2 Q^4}{4 n^2} + O(\a^6), \nonumber
\ea
where $Q$ is the electric charge of the heavy quark, i.e. $2/3$ for
toponium. Clearly even the 'leading' third term can only be
separated from the effect of the second one to the extent that
$\a(\m)$ and $\a_{QED}(\m)$ can be studied separately with sufficient
precision.

\subsection{Weak Corrections}

While also weak interactions usually can be neglected in QCD calculations,
this is not true in the high energy region, because the weak coupling
scales like $\sqrt{G_F m^2}$, becoming comparable to the strong
coupling if the fermion mass $m$ is large. Even bound states through
Higgs exchange are conceivable \pcite{Jain}. Therefore, we have to consider
weak corrections and especially the exchange of a single Higgs
or Z particle, assuming for simplicity the standard model with minimal Higgs sector.

The Higgs boson gives rise to the kernel
\beg
 -i H_{Higgs} = -i \sqrt{2} G_F m^2 \frac{1}{q^2-m_H^2}\approx i \sqrt{2} G_F m^2 \frac{1}{\q\,^2+m_H^2},
 \plabel{HH}
\ee
in an obvious notation.

Since we do not know the ratio $\a m/m_H$, which would allow some
approximations if that ratio is small,
we calculate explicitly the level shifts by transforming into
coordinate space. As in Appendix A, we express the Laguerre polynomials in terms of
differentiations of the generating function, do the integration and perform
the differentiation afterwards to obtain
\beg
\D M_{Higgs} = -m \frac{G_F m^2 \a}{4 \sqrt{2} \pi} I_{nl}(a_n),
\plabel{dMH}
\ee
valid for arbitrary levels and Higgs boson masses, with 
$$
  a_n = \frac{\a m}{n m_H}
$$
and
\begar \plabel{inl}
 I_{nl}(a_n) &\equiv& \frac{a_n^{2l+2}}{n^2 (1+a_n)^{2n}} \sum_{k=0}^{n-l-1}
            {n+l+k \choose k} {n-l-1 \choose k} (a_n^2-1)^{n-l-1-k}.
\ea
As an illustration some explicit results for the lowest levels are
given in tab.4.
\renewcommand{\arraystretch}{1.4}\\[.5cm]
\centerline{\begin{tabular}{c|cccccc}
 $n$ & 1 & 2 & 2 & 3 & 3 & 3 \\ \hline
 $l$ & 0 & 0 & 1 & 0 & 1 & 2 \\ \hline
 $I_{nl}(a_n)$ &
   $\frac{a_1^2}{(1+a_1)^2}$ &
   $\frac{a_2^2(2+a_2^2)}{4(1+a_2)^4}$ &
   $\frac{a_2^4}{4(1+a_2)^4}$          &
   $\frac{a_3^2(3+6a_3^2+a_3^4)}{9(1+a_3)^6}$ &
   $\frac{a_3^4(4+a_3^2)}{9(1+a_3)^6}$ &
   $\frac{a_3^6}{9(1+a_3)^6}$
\end{tabular}}

\vspace*{0.3cm}
\centerline{Tab. 4}
\vspace{.5cm}
It is evident that eq. \pref{dMH} will give a contribution of order $G_F m^2 \a^3$ if the Higgs
mass is comparable to the mass of the heavy quark and should therefore be
taken into account in a consistent treatment of heavy quarkonia spectra
and the related potential to numerical order $O(\a^4)$.
We will thus consider also a term corresponding to \pref{HH}
\begar
 \CV_{Higgs} = - \sqrt{2} G_F m^2 \frac{e^{-m_H r}}{4 \pi r}
\ea
as a correction in our potential.

Next we consider the contributions of the neutral current, the single Z-exchange
and Z-annihilation. For the Z-exchange in the SM we have
\begar
 H_Z &=& -\sqrt{2} G_F m_Z^2 [\g^{\m}(v_f-a_f \g_5)]_{\s\s'} \frac{g_{\m\n}-\frac{q_{\m}q_{\n}}
         {m_Z^2}}{q^2-m_Z^2}[\g^{\n}(v_f-a_f \g_5)]_{\r'\r}
     \plabel{HZ}
\ea
with
\begar
 v_f &=& T^f_3-2 Q_f \sin^2 \Theta_w, \\
 a_f &=& T^f_3,
\ea
where $T_3^f$ is the eigenvalue of the diagonal SU(2) generator for the fermion
$f$. If $f$ is the top quark then $T^f_3=1/2$.
Because we cannot expect $\a m$ to be much smaller that $m_Z$ we
use the exact expectation value of the Yukawa potential \pref{inl}.
$q_0^2$ may be dropped in the Z-propagator since this provides
at most a further correction of numerical $O(\a)$:

\begar
 \D M_Z &=& -\sqrt{2} G_F m_Z^2 \< \frac{a_f^2(3-2 \vec{S}\,^2)(1+
                 \frac{\q\,^2}{3 m_Z^2}) -v_f^2}{q\,^2-m_Z^2} \> = \\
      &=& m \frac{G_F m_Z^2 \a}{2 \sqrt{2} \pi} [a_f^2(1-\frac{2}{3}
    \vec{S}\,^2) ( I_{nl}(\frac{\a m}{n m_Z}) +\frac{(\a m)^2}{2 m_Z^2 n^3}
                \d_{l0} ) -\frac{v_f^2}{2} I_{nl}(\frac{\a m}{n m_Z} ) ]
       \nonumber
\ea
$\vec{S}\,^2=S(S+1)$ is total spin of the quark-antiquark system. Therefore
this expression gives rise to a singlet-triplet splitting within the order of interest.

Z-annihilation may also yield a sizeable effect.
The corresponding energy shift is easily evaluated
\begar
 \D M_{S=0} &=& \frac{3 G_F m^2 a_f^2}{2 \sqrt{2} \pi} \frac{m \a^3}
    {n^3} \d_{l0} \\
 \D M_{S=1} &=& -\frac{v_f^2}{a_f^2}\frac{m_Z^2}{4m^2-m_Z^2} \D
   M_{S=0} \approx 10^{-2} \D  M_{S=0}
\ea
and also produces a singlet-triplet splitting.
It should be noted that the two last contributions yield  singlet-triplet
splittings which are as important as the usual Breit
interaction (cf. the first two lines in \pref{result}). The corresponding contribution to $\CV$ is
\begar
 \CV_{Z} &=& \sqrt{2} G_F m_Z^2 a_f^2 \Big\{ \frac{e^{-m_Z r}}{2 \pi r}
       \big[1-\frac{v_f^2}{2 a_f^2} - (\vec{S}\,^2 - 3 \frac{(\vec{S}\vec{r})^2}{r^2})
        (\frac{1}{m_Z r} + \frac{1}{m_Z^2 r^2})- (\vec{S}\,^2 - \frac{(\vec{S}\vec{r})^2}{r^2})\big]  +
           \nonumber     \\
         & & \qquad \qquad \qquad + \frac{\d(\vec{r})}{m_Z^2} (7- \frac{11}{3} \vec{S}\,^2 ) \Big\}
        \plabel{vz}
\ea

\subsection{Schwinger-Christ-Lee Terms}

As mentioned in chapter 1, nonlocal interactions
have to be added to the Lagrangian in Coulomb gauge.
We are not aware of any previous attempt
to look whether these terms
give contributions to bound state problems or to some effective
potential.

By analogy to the second ref. \pcite{Schw} we calculate the $v_1$ term
to $O(g^4)$
\begar
 v_1 &=& -g^4 \frac{9}{16} \int d^3r d^3r' d^3r'' A_i^c(\vec{r}\,') K_{ij}(\vec{r}-\vec{r}\,')
         K_{jk}(\vec{r}-\vec{r}\,'') A_k^c(\vec{r}\,'')\\
 K_{ij}(\r) &=& \frac{1}{4 \pi |\r|} \left[
          \frac{\d_{ij}}{3} \d(\vec{\r})-\frac{1}{4 \pi |\vec{\r}|^5}
         (3 \r_i \r_j - \vec{\r}\,^2 \d_{ij}) \right]. \nonumber
\ea
This corrects the gluon propagator by
\beg
 \d G_{\m\n}^{ab}(x_1,x_2) = -\frac{1}{Z[0]} \frac{\d^2}{\d J_{\m}^a(x_1) \d J_{\n}^b(x_2)}
              \frac{9 i g^4 }{16} \int d^4x \int d^3r d^3r' d^3r'' \frac{\d}{\d J_i^c}
                              K_{ij} K_{jk} \frac{\d}{\d J_k^c} Z_0[J].
\ee
In momentum space $\d G$ can be calculated by using dimensional
regularization to give
\beg
 \d G_{mn}^{ab} (q,q') = (2\pi)^4 i \d^{ab} \d(q-q') \frac{9 g^4 }{8^5}
                        \frac{\q^2}{q^2} \frac{1}{q^2}(-\d_{mn}+\frac{q_m q_n}{\q^2}),
\ee
which means that we have a mass shift with the same structure as the one
transverse gluon exchange (cf. \pcite{KM1}), further suppressed by two
more orders in $\a$.

Since the second term $v_2$ also represents a correction to the
propagator of the transverse gluon, it can be estimated by the same
method to contribute only in higher orders of $\a$ as well.
We thus find that both terms can be neglected even including terms of
$O(\a^5)$.

\vspace{.5cm}

Finally we will collect the results obtained in this section. Both, the
large mass \underline{and} the large width of the top quark provide a new
field for rigorous QCD perturbation theory: In contrast to the lighter
quarkonia, (unstable) toponium is a weakly bound system, to be treated by
Bethe-Salpeter methods in a systematic manner. The large width even
further reduces the effects of confinement. As shown first by Fadin and
Khoze \pcite{Khoze} $\G$ can be included in the (weakly bound) Green
function at the threshold in a straightforward manner by analytic
continuation of the total energy into the complex plane.

We first showed that a similar trick may be also applied to a different
zero order equation, the BR-equation. On the basis of this result we
describe how to obtain the proper potential $\CV$ for such a Green
function, rigorously derived from QCD in a perturbative sense. Although a
fully 4-dimensional formalism is used, which especially also allows the
inclusion of off-shell effects, our result allows an interpretation as a
correction to the Coulomb potential:
\begar
\CV &=&  \CV_1 +\CV_2 \plabel{potres}
\ea
with $\CV_1$ from eq. \pref{result} and
\begar
 \CV_2 &=&  -\frac{81 }{128}(3- \frac{\pi^2}{4})
\frac{\a^3}{r}
-  \frac{2 \a^3}{(16 \pi)^2 r} \left\{ 27^2[\frac{\pi^2}{6}+
2(\g+\ln \m r)^2] + 576(\g+\ln \m r)\right\}   \nonumber \\
  & & - \frac{8}{9} \frac{4\pi \a_{QED}(\m) \a}{r}- \sqrt{2} G_F m^2
\frac{e^{-m_H r}}{4 \pi r}+ \sqrt{2} G_F m_Z^2 a_f^2 \frac{\d(\vec{r})}{m_Z^2}
(7- \frac{11}{3} \vec{S}\,^2 )  \plabel{v2} \\
& & + \sqrt{2} G_F m_Z^2 a_f^2  \frac{e^{-m_Z r}}{2 \pi r}
      \big[1-\frac{v_f^2}{2a_f^2} - (\vec{S}\,^2 - 3
         \frac{(\vec{S}\vec{r})^2}{r^2})
  (\frac{1}{m_Z r} + \frac{1}{m_Z^2 r^2})- (\vec{S}\,^2 -
       \frac{(\vec{S}\vec{r})^2}{r^2})\big]             \nonumber
\ea
It has been obtained by checking contributions up to numerical order
$O(\a^4)$ to real energy shifts, calculated independently from the weak
decay $t \to b + W$. The abelian relativistic corrections and the one
loop contributions are collected in $\CV_1$.

The first term of $\CV_2$ is the nonabelian box contribution, followed
by the two loop corrections to the coulomb gluon propagator.
Here the effects of the quark masses are not known yet.
On the other hand, we show that
certain electroweak effects (QED,Higgs,Z) are numerically important to
$O(\a^4)$. Their respective contributions are listed in the last two
lines of \pref{v2}.

Within a rigorous field-theoretical philosophy it would be
incorrect to add, say, a linear term to \pref{result} in order
to describe confinement. At best \pref{result} could be supplemented
by a piece $\propto \< G^2 \> r^3$
which mimics the tail of confinement effects by gluon condensate
\pcite{Leut,KumWL}.

In the derivation of our potential we have not only used the level
shifts, but also have described in much detail new closed forms
for such shifts etc. The reason for that has been that on the one hand
we hope to have given new useful methods to be applicable also
for future treatments of 
nonabelian $O(\a^5)$-effects. On the other hand certain
computations of level shifts may be useful in conjunction
with semi-phenomenological approaches to the lighter quarkonia.
However, even in the case of the bottom quark the nonperturbative 
corrections become quite large and their perturbative inclusion seems
rather dangerous.


\section{ Relativistic and Gauge Independent Off-Shell
 Corrections to the Toponium Decay Width}

\plabel{decay}

As already mentioned in the previous chapters, the top quark decays almost
exclusively into $W^+b$. Clearly the tree level SM decay  width has to be
corrected by QCD and weak corrections. QCD corrections to $O(\a_s)$
\pcite{kuehn} and the leading electroweak corrections \pcite{Denner}
have already been calculated. The width is also strongly influenced by 
supersymmetric corrections \pcite{Dabel}.
For alle these aforementioned corrections it suffices to
consider the t-quark as a free on-shell particle since it is off-shell
near threshold only to $O(\a_s^2)$. This means also that near threshold
not only the $O(\a_s^2)$ corrections to the free decay are needed but
also bound state corrections to the decay width play a role.
The calculation of these
corrections will be the subject of this chapter.
Concerning the $O(\a_s^2)$ corrections to the free decay only a partial
result is known yet \pcite{zalews}.

\subsection{Narrow Width Approximation}
\plabel{topdec}

In this section we will calculate off-shell and relativistic bound-state
corrections for the decay $t \to b+W$ to $O(\a_s^2)$ making full use
of the Bethe-Salpeter formalism for weakly bound systems \pcite{topdec}.
As a first step,
we will use the original BR equation as zero order equation.
Therefore the calculations in this section are performed in the narrow
width approximation. However, we will convince ourselves in the next sections
by using the BR equation for decaying particles developed in sect. 2.1
that this actually gives the correct result.

We are able to take into account all terms to that order in
a systematic and straightforward manner. One of the previously not
considered contributions is obtained by explicit calculation
and cancels precisely gauge dependent terms
which appeared in an earlier \pcite{Sumino} off-shell calculation. We will see that
important cancellations also determine the gauge-independent part.

As proposed first in ref. \pcite{Fadin} the bound state effects
for the inclusive reaction $\s_{e^+ e^- \to t\bar{t}}$ near the
$t \bar{t}$ threshold could be taken
into account by including the width $\G_0$ for a freely decaying top quark
\pcite{Bigi} in the absorptive part of the vacuum polarization,
through the replacement of the total energy $E$ by
$E+i\G_0$. Subsequently this approach was applied in other detailed
numerical studies involving potentials with the running strong
coupling constant
for the short distance Coulomb term and also including a confining part
\pcite{Peskin,Sumino,Kwong}. However, as noted recently in \pcite{Sumino}, 
the introduction
of a constant decay rate $\G_0$ in the Breit-Wigner resonance part of
this formalism leads to problems with unitarity. Therefore a "running"
width $\G(p^2)$ for a virtual top
was proposed with $p^2 \ne m_t^2$.
As expected, a naive off-shell extrapolation of $\G_0$ leads to
ambiguous and gauge dependent results. $\G(p^2)$ was
determined indirectly from the vacuum polarization function
making certain approximations \pcite{Sumino} . Furthermore it was claimed 
that the running width would give rise to effects of the order of 10\% to 
the total cross section.
{\it Qualitatively} the main physical effects from $t$ decaying in
toponium seem to be quite well understood by now \pcite{Jeza}:
On the one hand, $\G$ decreases by time dilatation and by
reduction of phase space, caused by the decay below the mass shell.
Cancellations should occur with an increase of $\G$ due to the
Coulomb-enhancement induced by the Coulomb interaction of the
relativistic $b$-quark, because this overlap effect is known to be
important from the qualitatively similar muonium system \pcite{Ueber}.
The purpose of this chapter is to show that existing quantum
field theoretical technology, well tested e.g. in the (abelian) positronium
case, may be used also here to solve the problem of a gauge-independent
"running width", including the qualitative effects listed above.
Even the Coulomb enhancement can be taken into account properly in this
context, which to the best of our knowledge has not been possible before.

The correction to the decay width is obtained
from the imaginary part of the expectation values, receiving contributions
from the graphs in fig.\pref{dtopdec}.

\begin{center}
\leavevmode
\epsfysize=10cm
\epsfbox{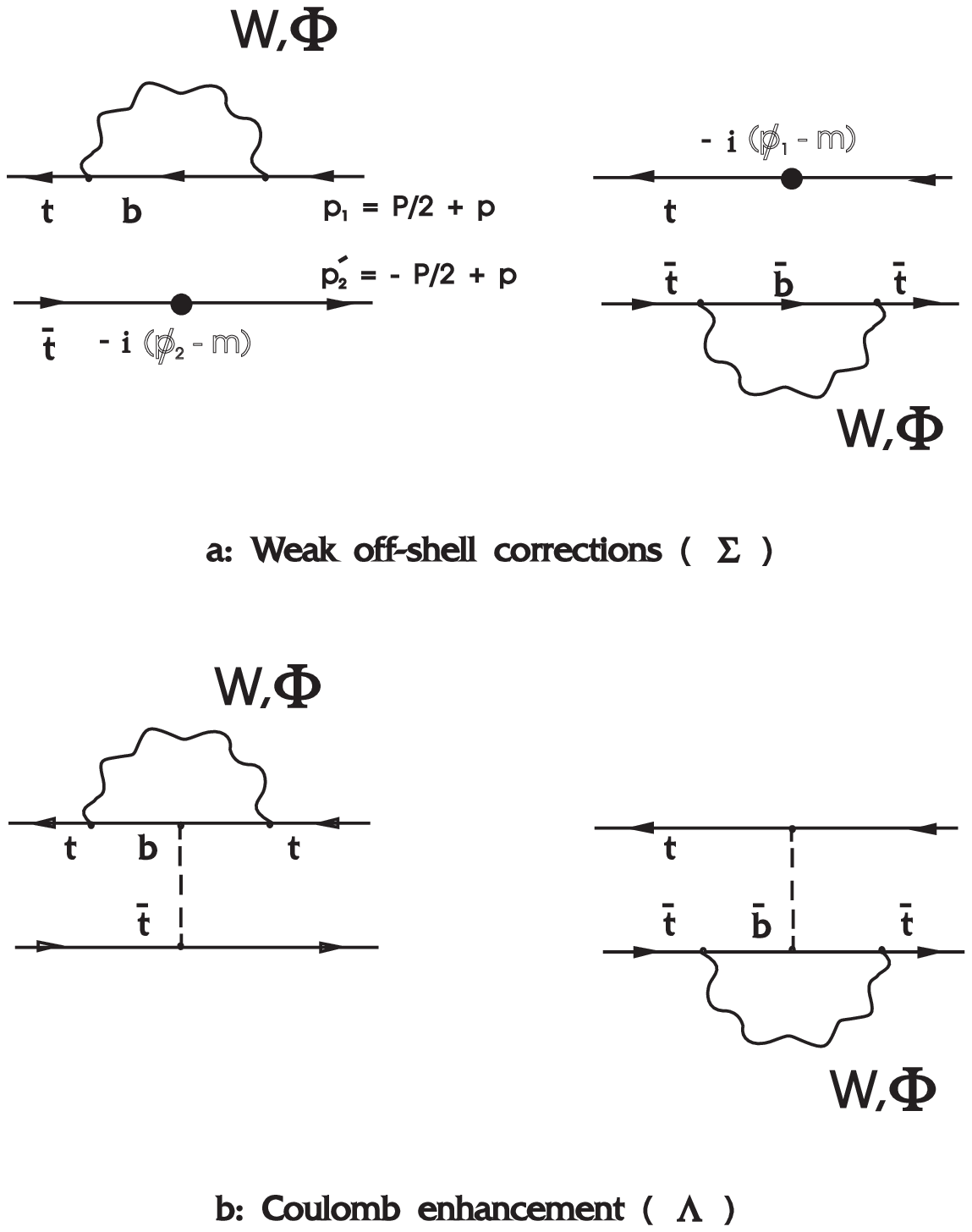}
\\
\centerline{ Fig. \refstepcounter{figure} \plabel{dtopdec} \thefigure}
\end{center}

It is important to note that eq. \pref{dM} is not a power series in $\a_s$.
In \pref{dM} it will be essential that $O(\a_s^2)$ corrections
arise from all three terms denoted here if the BR wave-functions are used
in lowest order.
It should be emphasized that starting from the Schr\"odinger equation
instead, the inclusion of the proper relativistic
corrections to the free fermion propagators would have necessitated
the inclusion of complicated higher order terms in $h$.
We also anticipate that in the present application of this formalism
for the imaginary part of \pref{dM}, $- \G/2$,
renormalization effects need not be considered.

The graph fig.\pref{dtopdec}.a gives rise to the BS perturbation  kernel
\beg \plabel{h1}
         H^{\S} = i [ \Sigma (p_1) \otimes (\ps_2-m) + (\ps_1-m) \otimes  
\Sigma (p_2)] (2\pi)^4 \d(p-p')
\ee
where $p_1=\frac{P_0} {2}+p$ and $-p_2=\frac{P_0} {2}-p'$ denote the
four momentum of the quark and the antiquark, respectively. The direct
product refers to the $t \otimes \bar{t}$ spinor space. The factors
$(\ps_1-m)$ and $(\ps_2-m)$ compensate the superfluous propagator on the
line without $\S$.

Since we are calculating higher order effects we neglect the mass of
the bottom quark although it may be included in principle.
For the electroweak theory we use the $R_{\xi}$-gauge with the
gauge fixing Lagrangian ( $M$ denotes the mass of the W-Boson)
$$
 \CL_{gf} = - \xi |\6^{\m} W_{\m}^+ - i \frac{M}{\xi} \phi^+ |^2
$$
because it eliminates mixed $W-\phi$ propagators. The gauge parameter
$\xi$ will not be fixed in the following.
We obtain for the imaginary parts of $\S$ in \pref{h1} ($s=\sin \theta_W$,
$\CP_{\pm} = (1 \pm \g_5)/2 ) :$
\begar
- \Im \S &=& \frac{e^2}{s^2} \frac{\ps}{16 \pi} \Th (p^2-M^2)
   [\CP_+ A^{(W)} +  \CP_- A^{(\phi)} ] \plabel{SW}
\ea
where (for $p^2 > M^2,\x >M^2/p^2$)
\begar
A^{(W)} &=& \t(p^2) + \r(p^2,\xi)  \nonumber   \\
A^{(\phi)} &=& m^2 ( \frac{\t(p^2)}{2 M^2} - \frac{\r(p^2,\xi)}{p^2} )  \plabel{As} \\
\t(p^2) &=&  \frac{(p^2-M^2)^2}{2 p^4}, \quad \r(p^2,\xi)=
 \frac{1-\x}{2\x} ( 1- \frac{M^2}{2p^2}\frac{1+\x}{\x}) \nonumber
\ea
In evaluating the expectation value according to the rule \pref{erww}
we encounter for both terms the same trace (cf \pref{L} and \pref{xBR} )
\begar
 T &:=& \2 \mbox{tr}[\g_0 \L^- S^{\dagger} \L^+\g_0 \CP_{\pm} \ps_1
                   \L^+S \L^- (\ps_2-m) ] = \nonumber \\
   &=& \2 (p_0+\o)[ m(1-\frac{\p\,^2}{m^2})+(p_0-\o)] + O(\a_s^4). \plabel{S}
\ea
It is seen to gives rise to (spin independent) relativistic corrections of the
same order as the off-shell ones from $\S$. Inserting \pref{As} into
\pref{SW} we may expand in powers of $(p^2 -m^2)$, because terms of
order $(p^2 -m^2)^2$ already yield negligible orders in $\a_s$:

\begar
- \Im \S^{(W)} &=& \frac{e^2m}{32 \pi s^2} \left[ (2+\frac{m^2}{M^2})
             \t(m^2) + C \frac{p^2-m^2}{m^2}  \right] + O( (p^2 -m^2)^2 ) \\
  C &=& ( 2 + \frac{m^2}{M^2} ) \sqrt{ 2 \t(m^2)} \frac{M^2}{m^2} +
            2 \r(m^2,\xi) \nonumber
\ea
Here $\Im \S^{(W)}$ denotes $\Im \S$ with $\ps \to m$ and without 
$\CP_{\pm}$.
Thus the total contribution of fig.\pref{dtopdec}.a to the decay width becomes
($h_0^{\S} = H^{\S} |_{P_0=M_{n,0}} $)
\begar \plabel{Gam1}
 \G_1 &=& -2 \Im \< \< h_0^{\S} \> \> = - 2  \< (1-\frac{5\p\,^2}{4 m^2} - \frac{3}{4} \s_n^2)  \Im \S \> = \\
      &=& \G_0 - 2 \G_0 \s_n^2 + \frac{e^2m}{16 \pi s^2} C  \< \frac{p^2 -m^2}{m^2}  \>,\nonumber
\ea
where $\G_0$ is the well known zero order result which is twice
the decay width of a free top quark \pcite{Fadin,Bigi}. A factor 2 arises
here and in the following, counting both self energy contributions
of the $t$ and $\bar{t}$ line:
\beg \plabel{Gam0}
 \G_0 = - 2 \Im \S^{(W)}(m) = \frac{e^2m}{16 \pi s^2} (1+\frac{m^2}{2M^2})
          \frac{(m^2-M^2)^2}{m^4}
\ee
The expectation value $\< \frac{p^2-m^2}{m^2} \>$, left over after
the $k_0$ integration, is taken in the ordinary sense, i.e.
between the (normalized) Coulomb wave functions $\phi_{nlm}(\k)$
in \pref{xBR}.

In the graphs shown in fig. \pref{dtopdec}.b to the required order it proves sufficient
to evaluate the leading contribution by applying the nonrelativistic
wave functions \pref{xapp}.
After a straightforward but tedious calculation we obtain for the
imaginary part of the vertex function
$\L_0^{(W)}$ ($ \xi > M^2/m^2$ and $1>M^2/m^2$):
\begar
 \Im \L_0^{(W)}&=& \frac{e^2}{32 \pi s^2 } \g_0 F \plabel{dirLa} \\
           F &=& (1+\frac{m^2}{2M^2})
            (1- \frac{M^2}{m^2})(1+ 3 \frac{M^2}{m^2}) + 4 \r(m^2,\xi).\nonumber
\ea
In the BS expectation value corresponding to the first term on the
r.h.s of \pref{dM} the $k_0$ integration to this order allows the
simple replacement $k_0 \to -\o$.
With the spinor trace  ($-\2$) after including the nonrelativistic
projectors in \pref{xapp} we arrive with $\Im \L_0^{(W)}$ at
($\q =\p- \p\,'$)
\begar \plabel{Gam2}
 \G_2 = \frac{e^2}{16 \pi s^2} \< \frac{4\pi\a}{\q^2} \> F.
\ea
Remembering that $\p = O(\a_s)$, the last expectation value of
eq. \pref{Gam1} can be rewritten
\begar
 \< \frac{p^2-m^2}{m^2} \> \approx -2 \< \s^2 +\frac{\p\,^2}{m^2} \>
                          = - \< \frac{2}{m} \frac{4\pi\a}{\q^2} \>, \plabel{20}
\ea
where the Schr\"odinger equation with reduced mass $m/2$ has been used in the
last equality. We note already here that with eq. \pref{20} the gauge
dependent terms proportional $\r(m^2,\xi)$ exactly  cancel in \pref{Gam1}
plus \pref{Gam2}!

So far only the first term in \pref{dM} has been
considered. The second order terms in $h$ for the leading correction
$O(\a_s^2 \a)$ can be summarized as
\begar
 \G_3 &=& -2 \Im [ \< \< h_0^{QCD}  g_1 h_0^{\S}\> \>+ \< \< h_0^{\S} g_1 h_0^{QCD} \> \>]  \plabel{Gam3}  \\
 \G_4 &=& -2 \Im [ \< \< h_0^{QCD}\> \> \< \< h_1^{\S}\> \> + \< \< h_0^{\S}\> \> \< \< h_1^{QCD}\> \> ] \plabel{Gam4}
\ea
For $g_1$ (cf. eq. \pref{gent}) in \pref{Gam3} we find it convenient to
write
\beg \plabel{g1}
 g_1 := \left[ G_0 - \frac{ \bar{\chi} \otimes \chi}{P_0-M_{n,0}} ) \right]_{P_0 = M_{n,0}}
\ee

We first evaluate the contribution $\G_{3}^{(1)}$ of the Green function $G_0$
expanded around
$P_0-M_n =2m \sqrt{1-\s^2} - 2m \sqrt{1-\s_n^2} \approx m (\s_n - \s^2)$.
For our purpose it suffices to calculate the leading nonrelativistic
contribution since relativistic corrections are
suppressed by an additional factor $O(\a_s^2)$. Therefore we can use the
nonrelativistic Green function $G_{nr}(k'',k',\s)$, i.e. essentially
Schwinger's solution $G_{nr}(\k\,'',\k\,',\s)$ to the Coulomb problem
\pcite{Schw}, supplemented
by appropriate factors $(k_0^2 - \o^2)^{-1}$ so as to include also the
zero component of relative momentum. For $P_0 \ne M_{n,0}$
and thus $\s \ne \s_n$ the first term of eq. \pref{g1} yields a contribution
(cf. \pref{xapp})
 \beg
 \G_{3}^{(1)} = -4  \int dk dk' dk'' \bar{\chi}_{nr}(k'') (k_0''+\omega''_n)
  ( i \Im \S) \, G_{nr}(k'',k',\s) \, h_0^{QCD}(k',k) \chi_{nr}(k)
      \nonumber
\ee
where again $-(k_0''+\omega''_n)$ represents the inverse nonrelativistic
propagator in fig. \pref{dtopdec}.a.
Carrying out the $k_0$ integrations leads to ($ d\k := d^3k/(2 \pi)^3$ etc. )
\beg
 \G_{3}^{(1)} = -2 \G_0 \int d\k d\k\,' d\k\,'' \Phi^*(\k\,'') [1-\frac{m(\s^2-\s_n^2)}{4\o''}]
   G_{nr}(\k\,'',\k\,',\s) h_0^{QCD}(\k\,',\k)  \Phi(\k),   \plabel{2}
\ee
if we use the fact that for $h_0^{QCD}$ to $O(\a_s^2)$
the QCD corrections in $H$ from  $K_{C} - K_{BR}$, Coulomb gluon exchange etc.
are real and independent of $k_0$.

The factor proportional to 1 from the square bracket in \pref{2} exactly reproduces the
second part of \pref{g1}, to be subtracted out before the limit $\s \to \s_n$
is taken. The limit $(\s^2-\s_n^2) \to 0 $ just selects the corresponding
pole term $\phi_n^* \phi_n [ \frac{m}{2} (\s^2-\s_n^2) ]^{-1} $ in $G_{nr}(\k\,'',\k\,',\s)$.
Thus from \pref{Gam3} only
\beg  \plabel{Gam3e}
  \G_{3} = \frac{\G_0}{2} \< h_0^{QCD} \> \< \frac{1}{\o} \>
\ee
remains. $\< h_0^{QCD} \>$ could be taken from refs. \pcite{KM1,Duncan}.
Its explicit form will turn out not to be relevant here.

For the first term of \pref{Gam4} $h_1^{\S}$, i.e. the derivative of
\pref{h1} with respect to $P_0$ at $P_0 = M_{n,0}$ is needed.
Again the factor $\< h_0^{QCD} \>$ provides additional powers of $\a_s$
and hence allows the use of the mass shell condition
$P_0 \to 2m$ in
\begar
 h_1^{\S} &=&  (2\pi)^4 i \d(p-p')[  \g_0  \S \otimes  (\2 \g_0) +
  (\2 \g_0)\otimes  \g_0  \S ].
\ea
Thus from this term
\beg \plabel{Gam41}
    \G_{4}^{(1)}  = -2 \< \< h_0^{QCD} \> \> \Im \< \< h_1^{\S}  \> \> \approx
     - \frac{\G_0}{2}  \< h_0^{QCD} \>   \< \frac{1}{\o} \>
\ee
follows.

In the second part of eq. \pref{Gam4} $h_1^{QCD}$, i.e. the first derivative
of $H^{QCD}$, will only receive contributions from
$H = -K_{C}+K_{BR}$ to $O(\a_s^2)$ . The instantaneous Coulomb kernel $K_{C}$
does not depend on $P_0$, nor does the  vacuum polarization \pcite{Duncan}
correction which had provided the leading contribution to $\< h_0^{QCD} \>$ above:
\beg
  \G_{4}^{(2)} = -2 [\Im \< \< h_0^{\S} \> \>] \< \< \frac{\6}{\6 P_0} K_{BR} \big|_{P_0 = M_{n,0}} \> \>.
\ee
Inserting the explicit (real) expression for the BR kernel $K_{BR}$, (eq. \pref{K0}
with $\G \to 0$ ), we obtain
\beg
\< \< \frac{\6}{\6 P_0} K_{BR} \> \>= \int \frac{d^3k}{(2 \pi)^3} | \phi (\k) |^2
              \frac{ P_0 - 2 E_k}{2 P_0} \approx -\frac{\s_n^2}{2}
\ee
and thus
\beg
  \G_{4}^{(2)} = - \2 \G_0 \s_n^2.
\ee

Before adding all contributions we note that precisely the contribution
$\G_1$ in \pref{Gam1} has been quoted already in ref. \pcite{Sumino}
as a result as a specific off-shell extrapolation of the free decay width.
It was obtained there by the intuitive replacement $\ps \to \sqrt{p^2}$ in
the on-shell expression. However adding $\G_2$ of \pref{Gam2}, with \pref{20} {\it all} gauge dependent
terms are found to cancel. Moreover a striking feature of the sum
$\G_1 + \G_2$ is that even the {\it gauge-independent } terms precisely
sum to zero as well, which is by no means "natural" in view of the
explicit form of $F$ and $C$, as compared to $\G_0$ of \pref{Gam0} occuring
in the second term of \pref{Gam1}.
In a similar manner also $\G_3+\G_4^{(1)}$ cancel each other
(cf. eq. \pref{Gam3e} and \pref{Gam41}), leaving only $\G_4^{(2)}$
in the final result
\beg  \plabel{erg}
  \G_{bound state} = \G_{0}(1-  \2  \s_n^2 + O(\a_s^3)).
\ee
Recalling that $1-  \s_n^2/2 \approx (1 - \< \k^2/m^2 \>)^{1/2} $, the
residual effect may be interpreted as an approximation of the
$\g$-factor for the time dilatation for the weakly bound top.
The origin of that term, the relativistically generalized Coulomb-kernel
$K_{BR}$ gives some credit to this interpretation.
For the complete cancellation of the other terms we will present
a convincing explanation in the next section.

Among the cancelled terms the other expected effects can be more or less
clearly
isolated: $\G_1$ (fig. \pref{dtopdec}.a) includes the off-shell corrections and the decrease
due to the phase space reduction. But this is here completely cancelled by the
"Coulomb enhancement" (fig. \pref{dtopdec}.b). It should be emphasized, though, that
bound-state effects of the $b \bar{t}$ or $b \bar{b}$ systems are, of course, irrelevant
for total energies near the $t \bar{t}$ threshold.
Our systematic perturbative approach also correctly describes the toponium
decay to $O(\a_{weak})$ as a sum of the decays widths $t \to b+W^+$ and
$\bar{t} \to \bar{b}+W^-$ (cf. the sum of the terms in fig.\pref{dtopdec}).

A reader used to phenomenological calculations may wonder why no running
coupling in $\a_s$ or even a more "realistic" potential involving
a confining piece has been used. The simple answer is that any generalization
of this type could in principle completely ruin the strictly perturbative approach
(and also the built-in gauge-independence) advocated her, where e.g.
the gluon vacuum polarization (cf. $h_0^{QCD}$ above ) must be accounted for
as a separate perturbation and must under no circumstances be mixed
with higher order leading logs in perturbation theory, as summarized in
a running $\a_s$. However, we will show in the next section, that this
limitation can be circumvented in the case at hand.

Anyhow, since the real part of the weak bound-state corrections
can be estimated to be small, the net effect of our result \pref{erg}
for the discrete levels of toponium allows a simple interpretation:
It consists of the (level-independent) free decay width $\G_0$ and a very
small imaginary correction factor $1+i \G_0/2m$ to the Rydberg energy.


\subsection{Including the Width in the Zero Order Equation}

In this chapter the bound state problem
for a fermion-antifermion system is considered
including a finite decay width of the constituents already in the zero order
equation as outlined in chapter 2. We focus especially on the $t\bar{t}$
system for which we reconsider the explicit calculation of the bound state
corrections to the toponium width of the last section, which was performed
in the narrow width
approximation and needed the use of second order Bethe-Salpeter perturbation
theory.  We find that one obtains the same result already in first order BS
perturbation theory if one uses our present approach. This also explicitly
shows the dependence of any result at a fixed order of BS perturbation
theory on the chosen zero order equation.

Furthermore the extensive
cancellations of gauge dependent terms turn out to be a consequence of a
Ward identity. This cancellation mechanism is shown to be valid for
general fermion-antifermion systems in the next section.

The correction to the decay width is obtained from the imaginary part of the
expectation values, receiving now contributions from the graphs in fig.\pref{dbrig}
subtracted by the zero order propagator and from fig. \pref{dtopdec}.b.

\begin{center}
\leavevmode
\epsfysize=5cm
\epsfbox{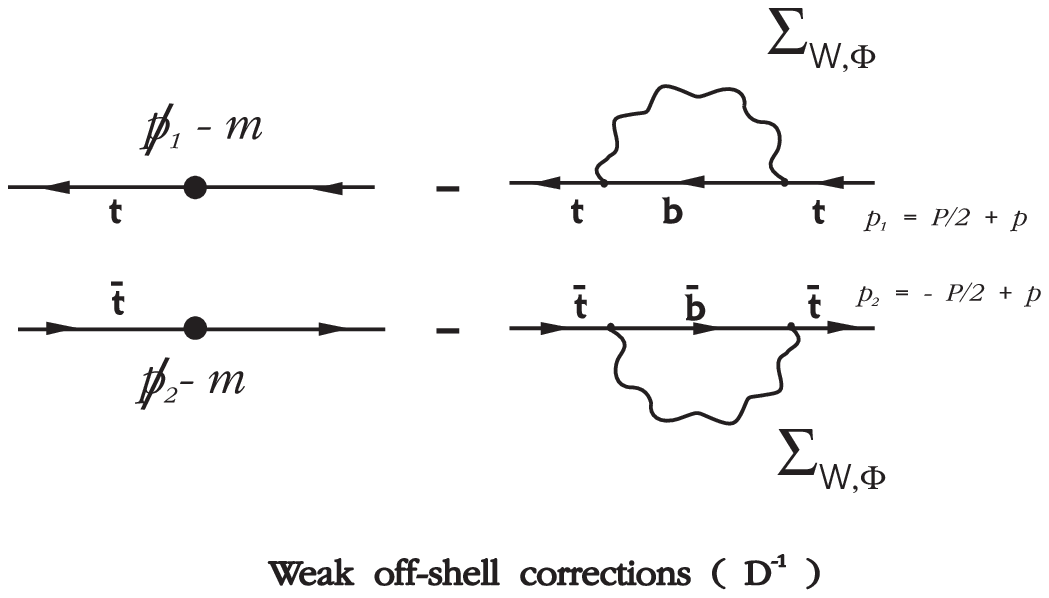}
\\
\centerline{ Fig. \refstepcounter{figure} \plabel{dbrig} \thefigure}
\end{center}

Corrections to the fermion propagator have to be included via the perturbation
kernel
\beg
h_0^{(1)} = ( i D^{-1} -i D_0^{-1} ) \Big|_{P_0 = M_n - i \G}.
\ee
We emphasize that the perturbation kernel now has to be evaluated at
$P_0 = M_n - i \G$.
With eq. \pref{H} and $D^{-1}_0 \big|_{P_0 = M_n - i \G} = D^{-1}_0
\big|_{\G=0}$  we find
\beg \plabel{h1g}
h_0^{(1)} = - i [ (\Sigma (p_1)+i \frac{\G}{2} \g_0 ) \otimes (\ps_2-m) + 
(\ps_1-m)
\otimes  (\Sigma (p_2)-i \frac{\G}{2} \g_0 ) ] (2\pi)^4 \d(p-p')
\ee
where $p_1=\frac{P_0} {2}+p$ and $-p_2=\frac{P_0} {2}-p$ denote the four momentum
of the quark and the antiquark, respectively. The direct product refers to the
$t \otimes \bar{t}$ spinor space. The factors $(\ps-m)$ and $(\pss-m)$
compensate the superfluous propagator on the line without $\S$. We have
neglected terms that will contribute to $O(\a^2 \G^2 /m)$.

Now we observe that the terms in \pref{h1g} which include $\S$ have already been
calculated in the last section except for the imaginary parts occurring in the
trace
\begar
T &:=& \2 \mbox{tr}[\g_0 \L^- S^{\dagger} \L^+\g_0 \CP_{\pm} \ps_1 \L^+S \L^-
(\ps_2-m) ] = \nonumber \\ &\approx& \2 (p_0+\o)[
m(1-\frac{\p\,^2}{m^2})+(p_0-\o)-i \frac{\G}{2}].  \plabel{Spur}
\ea
and in
\begar
p^2-m^2 \approx 2m p_0 -\p\,^2 - m \s^2 -i m \G. \plabel{pmm}
\ea
We will be able to neglect these imaginary parts in the following since they
either give rise to real corrections ( which we are not interested in) or
to imaginary ones from Re$\S$, both of which are of $O(\a_{weak}^2)$.
Of course, these corrections would be required a in
complete calculation of the toponium decay width to this order in order to
obtain a gauge independent result.  This goes beyond the scope of the present
work. Furthermore, these corrections are state independent, expected to be
very small and therefore of minor practical interest, anyhow.
As can be seen from \pref{Spur} the axial part of $\S$ does not contribute within
the expectation value.

Therefore, the contribution from first order BS perturbation theory to the decay
width to $O(\a_s^2 \G)$ reads
\beg
\< \< h_0^{(1)} \> \> = \G_1 + \G_3
\ee
with $\G_1$ as defined before
\begar
\G_1 &=& -4 \Im \< \< - i \Sigma (p_1) \otimes (\ps_2-m) (2\pi)^4 \d(p-p') \>\>=\\
&=& \G_0 - 2 \G_0 \s_n^2 + \frac{e^2m}{16 \pi s^2}
[ ( 2 M^2 + m^2 ) \frac{m^2-M^2}{m^4} + 2 \r(m^2,\xi)]
\< \frac{p_1^2 -m^2}{m^2}\>,\nonumber
\ea
and a new contribution from the inverse zero order propagator
\begar
\G_3 &=& -4 \Im \< \<  \frac{\G}{2}  \otimes (\ps_2-m) (2\pi)^4 \d(p-p') \>\> =
- \G_0 ( 1+ \frac{\s_n^2}{2})
\ea
Because of the inclusion of $\G$ in the zero order equation now
in the sum $\G_1 +\G_3$ the leading $\G_0$ terms cancel and the resulting
correction is already of $O(\a_S^2 \G_0)$.

Now we turn to the calculation of the vertex correction (fig.
\pref{dtopdec}.b). This correction remains unchanged. However,
in sect. \pref{topdec} we calculated $\Im \L_0 := \lim_{q \to 0, p^2 \to m^2}
\L(p,q)$ directly. Here we will show
that it is possible to use a Ward identity to determine this correction.
This also provides a good check for our explicit calculation.

We observe that
\begar
\L_0^a(p,q) \Big|_{q \to 0 } &=& - g_{QCD} T^a \frac{\6}{\6 p_0} \S (p)
\plabel{WI}
\ea
holds at least at the one-loop level. This identity is believed to
hold to all orders if one takes into account the equivalence of
Coulomb gauge calculations in the usual and in the background field
method and the Ward identities derived in the latter approach \pcite{Kreu}.

In the graphs shown in fig. \pref{dtopdec}.b to the required
order it proves sufficient to evaluate the leading contribution by putting the
fermions on the mass shell and applying the nonrelativistic wave functions
\pref{xapp}.
Using the explicit expression \pref{SW}  for $\Im \S$ we obtain with the
help of eq. \pref{WI}
\begar
\Im \L_0 &=& \frac{e^2}{s^2 16 \pi} \g_0 [ A^{(W)}(m^2) + A^{(\phi)}(m^2) +
2 m^2 \frac{\6}{\6 p^2} ( A^{(W)}(p^2) + A^{(\phi)}(p^2) ) \Big|_{p^2=m^2} ]
\nonumber \\
&=& \frac{e^2}{32 \pi s^2 } \g_0 F \\ F &=& (1+\frac{m^2}{2M^2}) (1-
\frac{M^2}{m^2})(1+ 3 \frac{M^2}{m^2}) + 4 \r(m^2,\xi).
\nonumber
\ea
This result agrees with that of \pref{dirLa} and one obtains a
correction to the decay width:
\begar \plabel{Gam22}
\G_2 = \frac{e^2}{16 \pi s^2} \< \frac{4\pi\a}{\q^2} \> F.
\ea

The fact that the gauge dependent terms in the sum $\G _1 + \G_2$ cancel is
thus traced back to the identity \pref{WI}. The gauge independent contribution
from first order BS-perturbation theory is now different from that of
\pcite{topdec} (actually it gives now the net
result) in accordance with the different zero order equation, used in our
present context.

To complete our calculation we finally check the contributions of second order
perturbation theory. It can be shown that due to the fact that the first weak
correction now is already of order $O(\a_s^2 \G)$, this corrections do not
contribute to the required order.  Thus our present approach  
considerably simplifies the
calculation of bound state corrections to the decay width.


Summing up all contributions we get the result
\beg \plabel{result2}
\D \G_{boundstate} = \G_1 +\G_2 +\G_3 = - \G_0  \frac{\s_n^2}{2}
\ee
in agreement with our previous calculation in the last section and
\pcite{topdec}.

\subsection{A General Theorem on Bound State Corrections}

The above considerations suggest that the result \pref{result2} is only a
special case of a more fundamental statement, with broader range of
applicability, which we will now derive \pcite{KM2}.

Consider a fermion whose decay can be described by the imaginary part of a self
energy function which will have the general form (in any covariant gauge in the
relevant sector of the theory)
\beg
\S(p) = \S_S(p^2) + \ps \S_V(p^2) + \g_5 \S_P(p^2) + \ps \g_5 \S_A(p^2).
\ee
Within the expectation value \pref{erww} the pseudoscalar and axial vector
parts vanish. Furthermore we will drop all factors $\Theta(p^2-\m^2)$, which is
a valid approximation in all cases where $\a^2 m \ll m- \m$. If this is not the
case, either the whole calculation makes no sense because then one would have to
include the entire rung of Coulomb interactions of the remaining particle with
the heavy decay products or the decay width becomes very small, or both.
Let us therefore consider the correction to the decay width which results from
the remaining parts of the self energy in an on-shell renormalization scheme (
$m$ is the pole mass of the fermion). We expand $\S$ around that mass shell:
\beg \plabel{sigall}
\S(p) = i (\Im \S_S + \ps \Im \S_V) + (\S_S'+\ps \S_V')(p^2-m^2) .
\ee
Terms of $O(\a_{weak}^2)$ are understood to be neglected (cf. remarks after
eq. \pref{pmm} ) and we denote
\begar
\S_X &:=& \S_X(m^2), \qquad X = S,V \nonumber \\ \S_X' &:=& \frac{\6}{\6 p^2}
\S_X(p^2) \Big|_{p^2=m^2}. \nonumber
\ea
From eqs. \pref{WI} and \pref{sigall} we can easily calculate the vertex
correction, say, for the upper particle line:
\beg
\L_0 \l^+ = g T^a \g_0 [i \Im \S_V + 2m (\S_S'+m \S_V')] \l^+
\ee
After some algebra we can write to the required order
\begar
h_0^{(\S)} &=& (\Im \S_S + m \Im \S_V + \frac{\G}{2}\g_0 ) \otimes (\ps_2-m) - i
D_{0,\G=0}^{-1} [i \Im \S_V + 2m (\S_S'+m \S_V')] \nonumber \\
\\
h_0^{(\L)} &=& -\frac{4 \pi \a}{\q\,^2} [i \Im \S_V + 2m (\S_S'+m \S_V')] \g_0
\otimes \g_0 \nonumber
\ea
Since
\beg
\G = -2 \bar{u} \Im \S u \Big|_{p = (m,\vec{0})}= - \Im \S_S(m^2) - m \Im
\S_V(m^2)
\ee
and $ (i D_{0,\G=0}^{-1} - K_{0,\G=0}) \chi = 0$ we obtain for the bound state
correction to the decay with to $O(\a_s^2 \a_{weak})$
\beg  \plabel{erg2}
\D \G_{boundstate} = -4 i \G \< \< \l^- \otimes (\ps_2-m) \>\>=
-\G_0 \frac{\< \p\,^2 \>}{2 m^2}
\ee
Now it is clear from the preceding argument that it applies not only
to the toponium system but also to all other
systems which can be described as weakly bound fermion- antifermion
system with unstable components. Another example which is known for a long
time is provided by the bound muon system where the above result has also been
obtained first by explicit calculation \pcite{Ueber}.  
This result is easily reproduced from \pref{erg2} by replacing $\G_0$ by
the free muon decay width and $m$ by the muon mass. The effect of the
reduced mass (in our case $m/2$, negligible for the bound muon ) is
entirely included in the expectation value.
We conclude that the
formalism developed here and especially the use of the identity \pref{WI}
simplifies the problem of the bound state
correction to the decay width in a profound way. It is now possible to
clearly isolate the underlying cancellation mechanism which automatically
gives a gauge independent result which can be interpreted as time dilatation
alone. However, because we may now also include the scalar part of the
self energy function and since $K_0$ can be any kernel, the result
\pref{erg2} is found to comprise \pcite{topdec} and \pcite{Ueber} as
special cases of a more general theorem: The leading bound state corrections
for weakly bound systems of unstable fermions (with decays like
$t \to b + W^+, \mu^- \to e^-+\bar{\n}_e+\n_{\mu}$) are {\it always}
of the form \pref{erg2}.
 Among the consequences we especially note that \pref{erg2}
may be safely applied to toponium even if one uses e.g. a
renormalization group improved potential.
While we have presented a field theoretically consistent picture of the
"smearing" of the individual levels in toponium below threshold, it must be
stressed that the treatment of the entire threshold region cross
section also requires an extension of our study to the
"continuum" $P_0 > 2m_t$ for which the existing BS formalism must be
adapted.

\chapter{Heavy Quark Production in $e^+ e^-$ Colliders}

In the last chapters we calculated the potential and bound state
corrections to the decay width with rigorous field theoretical methods.
These methods will be necessary to compare theory and high precision
measurements. However, for the near future \pcite{MAH}
it will probably be sufficient
to apply existing less rigorous methods \pcite{Fuji}.
In the case of quickly decaying particles near threshold the Green
function method \pcite{Khoze,Peskin,Sumino} seems appropriate.
Within this approach the four point function is approximated by a
nonrelativistic Green function fulfilling an ordinary Schr\"odinger
equation with an appropriate potential. This method works to $O(\a_s)$
since the respective corrections to the decay width and potential as
well as the hard corrections decouple to this order \pcite{Melni,Sumith}.
At $O(\a_s^2)$ this decoupling can be expected to break down  and thus
it turns out that
the calculation of the cross section of decaying particles
to order of 1\% requires a deeper understanding of the relativistic
effects near threshold. A completely satisfying and practically
tractable method does not yet exist.

In this chapter we will investigate the production cross section of a
$t\bar{t}$ pair near threshold. Whenever possible we will use the
results of the foregoing chapters.

The cross section for the process $e^+ e^- \to X $, where $X$ is a set of
final states $X_n$ can be written as
\begar
\s &=& \frac{1}{2 K(P^2)} \sum_{n} |\< X_n|T|e^+ e^-\>|^2
\ea with
\begar
K(P^2)&=& \sqrt{P^2(P^2-4m_e^2)}.
\ea
If we neglect initial state interaction and calculate the matrix element
to the first order in the electromagnetic coupling we can write for the
s-channel:
\begar
M = \< X|T|e^+ e^-\> &=& (2\pi)^4 \d(P-t-t') \bar{v}_\a(t') \left[  e \g_{\mu}
      \frac{g^{\m \n}}{P^2} \< X|J_{\n}^{(\g)}(0) |0 \> \right. -
           \plabel{Me} \\
        & & - \left. \frac{e}{2 s c} \g_{\mu} (v_e-a_e \g_5) \frac{g^{\m \n}}
            {P^2-M_Z^2+i M_Z \G_Z} \< X|J_{\n}^{(Z)}(0)|0 \> \right] u_{\beta}(t)
           \nn
\ea
The SM values for the vector and axial coupling of the Z-boson
read
\begar
v_f &=&  T^3_f - 2 Q_f \sin^2 \th_w \\
a_f &=& T^3_f
\ea
where $T^3_f$ is eigenvalue of the third component of the weak isospin
operator. For the top quark we have $T^3_t = 1/2$ and $Q_t=2/3$.

We are only interested in the contributions of one flavor to the
total cross section.  Investigating the top quark contribution, we
have to replace the currents which in general include a sum over all
kinds of particles by the current involving only the top quark
field operator $\Psi(x)$. Splitting the currents into vector and
axial vector parts we get
\begar
J_{\n}^{(\g)}(x) &=& i q_t e J_{\n}(x), \qquad
J_{\n}^{(Z)} = \frac{-i e v_t}{2 s c} J_{\n}(x) + \frac{i e a_t}{2 s c}
       J_{\n}^5(x) \nn \\
J_{\n}(x) &=&  :\bar{\Psi}(x) \g_{\n} \Psi(x):
, \qquad  J_{\n}^5(x)=\;\;:\bar{\Psi}(x) \g_{\n} \g_5 \Psi(x):
  \plabel{js1}
\ea
where from now on we abbreviate
\begar
 s &=& \sin \th_w \\
 c &=& \cos \th_w \nn
\ea
Substituting eq. \pref{js1} into eq. \pref{Me} we get for the matrix 
element $M$
\begar
M &=& (2\pi)^4 \d(P-t-t') \bar{v}_\a(t') (-i e^2) \frac{\g^{\m}}{P^2} \left[
     (a^{(V)}+b^{(V)}\g_5) \< X|J_{\m}(0)|0 \> \right. \plabel{M} \\
 & & + \left. (a^{(A)}+b^{(A)}\g_5)  \< X|J_{\m}^5(0)|0 \> \right]
     u_{\beta}(t)  \nonumber
\ea
where
\begar
 a^{(V)} &=& q_t - \frac{v_t v_e P^2}{4 s^2 c^2 z}, \qquad
       b^{(V)}=\frac{v_t a_e P^2}{4 s^2 c^2 z}, \plabel{av} \\
 a^{(A)} &=& \frac{a_t v_e P^2}{4 s^2 c^2 z}, \qquad b^{(A)}=-\frac{a_t
 a_e P^2}{4 s^2 c^2 z},\plabel{aa} \\
 z &=& P^2 -M_Z^2+i M_Z \G_Z , \plabel{z}
\ea
and $t$, $t'$ denote the momenta of electrons and positrons, respectively.
Since the mass of the top quark is much larger than the mass of the
Z-boson we can neglect the width of the Z-boson in \pref{z}. This means
that we can treat the $a^{(X)}$ and $b^{(X)}$ as real valued.

For the production of heavy scalars the above formulas can also be used
, but only the vector current is present and has to be replaced by

\beg
J_{\m}(x) = i :( \Phi^*(x) \6_{\m} \Phi(x)-\6_{\m}\Phi^*(x)\Phi(x)):
\plabel{js2}
\ee
where $\Phi$ represents the scalar field for a given mass eigenstate.
To be definite we will identify this scalar field in the following with
the lightest scalar partner of the top quark as predicted by 
the minimal supersymmetric SM.
For a given value $\th_t$ of the L/R mixing angle, the vertices for the
lighter particle can be written as $ i e \tilde{Q} (p_1^{\mu}+p_2^{\mu})$
with \pcite{Been}
\begar
 \tilde{Q}_{\g} &=& - q_t, \\
 \tilde{Q}_{Z} &=& (\cos^2 \th_t- 2 q_t \sin^2 \th_W)/(2 s c).
\ea
The coefficients $a^{(V)}$ and $b^{(V)}$ have thus to be replaced by
\begar
a_s &=& -q_t + \frac{\tilde{Q}_{Z} P^2}{2 s c z}, \qquad
       b_s=-\frac{\tilde{Q}_{Z} a_e P^2}{2 s c z},
\ea
respectively.
The lepton trace for positrons with polarization $\r'$ and electrons
with polarization $\r$ ($m_e \to 0$) can be readily calculated to yield
\begar
L^{\m \n}&=& \4 Sp[(1-\r' \g_5) \g^{\s}\g^{\m}(a^{(X)}+b^{(X)}\g_5)(1+\r \g_5)
                   \g^{\t}\g^{\n}(a^{(Y)}+b^{(Y)}\g_5)] t'_{\s} t_{\t}=
        \plabel{Lmn}  \\
        &=& [ (a^{(X)} a^{(Y)} + b^{(X)} b^{(Y)})(1-\r \r')+(a^{(X)} b^{(Y)}
        + b^{(X)} a^{(Y)} )(\r-\r')](t'^{\m}t^{\n}+
              t^{\m}t'^{\n}- t'^{\s}t_{\s } g^{\m \n}) - \nonumber \\
        & & -i [(a^{(X)} b^{(Y)} + b^{(X)} a^{(Y)} )(1-\r \r')+(a^{(X)}
            a^{(Y)} + b^{(X)} b^{(Y)})(\r-\r')]
             \e^{\a\m\beta\n} t'_{\a}t_{\beta} \nonumber
\ea

This simplifies in the CM frame to a purely space like tensor.
Let us now focus on the total cross section. Since $P^i$ is zero in the 
CM frame,
the hadronic tensor $T_{i k}$ has to be proportional to $\d_{i k}$
and therefore the antisymmetric part in $L$ vanishes in the calculation
of the total cross section. However, this part of the lepton trace will
lead to a forward backward asymmetry as discussed below.
It is reasonable to expect that a future linear $e^+ e^-$ collider will
work with polarized electrons and unpolarized positrons. Therefore we will
assume this scenario in the following. The generalization to the general
case is obvious from \pref{Lmn}.

For unpolarized positrons ($\r'=0$) the symmetrical contraction of $L$
simplifies to
\begar
 L^{i k}\d_{i k} &=& [a^{(X)} a^{(Y)} + b^{(X)} b^{(Y)} + (a^{(X)} b^{(Y)}
        + b^{(X)} a^{(Y)} ) \r] P^2
\ea

The modulus squared of the hadronic matrix elements can be simplified 
\pcite{Muta} according to

\begar
 \sum_{X} (2 \pi)^4 \d(P-P_X) \< 0| j_{\m}^A(0) |X \> \< X| j_{\n}^B(0)|0 \> =
       2 \mbox{abs} T_{\m\n}^{AB}
\ea
where
\begar
T_{\m\n}^{XY} &=& i \int d^4 x e^{i P x} \< 0|T j_{\m}^X(x) j_{\n}^Y(0)|0 \>=
    \int \frac{d^4p}{(2\pi)^4} \frac{d^4p'}{(2\pi)^4} Sp[\G^X_{\m} 
    G(P,p,p') \G^Y_{\n}] = \nn \\
    &=:& (\frac{P_{\m} P_{\n}}{P^2} - g_{\m \n} ) G^{XY}(P)
\ea
and the indices $X,Y$ refer to the vector and axial currents as defined
in \pref{js1} and \pref{js2}.  Here we made explicit the connection to the four
point function $G(P,p,p')$ defined in \pref{Gfeyn}.
Due to parity conservation the
mixed V-A currents vanish to all orders in the strong coupling
constant. The electroweak interaction may well introduce parity
non-conserving effects but near threshold the heavy particles should
be almost on shell and thus such effects should be suppressed at least
by one order in the electroweak coupling for the total cross section.
Thus we can split off the total cross section into a purely vector
and axial-vector contribution.
\beg
  \s = \s_V +\s_A \plabel{stot}
\ee
For fermions the vector part is the dominant one near threshold, since
it allows the production of S-waves (with spin $S=1$). The axial contribution
is suppressed by two powers in $\a_s$.

For scalars we have only the vector contribution, but in the
nonrelativistic limit the coupling proportional to spatial components $p^i$
(in the CM frame) selects $l=1$ states. Thus the production of scalars
is also suppressed by two powers in $\a_s$.

Collecting all ingredients from above and multiplying by a color factor
$N=3$ we have for the contribution to the total cross section from
the flavor under consideration :

\begar
 \s_X &=& \frac{72 \pi}{P^2} [ c_X + d_X \r]  \Im G^{XX}(P) 
\s_{\m\bar{\m}}
   \plabel{stotx} \\
\s_s &=&  \frac{72 \pi }{P^2}  [c_s + d_s  \r]  \Im G_s(P) \s_{\m\bar{\m}}
\ea
with
\begar
  \begin{array}{ll}
   c_X := (a^{(X)})^2  + (b^{(X)})^2 & c_s = a_s^2  + b_s^2 \\
   d_X := 2 a^{(X)} b^{(X)} & d_s = 2 a_s b_s
  \end{array}
\ea
and
\beg
 \s_{\m\bar{\m}}= \frac{4 \pi \a_{QED}^2(m_t)}{3 P^2}
\ee

\section{Top - Antitop Production Cross Sections Near Threshold}
\plabel{ttcross}

\subsection{Total Cross Section}

For the calculation of the total cross section
for $t\bar{t}$ production in the threshold  region the large width
of the toponium system plays a crucial role. It has the disadvantage that
individual levels disappear since the width becomes comparable to the
$1s-2s$ splitting. While corrections to each pole, which
lie now in the unphysical sheet, as described above, are in principle
possible, such a calculation becomes untractable in practice.
However, as already mentioned in the introduction, the large width
provides an effective cut-off for nonperturbative corrections and
renormalons. This will be demonstrated explicitly in the following.
To circumvent the calculation of corrections separately for each level and 
afterwards summing those to obtain the Green function, it is convenient to 
calculate it numerically
and use eq. \pref{stotx} for the total cross section.

As a first approximation in the region near threshold it should be possible
to replace the exact Green function by the nonrelativistic one. This
Green function fulfills the Schr\"odinger equation
\beg \plabel{Schroed}
\{ -\frac{\Delta}{m} + V(r) -E- i\G \} \tilde{G}(\vec{r},\vec{r}\,') =
         -\d(\vec{r}-\vec{r}\,')
\ee
For the potential $V(r)$ we would like to use the potential \pref{potres}
since it is e.g. a systematic expansion in the strong coupling constant
and thus is expected to lead to gauge independent results. However,
the inclusion of higher order corrections will lead to some problems,
as we will see below.

The leading contributions to \pref{stot} arises from the vector coupling
(e.g. $X=V$ in \pref{stotx}) since it produces $t\bar{t}$ pairs 
with angular momentum zero (S-waves). Therefore we are interested in
$l=0$ solutions of eq. \pref{Schroed}. Due to the symmetry of the
Green function and the requirement of regularity at $r \to \infty$ and
at the origin, the general solution of eq. \pref{Schroed} can be written as
\begar
 \tilde{G}(\vec{r},\vec{r}\,') &=& \sum_{l=0}^{\infty} g_l(r,r')  \sum_{m=-l}^l Y_{lm}^*(\O')
             Y_{lm}(\O) \plabel{gsum} \\
      g_l(r,r')  &=& \frac{ g_<(r_<) g_>(r_>)}{r_< r_>}, \qquad
       r_> = \left\{ \begin{array}{c} r: r> r' \\ r': r'> r \end{array}
       \right. \qquad
       r_< = \left\{ \begin{array}{c} r: r< r' \\ r': r'< r \end{array}
       \right.
\ea
where $ g_<(r_<)$ and $g_>(r_>)$ are regular solutions of the homogeneous
equation
\beg \plabel{gl}
 \{ \frac{\6^2}{\6 r^2} - m(E+i\G-V(r)) -  \frac{l(l+1)}{r^2} \} g(r) = 0
\ee
 at $r=0$ or
$r \to \infty$, respectively. The actual behavior of the solution
$ g_<(r_<)$ (and also of the irregular one) for $r \to 0$  depends on
the potential and on $l$. Due to the presence of the width $i\G$ the
behavior of $ g_>(r)$ for $r \to \infty$ is given by
\begar
  \lim_{r \to \infty} g_>(r) \propto e^{-a_-r}
\ea
with
\beg
 a_- := \sqrt{\frac{m}{2}} \sqrt{ - E + \sqrt{E^2+\G^2}}.
\ee
This exponential damping behavior is the origin of the infrared cut
off mentioned already
in \pcite{Khoze}. Consider for simplicity the case $E=0$. Then we can
write
\begar \plabel{damp}
a_-^{-1} = \sqrt{\frac{2}{m \G}} =  \sqrt{\frac{E_B}{\G}} r_B
\ea
where $E_B=m \a^2/2$ and $r_B$ are the Bohr energy and the Bohr radius of
the system, respectively . This formula clarifies
that if the width $\G$ is approximately equal to $E_B$ the form
of the potential is only "tested" up to the order of magnitude of the
Bohr radius. The situation is improved for $E< -\G$ since here we
have 
$$ a_-^{-1} \le \sqrt{\frac{1}{m|E|}}, \quad  E< -\G.$$
This relation also holds also for small $\G$ and thus the Bohr radius
again becomes the relevant scale.

The matching condition for $g_<$ and $g_>$ is provided by the $\d$
distribution in eq. \pref{Schroed}:
\beg \plabel{Wronski}
 -m =  g_<(r) g_>'(r)- g'_<(r) g_>(r)
\ee
Numerically it is possible to obtain the regular solution at the origin
directly by imposing suitable boundary conditions. For the singular
solution we use the following method \pcite{Peskin,Sumino}.
Suppose you have two solutions:
$ g_<(r)$ as above and $u_2(r)$ an solution determined by an arbitrary
boundary condition. Then the solution $g_>$ is given by
\beg
 g_>(r) = c[ u_2(r) + B  g_<(r)  ]
\ee
with
\beg B = -\lim_{r \to \infty} \frac{u_2(r)}{g_<(r)}.
\ee
The constant $c$ can be determined from eq. \pref{Wronski}. For $l=0$ and
a Coulomb-like potential at the origin we arrive with the boundary conditions
\begar
 g_<(0) &=&0 \\
 g_<'(0) &=& 1 \nonumber
\ea
and arbitrary ones for $u_2$ at the result
\begar
 \tilde{G}_{l=0}(\vec{r},0) &=& \frac{m}{4 \pi w} \frac{u_2(r) + B  g_<(r)}{r},
   \plabel{sGr}\\
 w &=& u_2 g_<' - g_< u_2'.
\ea
According to eq. \pref{stotx} the total cross section is proportional to
$\Im \tilde{G}(0,0)$. With this knowledge it is possible to calculate the total
cross section near threshold for any given potential $V(r)$. We have
written a numerical routine for "MATHEMATICA$^{TM}$" which has been 
checked to give the correct answer for the known Green function of a 
purely Coulombic potential.

Before presenting some numerical studies, a short remark on the coupling
constant to be used seems in order. It is customary to take 
the strong coupling constant $\a_s$
in the $\overline{\mbox{MS}}$ scheme. Especially experimental determinations
are always given in terms of $\a_s^{\overline{MS}}(M_Z)$. But in bound state
calculations it is natural to use an $\a_s$ defined differently as we did in
the foregoing. Clearly it should be possible to relate the two schemes.
To avoid complications from heavy fermion masses we treat 4 flavors as
massless. The difference in the bottom quark contribution can be expected 
to be smaller than the experimental uncertainties in $\a_s$. Therefore
we ignore it in the following consideration.
The quark-antiquark potential  in the $\overline{\mbox{MS}}$ scheme can be 
extracted e.g. from
\pcite{Billoire} to $O(\a_s)$. 
\beg
 \CV = - \frac{4\pi \a}{\q\,^2} \left[ 1 - \a (3 \pi \b_0 \ln
          \frac{\q\,^2}{\m^2} + \frac{31}{16 \pi} -  \frac{10 n_f}{16
          \pi})  \right]
\ee
Comparison with \pref{result} for $n_f$ light flavors gives
\beg
 \a^{BS}=\a_{\overline{MS}}\left(1- \frac{\a_{\overline{MS}}}{16 \pi}(31-\frac{10
                 n_f}{3} )\right)
\ee
where $BS$ denotes our bound state scheme. Numerically this means that
our $\a$ is slightly smaller than usual which is advantageous for our
perturbative calculation. For $m_t = 180, \a_s^{\overline{MS}}(M_Z)
=.117\pm0.05$ get for a renormalization at $\m = 1/r_B(\m)$
\begar
 \m &=& 18.5\pm1 GeV \\
 \a^{\overline{MS}}(\m) &=& 0.22 \pm 0.01 \\
 \a(\m)&=&0.19 \pm0.01
\ea
Remember that the above values for $\a,\a^{\overline{MS}}$ differ by a factor
$4/3$ from $\a_s$. 

For the comparison of the two schemes the terms of the form 
$\a_s^n/\q\,^2$ were important. To compare to relative $O(\a_s^2)$ terms 
up to $\a_s^3/\q\,^2$ are needed in the potential.
While the only term of this form in our scheme
has been calculated in our present work (eq. \pref{H5a}), we are not aware of
a  calculation of the terms of the order $const \a_s^2/\q\,^2$
in the $\overline{\mbox{MS}}$ scheme.
Therefore, it is at present possible to
compare the strong coupling constant in these two schemes only to
$O(\a_s)$.

Let us first compare fixed order perturbation theory with the 
renormalization group improved one.
This corresponds to an investigation of the effect of the gluon loops
given in \pref{result} and \pref{v2l}. The leading term is of the form 
$(\g+\ln \m r)$. This term will appear in any higher order. Therefore 
we add to the pure Coulomb
exchange successively terms of the form $(\g+\ln \m r)^n$ coming from the
leading log of the $n$-gluon loop.

The results are shown in fig.\pref{vgluon}. The uppermost line corresponds
to a pure Coulomb potential while the line slightly below has been 
calculated including the contribution of the bottom quark. For all
other curves the length of the segments is a measure for the number 
of logarithmic terms (e.g. from top to bottom: $n=2,4,\infty,5,3,1$).
\begin{center}
\leavevmode
\epsfxsize=12cm
\epsfbox{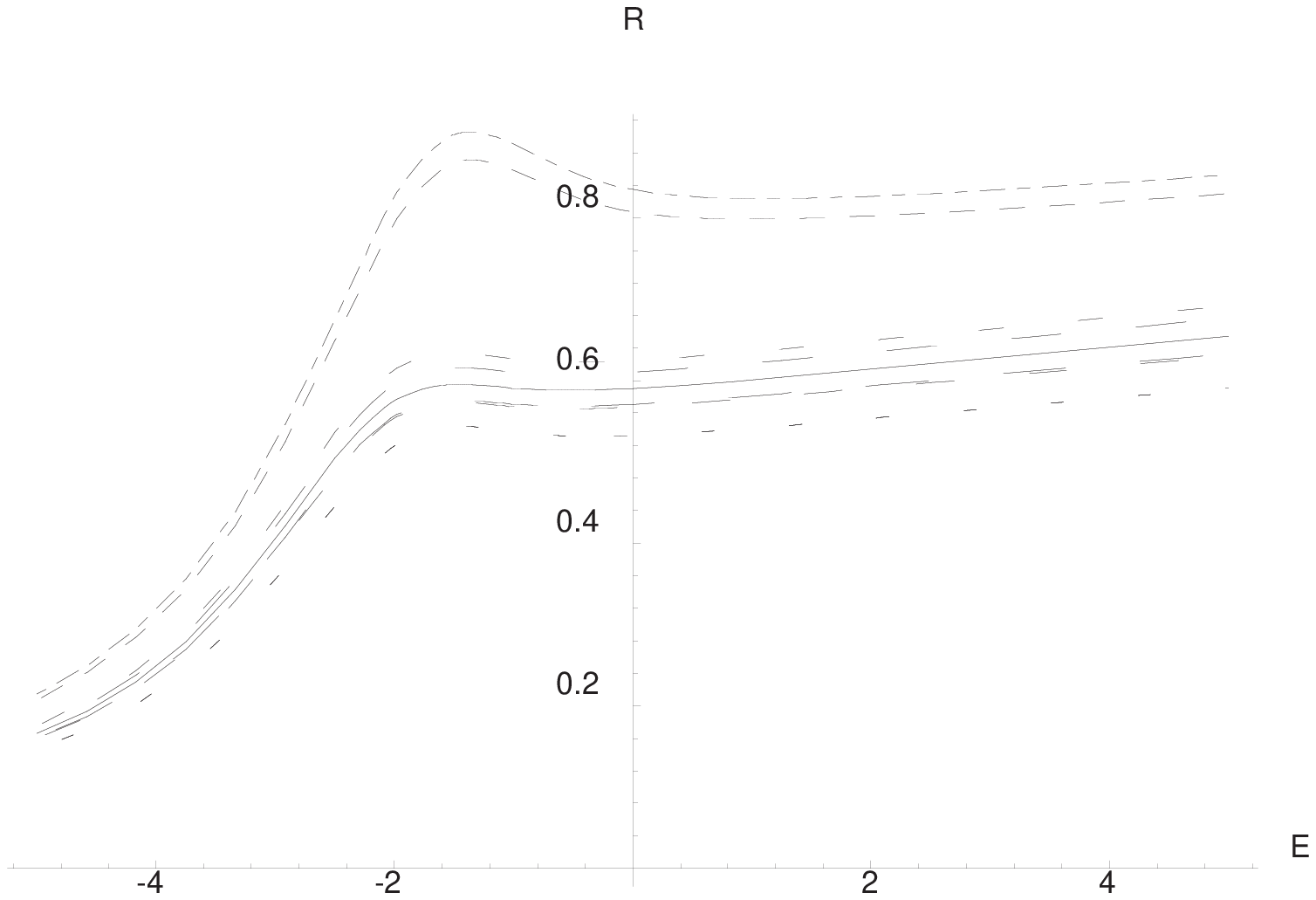}\\
\centerline{ Fig. \refstepcounter{figure} \label{vgluon} \thefigure}
\end{center}
Thus we conclude that the inclusion of the effects
of the gluon loops is
important. Therefore the strict application of perturbation theory is not
very useful. Instead one is forced to sum all leading logs from the
gluon loop corrections to the Coulomb line. Here one encounters the
problem that the resummed gluon propagator has a nonintegrable pole and
its Fourier transformation is thus not well defined.

We will impose the following recipe to obtain a resummed potential
\footnote{Another possibility frequently used in the literature is to use
the two loop renormalization group improved potential}.
We
calculate the corrections perturbatively in momentum space, perform the
Fourier transformation for each term and then sum up all terms
proportional to $(\g+\log \m r)$.
This leads to a geometrical series which serves as the definition
for the potential for values of $r$ where the sum does not converge.

The pole of the geometric series lies at
\beg
 r_{pole} = \frac{1}{\m} e^{\frac{2\pi}{(11-2/3 n_f) \a_s} -\g }.
\ee
We expect any regularisation which removes this apparently unphysical pole
to influence the potential at distances of that order of magnitude.
For a top quark of about 180 GeV we find
\beg
 r_{pole} \approx \frac{100}{\m}
\ee
which is about two orders of magnitude larger than the inverse damping constant
for this
mass (eq. \pref{damp}) if the renormalization scale is chosen as
\beg
 \m = \frac{1}{r_B}
\ee
or smaller, which is also suggested by the momentum scale relevant for
the bound state problem.
Therefore, we can safely sum the gluon loops and we will see that the
determination of the cross section is only sensitive to quantities well
below the pole.

Since the resulting potential is also the solution of the one-loop renormalization group
equation for massless loop quarks it is also one loop renormalization scale
independent.
The contributions from the nonleading log's in \pref{v2l} and from the
mass of the $c,b$ quarks are small and can thus be treated
in a perturbative manner. Since we have taken only the leading logs from a
definite set of diagrams it is clear which contribution has already been
taken into account and which one not.

Let us first briefly compare the numerical approach with the standard
bound state approach where individual bound states are considered and
corrections are calculated in a systematic way to the position of the
pole and to the wave functions as described in sect. \pref{BSpert}.\\
\begin{center}
\leavevmode
\epsfxsize=12cm
\epsfbox{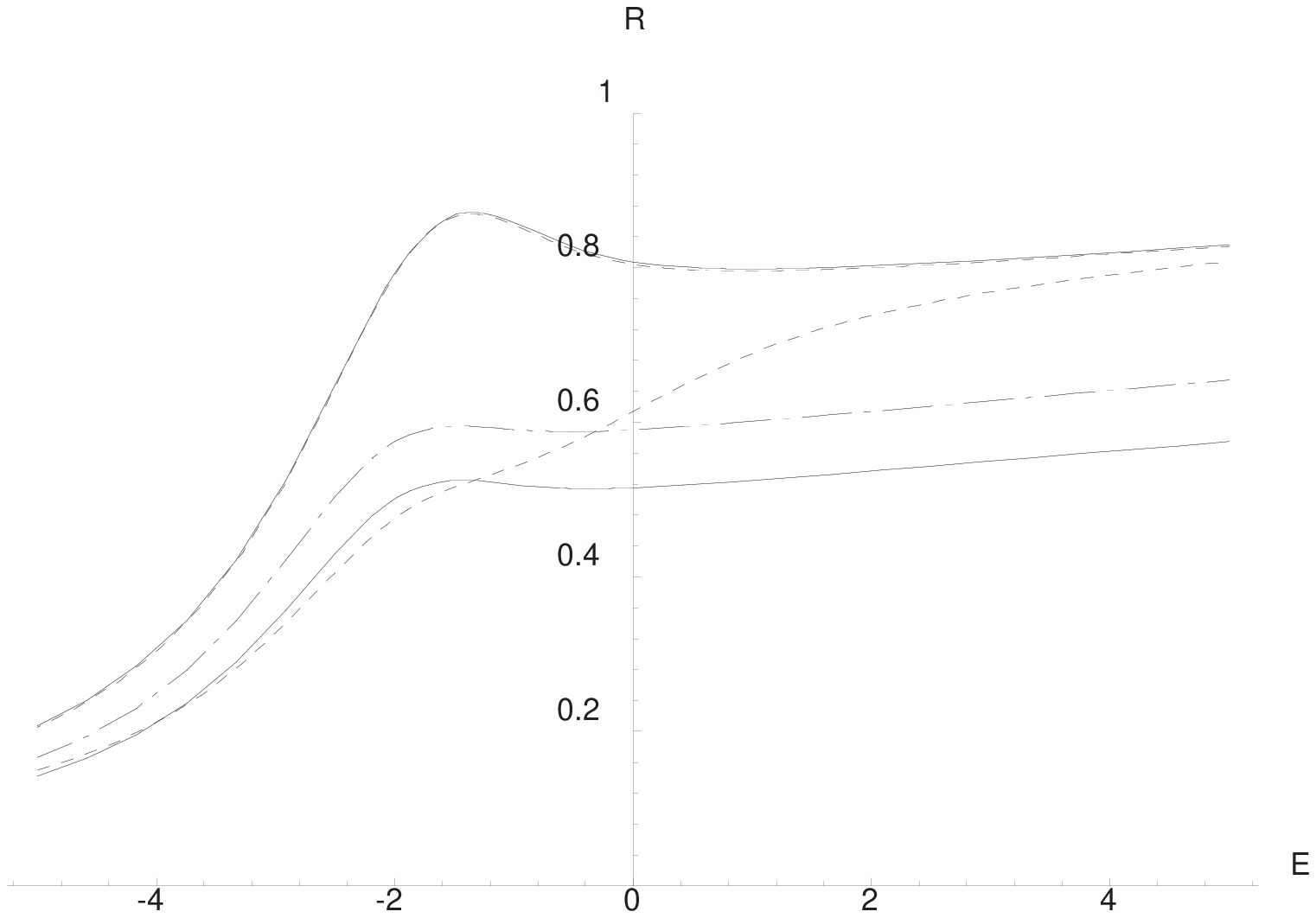}\\
\centerline{ Fig. \refstepcounter{figure} \label{vergl} \thefigure}
\end{center}

In fig. \pref{vergl} the 1s peak is corrected with the $O(\a_s)$
corrections for the position of the pole and the corrections of the
wave functions \pcite{KumW} (without correction to the continuum wave functions). 
Solid lines are calculated within the summation
approach , dashed lines with the Green function approach for $m_t =180$ 
and an electron polarisation of $0.6$.
The upper curves correspond to a pure Coulomb potential. The lower curves 
are from a potential including only the one loop correction. The 
dashed-dotted curve is the renormalisation group improved result.

The mass shift for the levels can be obtained from the relevant part of the 
potential by
performing the Fourier transformation into coordinate space,
where the integrations can be done analytically \pcite{KM1}.
The surprisingly simple result is
\beg \plabel{dMg}
 \D M_g = \<H_g\> = - m \a^3 \frac{11N}{16 \pi n^2 } [ \Psi_1(n+l+1)+\g + \ln \frac{\m n}{\a m}] + O(\a^5)
\ee
where $\Psi_n$ is the n-th logarithmic derivation of the gamma function
and $\g$ denotes Euler's constant.

The picture shows that this approach would give a good result for the
position of the rising edge of the cross section as well as for the
position of the 1s peak if it were not necessary to sum up the leading
logs.

In view of the above mentioned complications we will always include
the resummed potential in the following.

Let us now return to the bound state correction  for the toponium width.

As can be seen by inspection of eq. \pref{erg2} we could have obtained the
correct results for the level shifts simply by calculating the
expectation value \pref{erg2} between a wave function of a Schr\"odinger
equation with a QCD potential. The term
\beg \plabel{H1}
          H_1 =i \G \frac{\p\,^2}{2 m^2}
\ee
can thus be viewed as an absorptive part of the Hamiltonian. Adding this term is
equivalent to the introduction of a "momentum dependent width"
\pcite{Sumino,Jeza}. It has been shown \pcite{Sumino,Melni} that
the $O(\a_s)$ corrections to the absorptive part of the Hamiltonian
vanish.

Following this philosophy to estimate the effect of the bound state
corrections to the width we use the Hamiltonian
\beg \plabel{Hami}
 H = \frac{\p\,^2}{m}+i\G \frac{\p\,^2}{2 m^2} +V_{QCD}(r)
\ee
where $V_{QCD}$ is the one loop resummed potential.

The results of our numerical studies for $m_t =180,\r=0$ are shown in 
fig.\pref{brnum} and \pref{brcross}.

\begin{center}
\leavevmode
\epsfxsize10cm
\epsfbox{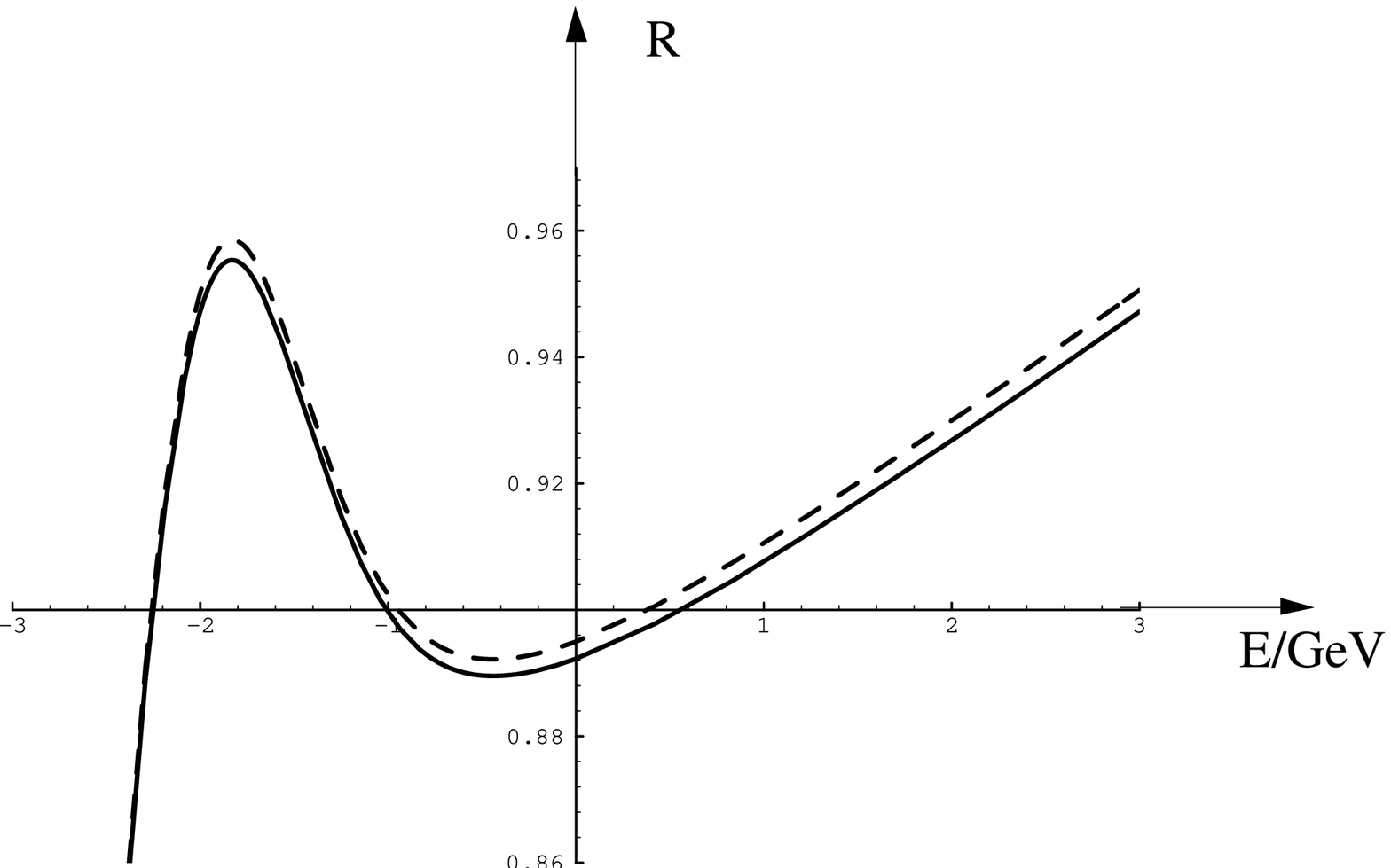}~\\
Fig. \refstepcounter{figure} \plabel{brnum} \thefigure :
 Effect of the boundstate corrections to the toponium
decay width on the total cross section near threshold in units of
$\s_{\m \bar{\m}}$. Full line: without, dashed line with
bound state corrections to the decay width.
~\\
\leavevmode
\epsfxsize10cm
\epsfbox{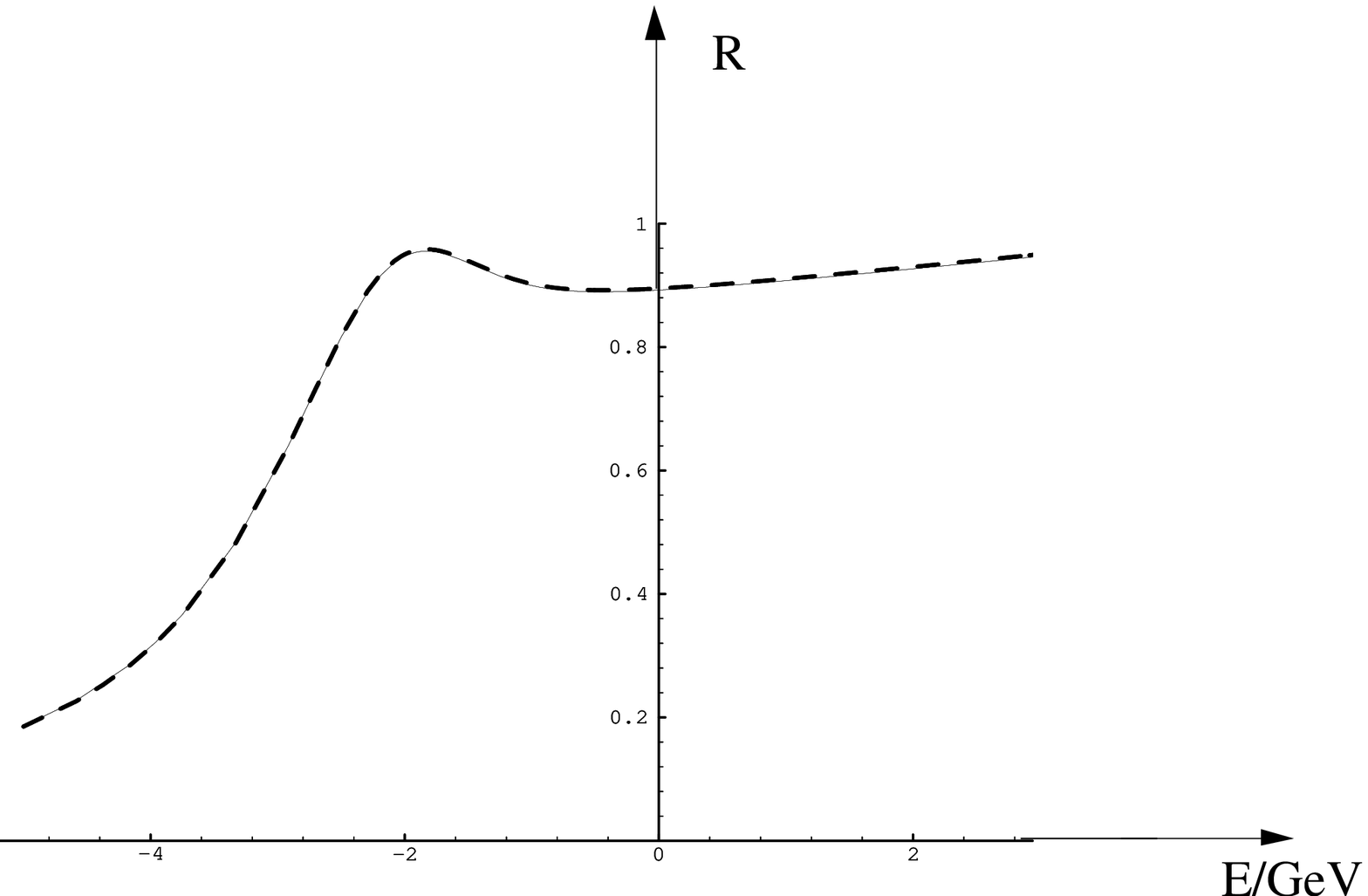}~\\
Fig. \refstepcounter{figure} \plabel{brcross} \thefigure :
The same as Fig. \pref{brnum} with a different scaling -
demonstrating the smallness of the effect.
~\\
\end{center}

The cross section is slightly enhanced and the effect is of the order of
less than 1\% and thus "really" an $O(\a_s^2)$ effect. However, we would like to indicate
that this enhancement is not only at the position of the 1S resonance
as it would be expected from a simple summation over resonances and from
the model calculation in ref. \pcite{Jeza}.
Furthermore, it is interesting to note that the Hamiltonian \pref{Hami}
gives the same
result as a similar one with the second term missing but with the replacement
$m \to m-i\G/2$. This convinces us that $H_1$ can be considered as small.

While the above result in the Green function approach shows
reasonable results,
we nevertheless want to add some words of caution.
One observation concerns the definition of the perturbed resolvent.
Assuming that relativistic effects are of the same order of magnitude
and can be included in a perturbative
manner, it seems necessary to go back to the perturbative
expansion and subtract the already included parts.
But due to our nonrelativistic approximation $H_1 \propto \vec{p}\,^2$
the second term in the series
$$\Im G = \Im G_0 + \Im ( G_0 H_1 G_0) + ... $$
is divergent while the relativistic expression is clearly well
behaved. Also the cancellation of gauge dependent contributions
coming from the high momentum region seems to be more involved
than it is for the level corrections.

The second observation concerns the comparison of the results
of fig \pref{brnum} with an approach where one calculates corrections to the wave
functions and to the position of the poles and afterwards takes the
sum over the whole spectrum to obtain the Green function \pcite{Kwong}.
This, in general, cannot lead to a correct result since $H_1$ makes the
Hamiltonian non hermitian
and thus the latter need not have a spectral representation.

We conclude that certain electroweak and strong corrections of the same order
of magnitude must be still missing or, even worse, it is not yet clear how to
correctly add them. Thus a better theoretical understanding of such
effects is desirable.

However, these effects will be of
the same order of magnitude and we thus may conjecture that the theoretical uncertainties of the calculation
of the $t\bar{t}$ cross section near threshold remain at the percent
level.

An effect larger than that could be induced by a relatively light Higgs
boson. This assumption is in agreement with recent electroweak data and
the corresponding fit to the MSSM. If the MSSM and the Higgs mechanism
turn out to be realized in nature the lightest higgs has to lie in the
range $90$ to $130$ GeV. Recent numerical studies \pcite{Harl} show an
enhancement of the cross section at the 10\% level for $m_H \approx 100$GeV
which seems to be the value presently favoured by the analysis of
radiative corrections to the SM and experimental limits.

\subsection{Forward-Backward Asymmetry}

In the last subsection we investigated the total cross section
for $t\bar{t}$ production in a future $e^+e^-$ collider. This should be
rather good observable to determine the mass of the top quark. But
since the total cross section at threshold is also significantly
determined by other quantities like the strong coupling constant and the
top decay-width, it is necessary to have some independent observables.
It was first proposed in \pcite{Sumino2} that the forward backward (FB)
asymmetry would provide such a quantity.
It has the additional advantage that it is independent of the absolute
normalization of the cross section measurement and is therefore a quantity
which can be determined experimentally with high accuracy.
However, treating this problem
in a nonrelativistic context, one encounters unphysical divergences which
made necessary the introduction of an unnatural cut-off. In
this subsection we will show a systematic way how to avoid such
divergencies.

The FB asymmetry is defined as the relative difference of the
cross section for particles produced in the forward and in the backward
direction, with respect to the direction of the $e^-$ beam.

\begar
 A_{FB} := \frac{\s_{FB}}{\s_{tot}} = \frac{ \int_{0}^{\frac{\pi}{2}} d\th
\frac{d\s}{d\th}- \int_{\frac{\pi}{2}}^{\pi} d\th \frac{d\s}{d\th}}{\s_{tot}}
\ea

Since the top quarks will be identified by their main decay products
$W$ and $b$ let us consider the cross section for the process
$e^+ e^- \to t\bar{t} \to b W^+ \bar{b} W^-$. Production of
 $W^+bW^-\bar{b}$ via other channels can be treated as in
\cite{Balle}.
\begar
 d\s_{e^+ e^- \to bW^+\bar{b}W^-} = \frac{1}{2 P^2} |M_A+M_V|^2 d\Phi_{W^+b}
 d\Phi_{W^-\bar{b}} (2\pi)^4 \d(t_1+t_2-k_1-k_2-b_1-b_2)
\ea
with
\begar
d\Phi_{W^+b} =  \frac{d^4b}{(2 \pi)^4}\frac{d^4k}{(2 \pi)^4}  (2\pi) \Th(k_{10}) \d(k^2-M^2) (2\pi)
      \Th(b_{10}) \d(b_1^2)  \plabel{dPhi}
\ea
and similarly for the decay of the antitop.

\begin{figure}
\begin{center}
\leavevmode
\epsfxsize=16cm
\epsfbox{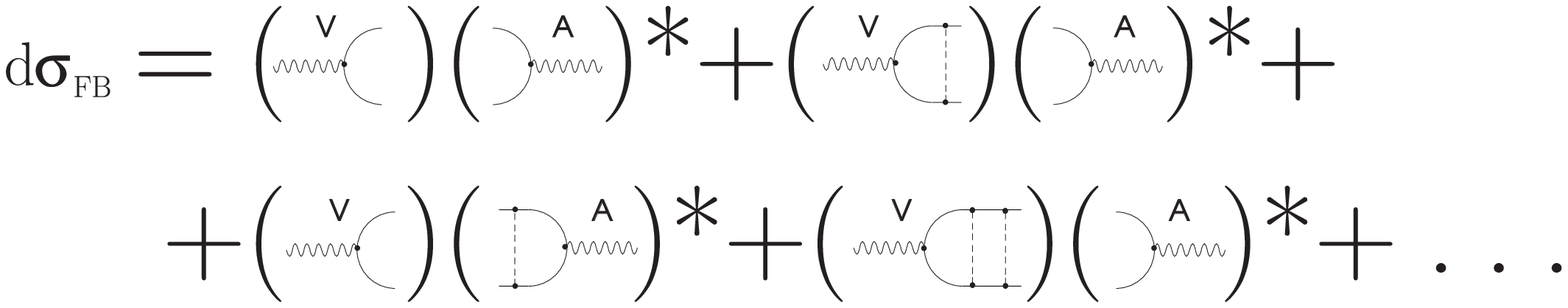}\\
Fig. \refstepcounter{figure} \label{fbf1} \thefigure 
\end{center} \end{figure}

The FB asymmetry originates in the interference terms of the
vector coupling ($M_V$) and the axial coupling ($M_A$).
This is depicted in figure \pref{fbf1} where V indicates a 
vector and A an axial vector coupling, the star denotes complex
conjugation.
The matrix element $M$ is essentially the vertex $\g t \bar{t}$ and $Z t
\bar{t}$. Near threshold the perturbative treatment of the Coulomb
interaction is no longer valid. Instead one has to include the whole rung
of Coulomb interactions and even worse, it is necessary to include the
leading logarithmic contributions from the gluonic self energy
corrections, as we have seen in the last section. We can sum the relevant
set of graphs by means of the equation
\beg
 \G^{\m} = \g^{\m} -i KD\G^{\m}
\ee
for the vertex function $\G^{\m}$. Comparison with the BS-Equation
\pref{bsf} shows that it has the solution
\begar
 \G_V ^{\mu} &=& -i D^{-1} G \g^{\mu} \\
 \G_A ^{\mu} &=& -i D^{-1} G \g_5 \g^{\mu},
\ea
where it is understood, that the two legs of $G$ on the right hand side
are connected with the $\g$-matrices and a momentum integration is performed.

We can split the matrix element into
\begar
M_A^i &=& m_1 m_2 D \G_A ^{i} \\
M_V^i &=& m_1 m_2 D \G_V ^{i} \nonumber
\ea
where
\begar
m_1 &=& \e_{\m}^{(\s)}(k_1) \bar{u}_{\l}(b_1) (-i \frac{e}{\sqrt{2} \sin
      \theta_W} V_{33}^* \g^{\m} P_-) \\
m_2 &=& \e_{\n}^{*(\s')}(k_2)  (-i \frac{e}{\sqrt{2} \sin
      \theta_W} V_{33} \g^{\n} P_-) v_{\l'}(b_2).
\ea
Near threshold the propagators in $D$ and the Green function $G$ should be
dominated by small momenta and therefore it should be possible to replace
them by their nonrelativistic approximations. However, for $D$ the
approximation to $O(\a)$ is needed for the axial vertex, since the zero
order propagators would give zero due to $\l^-\g_5 \g^i \l^+ =0$.

\begar
 D &\to & S^+_{nr} \otimes S^-_{nr} \\
        S^{\pm}_{nr} &=& (\l^{\pm} - \frac{\vec{p}\vec{\g}}{2m})
           \frac{1}{\frac{E}{2} \pm p_0 -\frac{\p\,^2}{2m}+i\frac{\G}{2}}
\ea
This gives rise to the effective nonrelativistic axial vertex
\beg
 \g_5 \g^i \to \frac{p^k}{2m} [\g^i,\g^k] \g_5 \l^- \plabel{AVnr}.
\ee
We note already here that the term $\frac{\vec{p}\vec{\g}}{2m}$
introduces an additional power in $\p$ which is not present in the
relativistic propagator. Instead the factor $\frac{p^k}{2m}$ in
\pref{AVnr} would be replaced by $\frac{p^k}{2 E_p}$ which is finite for
$p \to \infty$.

Using the fact that the heavy quarks are on shell up to $O(\a^2)$
we can write
\begar
\int d\Phi_{W^+b} \sum_{\s,\l} |m_1|^2 = - 2 \Im \S(m^2) \l^+ + O(\a^2)
      \plabel{imsig}
\ea
in  agreement with section 2.5.

From the modulus squared of the propagators we obtain a factor
\begar
\int \frac{dp_0}{2\pi} |\frac{1}{p_0^2-\o^2}|^2 &=& \frac{2}{\G |2 \o|^2}, \\
 \o &=& \frac{1}{2m}(\p\,^2-mE-im\G)
\ea

Collecting everything from above, performing the trace in a frame where
the leptons move along the z-axis, and using
eq. \pref{Lmn}  we get for the FB-asymmetry

\begar
 \s_{FB,\L} &=& (c_{AV} + d_{AV} \r) \frac{18 \G}{ \pi^2 P^2} \int_0^{\L} p^2 dp
           d \O_{FB} Re[ G^*(\p) \int
           \frac{d^3q}{(2 \pi)^3} G(\p,\q) \frac{q^{(3)}}{m} ]
           \s_{\m\bar{\mu}} \plabel{sfbnr}
\ea
where
\begar
       c_{AV} &=& a^{(A)} b^{(V)} +a^{(V)} b^{(A)} \\
       d_{AV} &=& a^{(A)} a^{(V)}+b^{(A)} b^{(V)} \\
       \int d \O_{FB} &=& \int_0^{2\pi} d\varphi ( \int_0^{\frac{\pi}{2}} d\th
            - \int_{\frac{\pi}{2}}^{\pi} d\th)\sin\th
\ea
The SM values for $a^{(X)}, b^{(X)}$ are given in \pref{av} and \pref{aa}.
We introduced an intermediate cut-off $\L$ in \pref{sfbnr}
since the $p$ integration
is logarithmically divergent. It will be removed in the following.
The Green function $G(\p)$ in \pref{sfbnr} is defined by
\begar
 G^*(\p) &=& \int \frac{d^3q}{(2 \pi)^3} G^*(\p,\q) =
        \frac{4\pi}{p} \int_0^{\infty} dr r \sin pr \tilde{G}_{l=0}(r,0).
\plabel{Gs}
\ea
$\tilde{G}_{l=0}(r,r')$ denotes the S-wave Green function in 
configuration space, eq. \pref{sGr}. Let us now investigate the term
\begar
 \int \frac{d^3q}{(2 \pi)^3} G(\p,\q) \frac{q^{(3)}}{m} =
 \frac{-i}{m} \int d^3x e^{-i\p\vec{x}} \frac{\6}{\6 y_3}
 \tilde{G}(\vec{x},\vec{y}) \big|_{\vec{y}=0} \plabel{Gp1}
\ea
Using the representation \pref{gsum} one can show that only P-waves
($l=1,m=0$) contribute to the sum. Thus we need to evaluate eq. \pref{gl}
for $l=1$. The regular and singular solutions behave at the origin as
\begar
 g_<^{l=1}(r) &\to& c_{\scriptscriptstyle <} r^2,  \\
 g_>^{l=1}(r) &\to& c_{\scriptscriptstyle >} \frac{1}{r}, \nonumber
\ea
respectively. We now construct a singular solution with
$c_{\scriptscriptstyle >}=1$ out
of two arbitrary solutions $u_{p1}$ and $u_{p2}$ :
\begar
 g_>^{l=1}(r) &=&  a_p[u_{p2}(r) - B_p u_{p1}(r)]
\ea
by means of
\beg
 a_p := \lim_{r \to 0} r^{-1}[ u_{p2}(r) - B_p u_{p1}(r)]^{-1}.
\ee
$B_p$ is determined by the requirement of regularity at infinity
\beg
 B_p = \lim_{r\to \infty} \frac{u_{p2}(r)}{u_{p1}(r)}.
\ee
Then the condition \pref{Wronski} demands $c_{\scriptscriptstyle <}=m/3$. Performing the
differentation in the $z$-direction in the limit $y \to 0$ in \pref{Gp1}
leads to
\begar
 \int \frac{d^3q}{(2 \pi)^3} G(\p,\q) \frac{q^{(3)}}{m} &=&
 \frac{\cos \th}{p^2} \int_0^{\infty} dr ( \frac{\sin pr}{r} - p \cos pr)
         g_>^{l=1}(r). \plabel{Gp2}
\ea
While the expressions \pref{Gs} and \pref{Gp2} are well defined,
it has been observed in \pcite{Sumino2} and \pcite{jeza2}
that \pref{sfbnr} is logarithmically divergent. Thus cutoffs ($\L$) have
been introduced to make numerical predictions possible. Since in our
present case the divergence is only logarithmical, this leads to
phenomenologically reasonable results.
But the explanations
given in these references are different and so are the cut-offs.
It may, however, happen that for the comparison with experimental data
with some cuts a reintroduction of some kind of cut-off will be necessary
\footnote{The author is grateful to M.Je\.zabek for this information},
but we feel saver by first giving a satisfactory theoretically prediction
and leaving this possibility open to the experimentalists.

In any case, the introduction of these cut-offs leads to some numerical
uncertainty in the result, and
is clearly unsatisfactory from the theoretical point of view. Furthermore
we will see in the next section, that a similar divergence occurs
in the context of the stop--anti-stop production. Since this divergence
is linear, a better understanding of the origin of this kind of divergencies
is necessary to give a quantitative prediction.

The key point of our solution of the problem is to go back to the
perturbative sum of the graphs for $\s_{FB}$. This is illustrated in
figure \pref{fbf1}.

Consider now the leading (tree) contribution in the nonrelativistic
approximation (e.g. the first graph in fig. \pref{fbf1}). In this case 
both S- and P-wave Green functions
are replaced by
\beg
 G(\p,\p\,') = \frac{(2\pi)^{3} \d(\p-\p\,')}{\frac{\p^2}{m} -E -i\G}.
\ee
This leads to the logarithmically divergent expression (for $\L \to
\infty$)
\begar
 \s_{FB,\L}^{(1),nr} = \frac{9 \G}{2 m^3} \int d\O_{FB} \int_0^{\L} dp
        \frac{p^3 \cos \th}{|\frac{\p^2}{m} -E -i\G|^2} \plabel{sfb1nr}.
\ea
Since the remaining graphs in fig. \pref{fbf1} by power
counting can be shown  to give finite results we
conclude that the nonrelativistic approximation was not valid in the tree
graph due to the extra power in $p$ from the axial vertex. Furthermore
since \pref{sfb1nr} is the only divergent contribution in the
(infinite, but assumed to be convergent) sum  \pref{fbf1} representing $G$
we conclude that
the divergent part of \pref{sfbnr} is entirely contained in the first
(free) contribution \pref{sfb1nr}. We could now
remove this divergence by a replacement $p/m \to p/E_p$, as indicated
above. This, however, does not respect another qualitative difference
between the exact tree contribution and its nonrelativistic approximation.
Namely in the relativistic calculation the phase space is cut off
when the momentum squared of one quark falls below the invariant mass
of the decay products \pcite{Sumino}.

Therefore, we return, instead, to the relativistic expression for the tree
contribution. Since we are dealing with a decaying particle, we will
have to include the self energy contribution due to the decay in the
propagator. This will in general lead to gauge dependent results,
but we can use the constant on-shell width to obtain the leading
contribution in the weak coupling \pcite{Velt}.

This makes the quantities we are considering finite (for $\G\to0$) and gauge independent
and the remaining higher order contributions (e.g. gauge dependent if one considers
only one diagram) can be calculated
perturbatively. Therefore to leading order it seems reasonable to take the
constant width approximation for the top propagator.

Since we are only investigating processes
with $t\bar{t}$ intermediate states which will give the main contribution
to the cross sections we focus on the graph shown in fig. \pref{reltree}.
\begin{center}
\leavevmode
\epsfxsize=7cm
\epsfbox{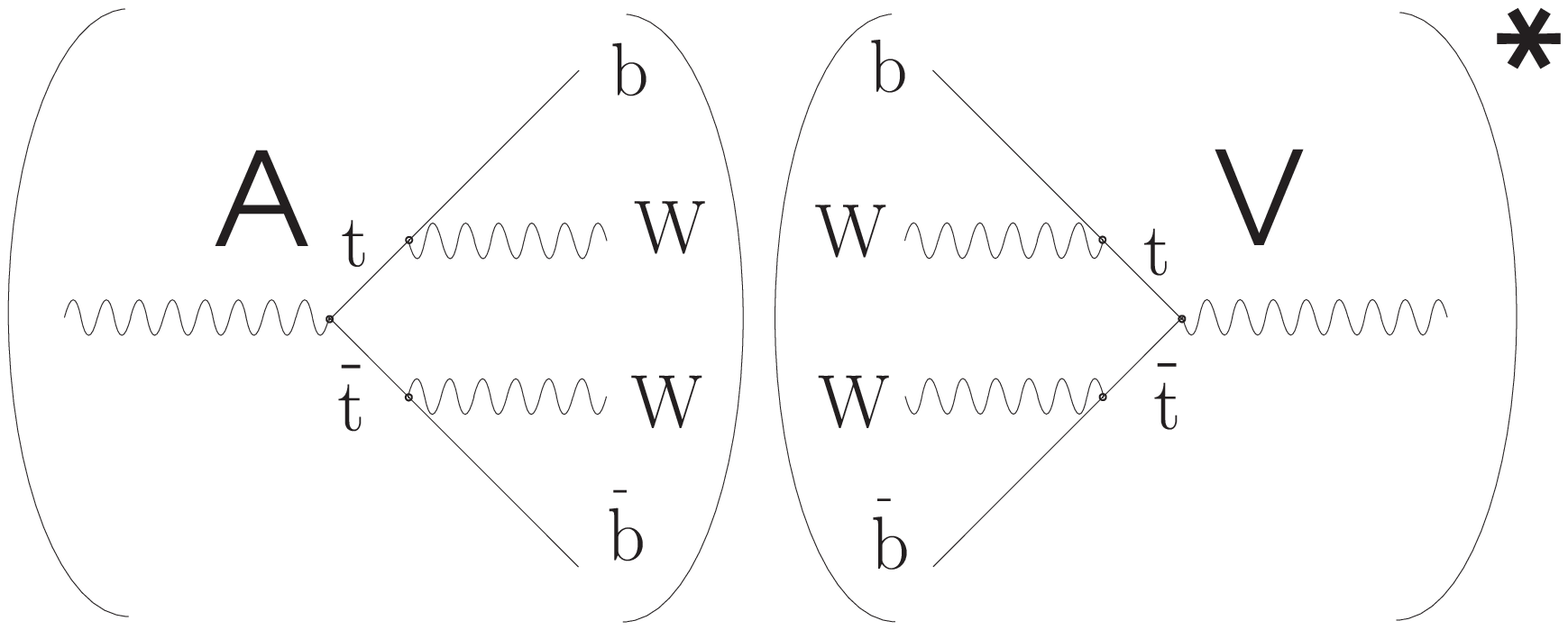}\\
Fig. \refstepcounter{figure} \label{reltree} \thefigure
\end{center}
A further advantage of this method is that now
the other - non resonant - graphs for the process $e^+e^- \to
W^+bW^-\bar{b}$ to leading order in the weak coupling \pcite{Balle}
need only to be added to yield the background contribution.

Performing the straightforward calculation for the relativistic tree 
contribution with a constant, but non-zero width $\G$, we arrive at
\beg
 \s_{FB}^{(1)} = \frac{18}{P^2} (c_{AV} +c_{AV} \r) \int d\m_1^2 \int d\m_2^2
             \D(\m_1^2) \D(\m_2^2) \left(1-\frac{(\mu_1+\m_2)^2}{P^2}
        \right) \left(1-\frac{(\mu_1-\m_2)^2}{P^2} \right)
\ee
where
\beg
\D(p^2) = \frac{m \G}{\pi[(p^2-m^2)^2+m^2\G^2]}
\ee

This could have been also obtained by replacing the fermion propagator by
\beg
 S = \int d\mu^2 \frac{\G}{\pi[ (m^2-\mu^2)^2+m^2 \G^2 ]}
         \frac{(\ps+\m)}{p^2-\m^2+i\e}.
\ee
and cutting it to effectively replace the  phase space element for a stable quark

$$ d\Phi_{stable} = \frac{d^4p}{(2\pi)^3} \Th(p_0) \d(p^2-m^2)$$ by
\begar
d\Phi_{unstable} &=& \frac{d^4p}{(2\pi)^3} \Th(p_0) \D(p^2).
\ea

To conclude we can say that our simple prescription to obtain  finite,
gauge independent results to the desired order is to replace the
nonrelativistic tree contribution by the relativistic one and leave the
(finite) rest unchanged:
\beg
 \s_{FB}= \s_{FB}^{(1)}+ \lim_{\L \to \infty}
   (\s_{FB,\L}-\s_{FB,\L}^{(1),nr})
\ee
In fig. \pref{fb1} the numerical results are shown for the different
contributions to $\s_{FB}$ for $\L=300$ GeV. This can be estimated to 
give a result for $ \lim_{\L \to \infty} (\s_{FB,\L}-\s_{FB,\L}^{(1),nr})$
lying only 2\% below the final answer.
The results of the purely nonrelativistic calculations are
compared to our approach in Fig. \pref{fb2} for $m_t=180$ and an electron 
polarization of 0.6. We also included the hard corrections as given in
\pcite{ph,jeza2}. 

We believe that our approach has the decisive 
advantage that it can be based upon
to the {\it original} set of graphs to be considered
and it is clear which graph has been
calculated to which accuracy. Thus at least in principle 
a systematic improvement
is possible. The present approach should also be applicable to the 
$O(\a_s)$ final state corrections calculated in \pcite{Sumith}.
\begin{figure}
\begin{center}
\leavevmode
\epsfxsize10cm
\epsfbox{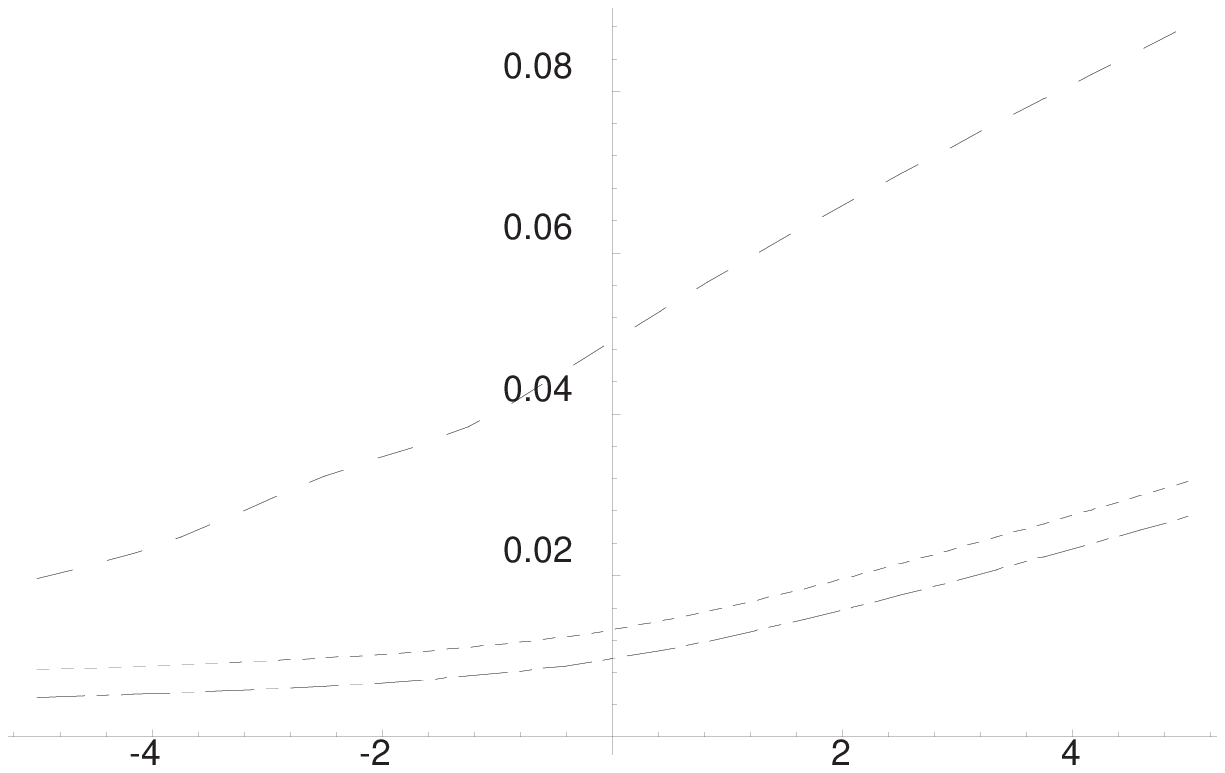}\\
Fig. \refstepcounter{figure} \plabel{fb1} \thefigure:
Different Contributions to $\s_{FB}$\\
dashed: $\s_{FB}^{\infty}$, dotted:$\s_{FB}^{(1),nr}$,
dashed-dotted: $\s_{FB}^{(1)}$
~\\
\end{center}
 \unitlength1mm
 \begin{picture}(0,0) \put(77,87){$\frac{\s}{\s_{\m\bar{\m}}}$} 
   \put(130,23){E/GeV} \end{picture}
\end{figure}
\begin{figure}
\begin{center}
\leavevmode
\epsfxsize10cm
\epsfbox{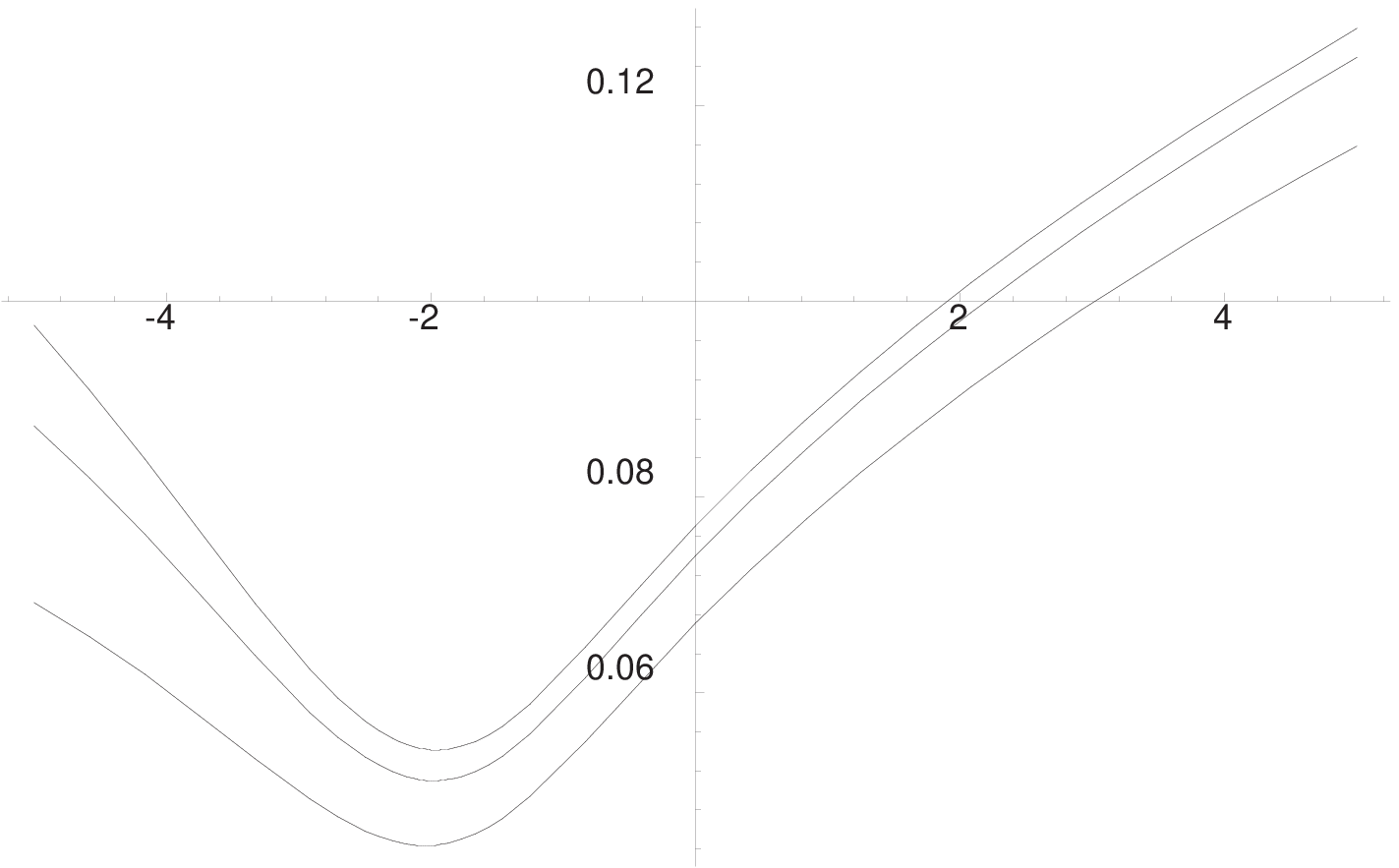}\\
Fig. \refstepcounter{figure} \plabel{fb2}\thefigure:
Forward-backward asymmetry $A_{FB}$;\\
curves from top to bottom: present work, ref.\pcite{Sumino2}, ref.\pcite{jeza2}
\\
\end{center}
\unitlength1mm
\begin{picture}(0,0) \put(77,86){$A_{FB}$} 
\put(130,61){E/GeV}
\end{picture}
 \end{figure}


\section{ Axial Contribution to the $t\bar{t}$ Total Cross Section
and the Production of Stop-Antistop near Threshold}

The total cross section for the
production of P-wave states near threshold gives the leading term
for the production of stop-antistop and a contribution of $O(\a_s^2)$
to the total cross section for $t\bar{t}$. Only a qualitative investigation
on the basis of the Coulomb Green function for the former has been
published up to now \pcite{Bigi2}. We will apply the method of
replacing divergent nonrelativistic graphs, implicit in the Green function approach,
by the relativistic, finite ones to get a quantitative reliable
result.

Considering the Coulomb green function
as a first approximation it was observed in \pcite{Bigi2} that the total
cross section for a stop-antistop pair near threshold develops an
unphysical linear divergence due to the nonrelativistic approximation.
Even a relativistic calculation of the imaginary part of the one loop
contribution using the commonly used propagator $1/(p^2-m^2+im\G)$
remains logarithmically divergent. This is due to the fact that this
propagator does not fulfill the requirements of quantum field theory
as does the propagator
\beg  \plabel{lehm}
 \int \frac{d\m^2}{\pi} \frac{m \G}{(\m^2-m^2)^2+m^2\G^2}
      \frac{1}{p^2-\m^2+i\e},
\ee
which is in agreement with the Lehmann representation of the full
propagator.
The tree contribution using this propagator reads
\beg
 \s_{1}= \frac{3}{2}(c_s+d_s \r) \int d\m_1^2 \int d\m_2^2 \D(\m_1^2)
         \D(\m_2^2)
\Th(P^2-(\m_1-\m_2)^2)\left[1-2\frac{\m_1^2+\m_2^2}{P^2}+
      \frac{(\m_1^2-\m_2^2)^2}{P^4} \right]^{\frac{3}{2}} \s_{\m\bar{\m}},
\ee
whereas we have for the nonrelativistic tree contribution
\beg
\s_{1,nr}^{\L}= \frac{3 \G}{\pi m^2}(c_s+d_s \r) \int_0^{\L} dp
           \frac{p^4}{(p^2-mE)^2+m^2\G^2} \s_{\m\bar{\m}}.
\ee
However, a single subtraction of the tree graph is not sufficient.
One should also subtract the logarithmically divergent one-gluon exchange
term. But since
we replaced the gluon propagator by the resummed one we should also
calculate the relativistic graph with a resummed propagator to get the
correct behavior at infinity. Unfortunately this leads to renormalon
ambiguities. Therefore, we conclude that the uncertainties in choosing
the right cut-off are of the same order of magnitude as higher QCD
corrections, and we circumvent these difficulties for the time being by
keeping the cut-off in the logarithmic divergent terms,
choosing its value $\L=\L_0=m$ from
the observation that this cut-off would have given the correct
answer in the case of the FB-asymmetry discussed in the last section.\\
\begin{center}
\leavevmode
\epsfxsize10cm
\epsfbox{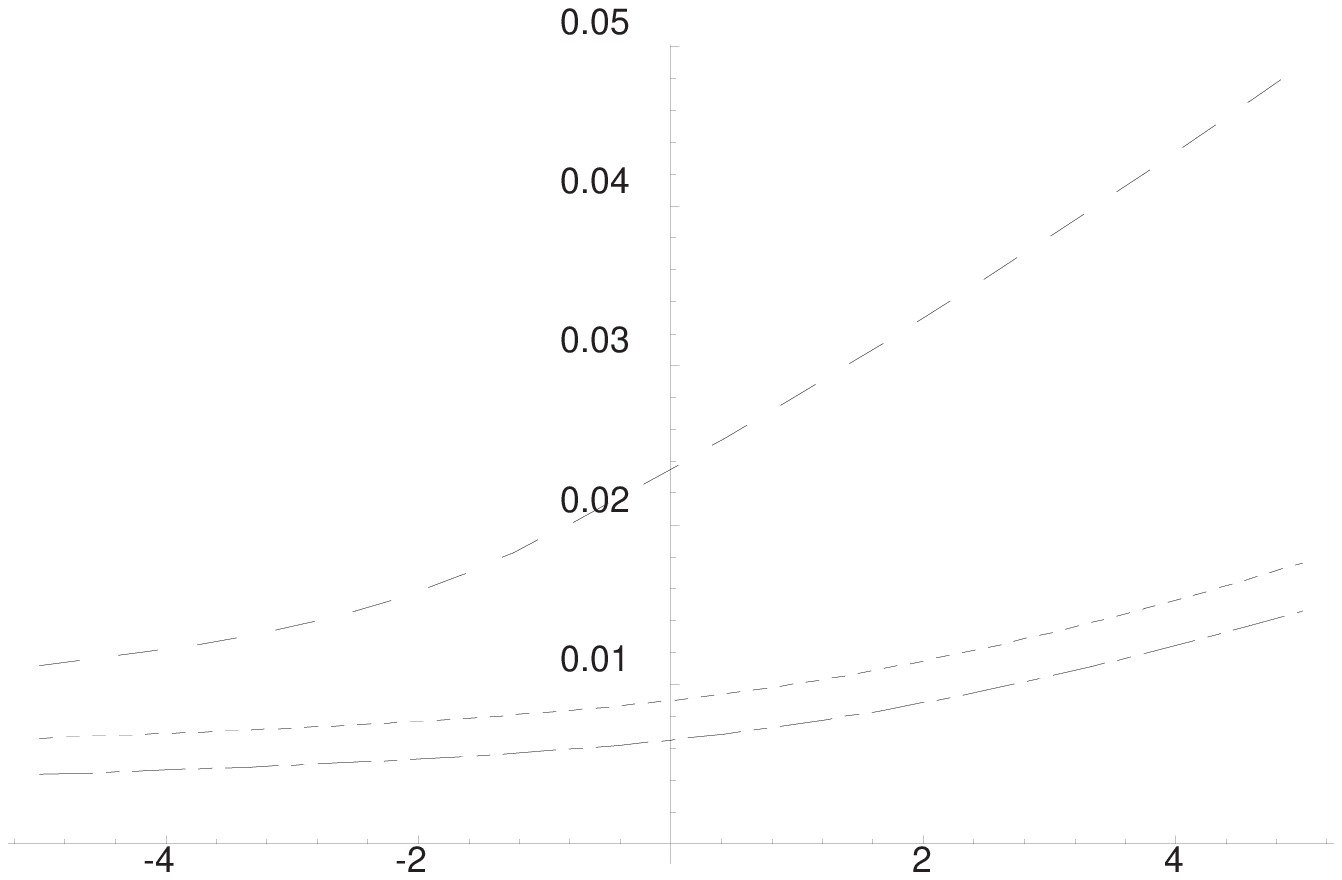}\\
Fig. \refstepcounter{figure} \plabel{stfig1}\thefigure:
Different Contributions to $\s_{\tilde{t}\tilde{\bar{t}}}$\\
dashed:$\s_{nr}^{\L_0}$,dotted:$\s_{1,nr}^{\L_0}$,
dashed-dotted: $\s_{1,rel}$
\\
\end{center}
\unitlength1mm
\begin{picture}(0,0) \put(77,85){$\frac{\s}{\s_{\m\bar{\m}}}$} 
\put(130,23){E/GeV}
\end{picture}
The same is true for the pure axial contribution to the total cross section
for $t\bar{t}$ production since each axial vertex contributes in the
nonrelativistic limit an extra power in $\p$ as explained in the last
section. The only change is to replace the factor
\beg \plabel{fs}
 f_s :=\frac{3}{2}(c_s+d_s \r)
\ee
by
\beg \plabel{ff}
 f_f := 6 (c_{A} +d_{A} \r).
\ee
Using the methods of the preceding sections we get within the Green
function approach
\beg
 \s_{nr}^{\L} = f_X \frac{3 \G}{ 2 \pi^2 m^2} \int_0^{\L} dp p^2 \int d\O
 \left| \int \frac{d^3q}{(2\pi)^3} \frac{q^{(3)}}{m} G(\q,\p)
 \right|^2 \s_{\m\bar{\m}},
\ee
where $f_X$ denotes either  $f_s$ or $f_f$.
The different contributions to the cross section for the scalar case
\beg
 \s = \s_{nr}^{\L_0}-\s_{1,nr}^{\L_0} + \s_1
\ee
are depicted in fig. \pref{stfig1} for $\cos^2 \th_t =
0.5,m=180,\G=\G_{top}$. The net result is shown in fig \pref{stfig2}
for the axial contribution to the $t\bar{t}$ cross section.

\begin{center}
\leavevmode
\epsfxsize10cm
\epsfbox{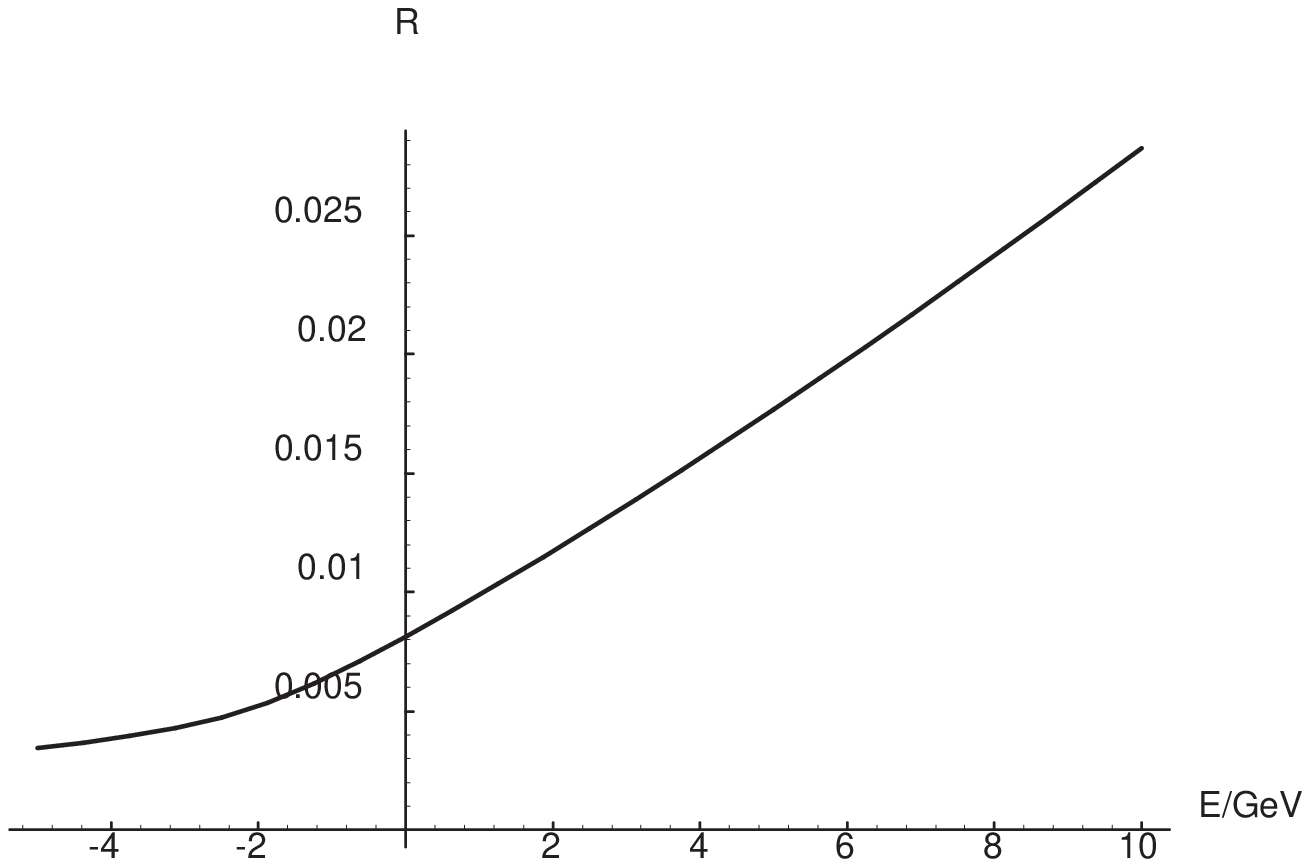}\\
Fig. \refstepcounter{figure} \plabel{stfig2}\thefigure:
Drell-ratio (R$=\s_A/\s_{\m\bar{\m}}$) for the production of $t\bar{t}$
\\
\end{center}

According to ref. \pcite{Gues} we included  the
hard corrections by a factor $(1-\frac{\a}{\pi})^2$. 
We conclude that the axial contribution gives a sizeable effect (of a
few percent ) to $\s_{tot}$ beginning at $\approx 5$GeV above threshold.
However, it is small enough in order to hide the uncertainty in $\L$ 
with respect to $O(\a_s^2)$ corrections to the vector contribution to
$\s_{tot}$. 

Finally in fig. \pref{stfig3} we compare the cross sections for the
production of $\t\bar{\t}$ near threshold for two different values of the 
decay constant ( $\cos^2 \th_t =0.5,m=180$).

\begin{center}
\leavevmode
\epsfxsize10cm
\epsfbox{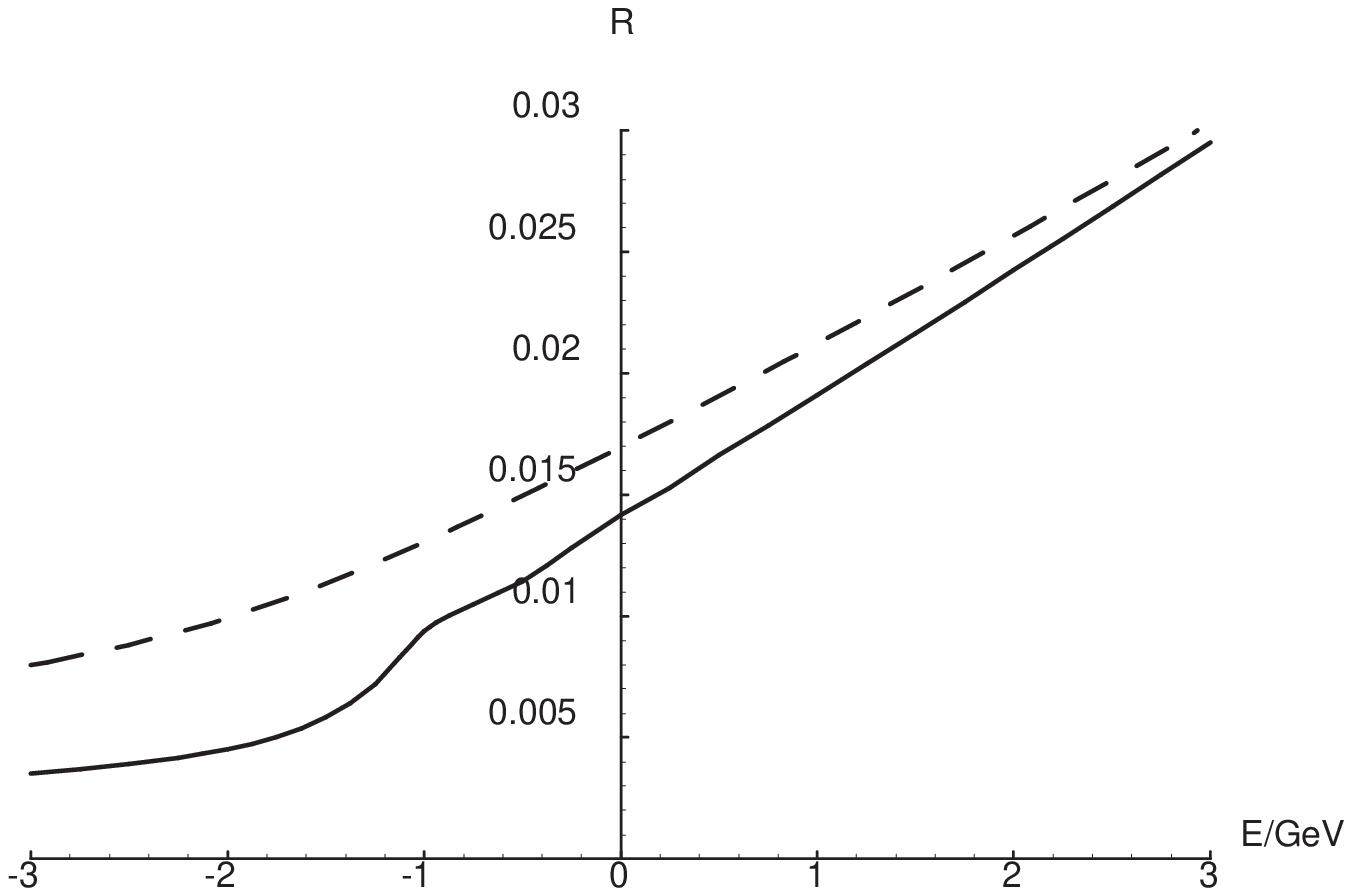}\\
Fig. \refstepcounter{figure} \plabel{stfig3}\thefigure:
dashed curve: $\G=1.5$GeV, full line: $\G=0.5$GeV
\\
\end{center}

\chapter{Conclusion}

In this thesis we considered the quantum field theory near thresholds of
ultra-heavy particles, where we especially focused on the $t\bar{t}$ and
$\tilde{t}\bar{\tilde{t}}$ systems. Due to the large mass and the large
width of the top quark it is possible to treat the former in a pure
perturbative manner although the top is a strongly interacting particle.
The large mass of the top close to the weak scale may be an indicator
of new physics. Therefore, it is of paramount
interest to predict physical observables within the framework of the
existing Standard Model in order to be able to single out possible
new effects.

In this context the question of the possible large corrections due to
a running width of the top quark could be solved in this work. 
It was shown that a
previously not considered contribution leads to large cancellations and
leads especially to a gauge independent result.
This was first achieved in the narrow width approximation and subsequently
improved by the introduction
of a new relativistic zero order equation already including the 
on-shell decay
width. With the help of that equation and a Ward identity it was further
possible to show that the bound state corrections to the decay width
for a generic fermionic system which does not decay by annihilation
is given by
\beg
 \D \G = -\G_0 \frac{\< \p\,^2 \>}{2 m^2}.
\ee
Furthermore a systematic derivation of the $t\bar{t}$ potential to
numerical order $O(\a_s^4)$ was given. Especially the contribution of
a QCD box graph and a thorough investigation of effects of the weak
interaction are new. Also the potential for scalar-scalar bound states
has been investigated with the help of a new zero order equation for
such systems. A deviation from a previous result was found.

In the second part of this work we considered within the numerical
Green function approach the cross sections for the production of
ultraheavy particles near threshold. First we derive the general formalism
and consider the total cross section. Here the large width is
demonstrated explicitly to be very important for hiding nonperturbative effects.
Then we investigate the forward-backward asymmetry near threshold. We
avoid the introduction of an unnatural cut-off as contained in previous
work. This is achieved by identifying the divergent graphs in the
nonrelativistic Green function and replacing them by the relativistic
ones. The same method can also be applied to the pure axial contribution
to the $t\bar{t}$ total cross section as well as to the stop-antistop
production. This enables us to give first quantitative predictions
of these observables.

While some aspects of the production of heavy particles near threshold
are well understood by now, there remain several unsolved problems.
Especially proceeding to higher orders in perturbation theory seems rather
difficult in view of the necessity of including the leading Coulomb
interaction to all orders. It has not been possible to completely
combine the
rigorous bound state approach of Chapter 2 with the practically
tractable numerical approach of Chapter 3 in a satisfactory way, yet.
Nevertheless, we hope that this work provides a basis for further investigations
in this direction.

          

\begin{appendix}
   
\chapter{Expectation Values}

In sect. 2.4 and 3.1 we needed the expectation values of logarithmic
potentials between Schr\"odinger wave functions. They can be obtained
by
\beg
 \< \frac{\ln^nr}{r} \> = \frac{d^n}{d\l^n} \< r^{\l-1} \> \big|_{\l=0},
\ee
if the expectation value $\< r^{\l-1} \> $ is known analytically.
In the latter the representation 
\beg
 L_{n-l-1}^{2l+1}(\r) = \lim_{z\to 0} \frac{1}{(n-l-1)!} \frac{d^{n-l-1}}{dz^{n-l-1}}
                    (1-z)^{-2l-2} e^{\r \frac{z}{z-1}}
\ee
of the Laguerre polynomials may be used. This allows an easy evaluation of the integrations 
and the remaining differentiations can be done with some care afterwards:
\beg
 \< r^{\l-1} \> = \frac{(\a m)^{1-\l }}{2 n^{2-\l } } \frac{(n-l-1)!}{(n+l)!}
                 \G(2l+2+\l) \sum_{k=0}^{n-l-1} {\l \choose   n-l-1-k }^2
                  {-2l-2-\l \choose k } (-1)^k
\ee
Using now the Fourier transformations \pcite{Gelf}
\begar
 F[ \frac{\ln\frac{\q\,^2}{\m^2}}{\q\,^2}] &=& -\frac{\g+\ln \m r}{2\pi r}\\
 F[ \frac{\ln^2\frac{\q\,^2}{\m^2}}{\q\,^2}] &=& \frac{1}{2\pi r} [\frac{\pi^2}{6} + 2(\g+\ln \m r)^2]
\ea
one arrives immediately at eq.\pref{dMg} and \pref{eln2}, respectively.


\end{appendix}

\end{document}